\documentclass[aps,
prd,
english,
nofootinbib,
reprint]{revtex4-2}

\usepackage{amsfonts,amsmath,amssymb}
\usepackage{mathrsfs}
\usepackage{graphicx}
\usepackage[utf8]{inputenc}
\usepackage{hyperref}
\hypersetup{hidelinks}
\usepackage{babel}
\usepackage[table]{xcolor}
\usepackage{booktabs}
\usepackage{bm}

\usepackage{lipsum}

\newcommand{\dd}{\mathrm{d}}
\newcommand{\ii}{\mathrm{i}}

\newcommand\be{\begin{equation}}
\newcommand\ee{\end{equation}}
\newcommand\bea{\begin{eqnarray}}
\newcommand\eea{\end{eqnarray}}

\newcommand{\Mbar}{\overline{\mathcal M}}
\newcommand{\Vcal}{\mathcal V}
\newcommand{\Kcal}{\mathcal K}
\newcommand{\Jcal}{\mathscr J}
\newcommand{\deltacal}{\mathscr \delta}
\newcommand{\Dcal}{\mathscr D}
\newcommand{\Rcal}{\mathcal R}
\newcommand{\Ecal}{\mathcal E}
\newcommand{\WP}{\mathrm{WP}}

\begin{document}
\begin{flushright}
\phantom{
{\tt arXiv:2026.$\_\_\_\_$}
}
\end{flushright}

{\flushleft\vskip-1.4cm\vbox{\includegraphics[width=1.15in]{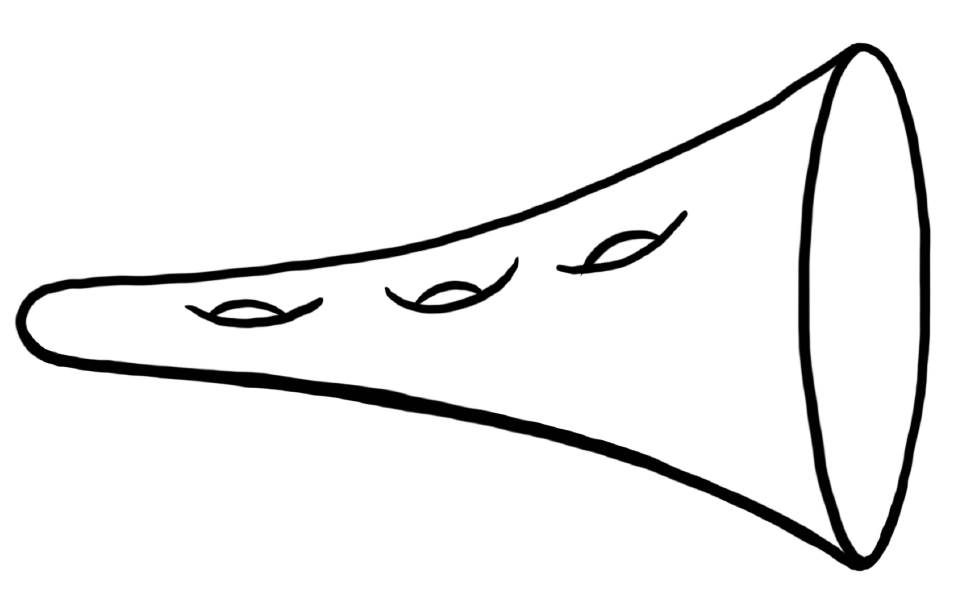}}}

\title{A boundary-addition 
identity for Weil-Petersson volumes\\ and their supersymmetric generalizations}
\author{Clifford V. Johnson}
\email{cliffordjohnson@ucsb.edu}

\affiliation{Department of Physics, Broida Hall,   University of California, 
Santa Barbara, CA 93106, U.S.A.}

\begin{abstract}
We consider $V_{g,n}(\{b_i\})$, the Weil--Petersson volumes of the moduli space of Riemann surfaces of genus~$g$ with $n$ geodesic boundaries of lengths $b_i$, $(i{=}1,\hdots,n)$, and various ${\cal N}{\geq}1$ supersymmetric generalizations of them, using a framework within which  they are specializations of a larger family of quantities naturally governed by the integrable KdV hierarchy. An exact boundary-addition identity is derived that directly relates $V_{g,n+1}(\{b_1,b_2,\hdots,b_n,b\})$, for any $b$, to the corresponding $n$-boundary KdV data underlying $V_{g,n}(\{b_1,b_2,\hdots,b_n\})$. For ordinary Weil--Petersson volumes, expansion about the special removable-cone value $b{=}2\pi\ii$ reproduces  three known results of Do and Norbury, organizing them as the first levels of a complete hierarchy. The higher levels are naturally built from additional KdV data, which we interpret geometrically  in intersection theory as a tower of $\kappa$-decorated volumes.  Several other recursive relations in the literature are illuminated by the identity, and various new ones are derived. The cases with ${\cal N}{=}1,2$ and (small) ${\cal N}{=}4$ supersymmetry are also covered by the identity, and we exhibit some of the striking special features and results arising in each case. Among many applications presented, we use the identity to uncover the intersection theory descriptions of ${\cal N}{=}2$ and small ${\cal N}{=}4$ Weil-Petersson volumes.
\end{abstract}


\maketitle

\section{Introduction}
\label{sec:introduction}
\subsection{Background}
Weil--Petersson (WP) volumes~\cite{Wolpert2010WeilPetersson,Do:2011WPsurvey,Do:2008Tourist} are key quantities that live at the intersection of several  fields of inquiry in both mathematics and physics, including  geometry, topology, string theory, random matrix models, black holes, and  quantum gravity. They concern the   geometry of moduli spaces of bordered Riemann surfaces that show up naturally in several physics problems. Integration over such  moduli spaces is a basic ingredient of perturbative string amplitudes, while in two-dimensional Jackiw-Teitelboim (JT) gravity~\cite{Jackiw:1984je,Teitelboim:1983ux}, emerging in the near-horizon geometry of higher dimensional extremal black holes at low temperature,  the volumes live at the heart of the organization of topological expansion of the correlators, in both the gravitational path integral and random matrix model approaches~\cite{Saad:2019lba}.  Natural supersymmetric generalizations, where the study is of the volume of moduli space of (suitably defined) super-Riemann surfaces, have also been studied considerably~\cite{Engelsoy:2016xyb,Stanford:2017thb,Stanford:2019vob,Norbury:2020vyi,Heydeman:2020hhw,Turiaci:2023jfa,Johnson:2023ofr,Heydeman:2025vcc,Ahmed:2025lxe,Johnson:2025oty},
with associated connections to various intersection theory problems, and supersymmetric JT gravity (and higher dimensional supersymmetric black holes).

A striking feature of the volumes (starting with the  ordinary case) is that the complete set of $V_{g,n}(\{b_i\})$ can be generated recursively, as first shown in the work of Mirzakhani~\cite{Mirzakhani:2006fta} where decompositions of Riemann surfaces in terms of pairs-of-pants embeddings played a central role.  Relations to intersection theory, Virasoro constraints,  the KdV hierarchy, random matrix models and topological recursion have also been established~\cite{Wolpert:1983,Mirzakhani2007WPVolumesIntersection,MulaseSafnuk2008,Eynard:2007fixed}.  

There are of course other kinds of relations between volumes that don't directly follow from Mirzakhani's recursion. Sometimes they can reveal glimpses of  geometrical and organizational structures that are potentially instructive and useful.  An example is the observation of  Do and Norbury~\cite{do:2009weil}. They noticed that after analytically continuing one geodesic length to the ``removable-cone'' value $b{=}2\pi\ii$, there are  direct identities relating $V_{g,n+1}$ and $V_{g,n}$, at that value of~$b$, together with relations connecting properties of $V_{g,n}$ to the first two $b$-derivatives of $V_{g,n+1}$ at that point~(see equations~(\ref{eq:intro-DN-value})--(\ref{eq:intro-DN-second})). They   constitute a set of ``boundary-addition'' identities, at fixed genus, valid at the cone point.

It is natural to wonder if it is possible  to find a boundary-addition identity that works at {\it any} value of~$b$. This paper presents such an identity, and in fact it works not just for Weil-Petersson volumes, but a wide range of generalizations, including various supersymmetric cases we will discuss in detail.
Specializing to the case of ordinary Weil-Petersson for the sake of introduction (although the general form~(\ref{eq:master-physical}) is very similar), it is:
 \begin{equation}
\begin{aligned}
 V_{g,n+1}(\bm b,b)
 ={}&\sum_{i=1}^{n}\int_0^{b_i}y\,
 V_{g,n}(b_1,\ldots,y,\ldots,b_n)\,\dd y
 \\
&\hskip1.6cm+\left.\Jcal_b\Vcal_{g,n}(\bm b;x)\right|_{\WP,x=0}\ .
\end{aligned}
\label{eq:intro-WP-identity}
\end{equation}
The added boundary has length $b$, and $\bm b$ is the set~$\{ b_i\}$, $i{=}1,\hdots,n$ of the $n$ previously present  boundaries. The use of regular Roman $V$ in $V_{g,n}$, {\it etc.} means the specific Weil-Petersson volume polynomial while the calligraphic~$\Vcal_{g,n}$ with $x$ dependence denotes an ``off-shell'' object that is naturally acted on by the simple operator $\Jcal_b$ (through that $x$-dependence), and the result is put ``on-shell'' at the end (by setting $x{=}0$). This will all be explained  in a moment, but a  key point worth making immediately is that this {\it single} identity, when reorganized exactly about the removable cone point $b{=}2\pi\ii$ in powers of the natural variable $T{=}b^2+4\pi^2$, produces (from the first three orders), the Do-Norbury relations~(\ref{eq:intro-DN-value})--(\ref{eq:intro-DN-second}), while the higher powers readily produce a hierarchy of new additional relations at that cone point. We will show that, in retrospect, the techniques of intersection theory can be used to explain this structure.

Many other interesting relationships among Weil--Petersson volumes also emerge rather straightforwardly from the identity, as we will see from some examples we explore.  Very recently,  relations for genus zero volumes with both cusps and finite geodesic boundaries were presented by Ancona and Gayet~\cite{ancona2026recursionvolumemodulispace}, generalizing the classic all-cusp results of Zograf~\cite{Zograf1993WeilPetersson}.  This setting provides a particularly useful laboratory for the boundary-addition identity.  The identity gives extremely compact alternative forms of these results and makes their underlying Bessel function structure especially transparent.

The same construction extends readily to arbitrary genus. The all-cusp generating functions for volumes and their multi-finite-boundary generalizations follow directly from the form of the boundary-addition operation, recovering and extending older work of  Zagraf~\cite{Zograf:1998ur,ZografP.G.1998Wvol}, and Zograf and Manin~\cite{Manin:1999wj}.  This also connects nicely with recent work of Hide and Thomas~\cite{HideThomas2025} on the large-number-of-cusps asymptotics of Weil--Petersson volumes. The universal Bessel factors appearing in their result emerge naturally from our form of the finite-boundary insertion result, together with the properties of a  critical point~\cite{Kaufmann:1996aa} that appears when inverting the Zograf representation of the genus zero all-cusp generating function.

There are many generalizations of all this to the ${\cal N}{=}1$, ${\cal N}{=}2$, and small ${\cal N}{=}4$ supersymmetric cases as well, following from tailoring the same master identity as appropriate. The resulting identities again allow key known results to be obtained in new (sometimes much more straightforward)  ways, and enable several new results to be uncovered.
 The identity even hints at aspects of the structure of the underlying intersection theory descriptions of the ${\cal N}{=}2$ and (small) ${\cal N}{=}4$ Weil-Petersson volumes (which currently do not seem to be available in the literature). Following the clues (and one contained in our earlier work~\cite{Johnson:2026plw}), and constraining the structure with a simple application of the identity, we uncover the  full intersection theory descriptions.

With all that (hopefully) serving as motivation, the next subsection explains a bit more about what underlies the identity, and how simply it works.

\subsection{Integrable and random matrix origins}
The identity comes from recasting a key part of the recent more general work of ref.~\cite{Johnson:2026twg} directly into the language of volumes. 
In the large $N$ random matrix model formulation used there ($N$ sets the size of the matrices) the central result is that quite general formulae can be written for  the natural observables of the model, the energy correlators, in terms of a single function $u_0(x)$ and its $x$-derivatives.  Here,  $u_0(x)$ is the leading (``dispersionless'') piece of the function $u(x)$ that plays a central role in the model, and which satisfies the integrable Korteweg-de Vries (KdV) hierarchy of flows, as shall be recalled later. The ``off--shell'' results for a general $u_0(x)$ apply to a large class of models, and going ``on-shell'' to define a particular model is done by specifying the function   $u_0(x)$ determined by the specific random matrix potential. In particular, what  determine the physics is the value of $u_0$ and its derivatives at a certain endpoint value of $x$, denoted $\mu$. (In the case of the identity given above in~(\ref{eq:intro-WP-identity}), $\mu{=}0$, but it can be more general, as we shall see.)
 
 The key observation~\cite{Johnson:2026twg} is that the boundary operator that adds a new energy to a  correlator  has a strikingly simple form when everything is written in terms of $u_0(x)$, making it straightforward to write general formulae for the correlators with any number of energies.
For a certain specific choice of random matrix potential, {\it i.e.,} specific choice of $u_0(x)$, the energy correlators are equivalent (after a slight repackaging and then a  Laplace transform) to   Weil-Petersson volumes, where $n$ distinct energies are exchanged for $n$ geodesic boundaries, and the order $N^{2-2g}$ contribution to the energy correlator in the large $N$ expansion gives volumes at genus $g$.
(The generality of the formulae means that results for a large class of  supersymmetric and other generalizations, as well as other models,  follow by simply changing the function~$u_0(x)$.)

Two simple examples of  the off-shell volumes from that work are:
\begin{align}
 \Vcal_{0,3}(x)=-\frac12u_0'(x)\ ,\quad
 \Vcal_{1,1}(b_1;x)=\frac{u_0''(x)}{24u_0'(x)}-\frac{u_0'(x)}{96}b_1^2\ .
 \label{eq:intro-V03-V11-off}
\end{align}
For  ordinary Weil-Petersson, $u_0(x)$ satisfies the equation: $ \frac{\sqrt{u_0}}{2\pi}\,
I_1\!\left(2\pi\sqrt{u_0}\right)+x{=}0$, and  $\mu$, the on-shell endpoint value of $x$,  is $0$ for this case, giving:
\begin{equation}
\begin{aligned}
 u_0(0)&=0\ ,\qquad u_0^{\prime}(0)=-2\ ,\quad 
 u_0^{\prime\prime}(0)=-4\pi^2\ ,\\
 u_0^{\prime\prime\prime}(0)&=-20\pi^4\ ,\qquad
 u_0^{\prime\prime\prime\prime}(0)=-\frac{488}{3}\pi^6\ .
\end{aligned}
 \label{eq:intro-JT-low-jet}
\end{equation}
The expressions in~\eqref{eq:intro-V03-V11-off} 
specialize to the familiar results:
\begin{align}
 V_{0,3}=1\ ,\quad
 V_{1,1}(b_1)=\frac{b_1^2+4\pi^2}{48}\ .
 \label{eq:intro-V03-V11-WP}
\end{align}

 Translating the key outputs of ref.~\cite{Johnson:2026twg}'s boundary-adding formalism through the Laplace transform to length space, results in the powerful general boundary-addition identity for volumes, and equation~(\ref{eq:intro-WP-identity}) is the specialization to ordinary Weil-Petersson volumes.
 So, in the last term, the calligraphic $\Vcal_{g,n}$ denotes the off-shell proto-volume still  in its natural form, built from the KdV function $u_0(x)$ and its   derivatives $u_0',u_0'',\ldots,$ {\it i.e.,} its ``jet'' data.

 All we now need to know is how   the operator $\Jcal_b$ acts on the $u_0(x)$ and its derivatives.\footnote{The $\Jcal$ stands for ``jet'', since it acts naturally on $u_0(x)$ and its derivatives. Then the subscript $b$ specifies   the length of the boundary being added.} This is straightforward to work out (the general case will be discussed with the full derivation in Section~\ref{sec:boundary-addition-derivation}), and a few examples are given below:
\begin{align}
 \Jcal_bu_0&=u_0'\ ,
 \nonumber\\
 \Jcal_bu_0'&=u_0''-\frac{b^2}{4}(u_0')^2\ ,
 \nonumber\\
 \Jcal_bu_0''&=u_0'''-\frac{3b^2}{4}u_0'u_0''
 +\frac{b^4}{32}(u_0')^3\ .
 \label{eq:intro-Jupp}
\end{align}

\subsection{How it works, by example}

\label{sec:WP-examples}

Let's do two  low-genus examples to show how the identity~\eqref{eq:intro-WP-identity} works. A key point here is that it will work for {\it any}  genus the same way, even though specific expressions might get more involved. Starting with making~$V_{0,4}$,  using the on-shell $V_{0,3}$ given in~\eqref{eq:intro-V03-V11-WP}, the integral term in the identity~(\ref{eq:intro-WP-identity}) yields $\frac12\sum_{i=1}^{3}b_i^2$. The next term asks us to act on the  off-shell expression for $\Vcal_{0,3}$ in~\eqref{eq:intro-V03-V11-off} with the operation from the second line of~\eqref{eq:intro-Jupp}, and   then using the endpoint data~\eqref{eq:intro-JT-low-jet}. This gives the residual term:
\begin{equation}
 \left.\Jcal_b\Vcal_{0,3}\right|_{\WP}
 =\frac{b^2+4\pi^2}{2}\ ,
 \label{eq:intro-03example}
\end{equation} 
and adding them  reproduces the known result:
\begin{equation}
 V_{0,4}=2\pi^2+\frac12\sum_{i=1}^{4}b_i^2\ .
\end{equation}
Slightly less trivially,  acting with $\Jcal_b$ on the off-shell genus one expression in equation~\eqref{eq:intro-V03-V11-off}, and going on-shell gives, after a little  algebra:
\begin{equation}
 \left.\Jcal_b\Vcal_{1,1}(b_1)\right|_{\WP}
 =\frac{T}{2}V_{1,1}(b_1)+\frac{T^2}{192},
 \qquad T=b^2+4\pi^2\ ,
 \label{eq:intro-J11-residual}
\end{equation}
where $V_{1,1}$ is the Weil-Petersson volume in~\eqref{eq:intro-V03-V11-WP}.  Equation~\eqref{eq:intro-WP-identity} therefore gives the known result:
\begin{align}
 V_{1,2}(b_1,b)
 &=\int_0^{b_1}yV_{1,1}(y)\dd y
 +\frac{T}{2}V_{1,1}(b_1)+\frac{T^2}{192}\\
 &=\frac{(4\pi^2+b_1^2+b^2)(12\pi^2+b_1^2+b^2)}{192}\ .
 \label{eq:intro-11example}
\end{align}
It really is that simple. Once we proceed to the  supersymmetric generalizations, there will  be additional special features that emerge, sometimes making things even simpler. For example, the fact that one can readily write closed form expressions  for all~$n$ at a given $g$ for $V_{g,n}$ in the ${\cal N}{=}1$ supersymmetric case (observed for low genus in ref.~\cite{Norbury:2020vyi} and extended to all genus in ref.~\cite{Johnson:2026twg}) will become very clear presently.

\subsection{Back to the removable cone point}

Turning to the  distinguished ``removable cone'' value $b{=}2\pi \ii$, something special happens.\footnote{\label{fn:conical}
For a
real geodesic boundary of length $b$ the holonomy is hyperbolic, or
boost-like, and obeys
\(
 |\operatorname{Tr}M|=2\cosh\frac b2\ ,
\) where  $M$ is an $SL(2,\mathbb R)$ representative of the
$PSL(2,\mathbb R)$ holonomy.  
Analytically continuing $b=i\theta$ gives the elliptic, rotation-like
holonomy of a cone point of angle $\theta$,
\( |\operatorname{Tr}M|
 =2\left|\cos\frac\theta2\right|.
\)
At $\theta{=}2\pi$ the holonomy is trivial in $PSL(2,\mathbb R)$, the
angular deficit vanishes, and the cone singularity is removable~\cite{TanWongZhang2006,do:2009weil,Wolpert2010WeilPetersson}.}  Notice from the examples~(\ref{eq:intro-03example}) and~(\ref{eq:intro-J11-residual}), that the result {\it vanishes} when $b{=}2\pi\ii$. In fact we will prove in Section~\ref{sec:removable-cone-point} that, quite generally for the ordinary WP data:
\begin{equation}
 \left.\Jcal_{2\pi\ii}\Vcal_{g,n}\right|_{\WP}=0
 \qquad (2g-2+n>0)\ .
 \label{eq:intro-residual-zero}
\end{equation}
This means that  equation~\eqref{eq:intro-WP-identity} reduces to the Do--Norbury identity~\cite{do:2009weil} at that point:
\begin{equation}
 V_{g,n+1}(\bm b,2\pi\ii)
 =\sum_i\int_0^{b_i}yV_{g,n}(\ldots,y,\ldots)\dd y\ .
 \label{eq:intro-DN-value}
\end{equation}
But now we can simply reorganize the exact identity about this point in the natural parameter
that has already appeared, $T{=}b^2{+}4\pi^2$. Experimenting with the examples given  
 yields at the next two orders the first two derivative identities of ref.~\cite{do:2009weil}:
\begin{widetext}
\begin{align}
 \left.\frac{\partial}{\partial b}V_{g,n+1}(\bm b,b)\right|_{b=2\pi\ii}
 &\!\!=2\pi\ii(2g-2+n)V_{g,n}(\bm b)\ ,
 \label{eq:intro-DN-first}\\
 \left.\frac{\partial^2}{\partial b^2}V_{g,n+1}(\bm b,b)\right|_{b=2\pi\ii}
 &\!\!=\left[\sum_i b_i\frac{\partial}{\partial b_i}-(4g-4+n)\right]V_{g,n}(\bm b)\ ,
 \label{eq:intro-DN-second}
\end{align}
 \end{widetext}
which will be  explicitly proven for all $g,n$ in Section~\ref{sec:removable-cone-point}. How the structures on the right hand side appear will be very natural once the operator $\Jcal_b$ is examined order by order. In the first derivative case it simply measures the overall degree of $\Vcal_{g,n}$ (and hence  $V_{g,n}$) through its~$u_0(x)$ content, and at second derivative order it measures a scaling dimension.

It should be clear that the exact expansion in~$T$ does not stop there, and our identity will continue to produce a tower of terms, generalizing the basic cone-point relations by constructing precise results for higher derivatives of $V_{g,n+1}$.\footnote{We will observe that, in intersection theory terms, the hierarchy is really a  tower of $\kappa$-class decorated volumes.} Of course, at a given $(g,n)$ the tower is finite since Weil-Petersson volumes are polynomials in the (squared) boundary lengths, and so only a finite number of derivatives are needed to  determine the complete dependence on the added boundary $b$. A key point (answering a question posed in ref.~\cite{do:2009weil}, which anticipated the possibility of higher derivative identities) is that eventually (at high enough genus) it becomes impossible to write those higher derivatives entirely in terms of ordinary geometric (``on-shell'') data about $V_{g,n}$. The full off-shell $\Vcal_{g,n}$ is really the natural object, and that is of course what the complete identity is written in terms of, for arbitrary values of the parameters.

\subsection{Outline}

Having illustrated the boundary-addition identity and some of its
consequences, we next derive it in its general form and then explore
what it reveals in several different backgrounds.
{\bf Section~\ref{sec:boundary-addition-derivation}} gives the derivation of
the universal  identity, starting with the simple form found in ref.~\cite{Johnson:2026twg} for the boundary adding operator  in energy space.

{\bf Section~\ref{sec:boundary-addition-WP}} specializes the result to
ordinary Weil--Petersson volumes.  There we recover the
Do--Norbury relations at the removable-cone point $b{=}2\pi\ii$, and extend them to a
full exact  hierarchy of relations in an expansion about that point in powers of $T{=}b^2{+}4\pi^2$. Using the intersection theory framework, we relate the resulting KdV
data in those relations   $\kappa$-decorated volume data.  We explain why ordinary geometric volume data about $V_{g,n}$ will eventually not be enough to reconstruct $V_{g,n+1}$ and where the  KdV data (or $\kappa$-decorated volume data) begins to be needed. Connections to some of the other various other known relations in the literature are also explored. We interpret the results in terms of intersection theory.
Several compact expressions for the generating functions of $V_{g,n}({\bm b},{\bm 0})$, where ${\bm b}$ is a set of $k$ finite lengths $b_i$, and~${\bm 0}$ is a set of $(n-k)$ cusps,  are derived in the later parts of this section. Some applications of them are explored.

{\bf Sections~\ref{sec:boundary-addition-WP-N1}--\ref{sec:boundary-addition-WP-N4}}
treat the ${\cal N}=1,2$, and small ${\cal N}=4$ generalizations.
The ${\cal N}=1$ theory is qualitatively different because the underlying random matrix model is hard-edged, which results in the special situation that the function $u_0(x)$ is identically zero. This results in certain remarkable simplifications for the identity that we explore, using ref.~\cite{Johnson:2026twg}'s general representation of the boundary operator (in energy space) in terms of the Gel'fand--Dikii resolvent  to derive $\Jcal_b$'s action, and show how to swiftly derive closed form formulae for $V_{g,n}$ at any $g$ (including results for Ramond insertions, following ref.~\cite{Johnson:2026jls}). We 
show how the removable-cone expansion matches an intersection theory interpretation in terms of Norbury's
$\Theta$-class insertion. 

The cases of ${\cal N}{=}2$ and ${\cal N}{=}4$, return to   soft edge behaviour (generically) and we  explore several analogous features to some of those seen for ordinary Weil-Petersson, but also find a richer set of phenomena.

A particularly exciting result is that for these extended supersymmetric cases we are able to  determine explicit intersection theory descriptions of the  Weil--Petersson volumes. Their ``gravitational dressings'' (the analogue of the Wolpert factor~\cite{Wolpert:1983} for the ordinary case, $\exp(2\pi^2\kappa_1)$) are  infinite towers of $\kappa$-classes, both of which we fully characterize.

{\bf Section~\ref{sec:discussion}} ends the paper with a brief discussion, gesturing at the many avenues of future research that suggest themselves.

\section{Derivation of the Boundary-Addition Identity}
\label{sec:boundary-addition-derivation}

We first state our conventions to be  used below.  We will use certain objects, denoted $R_k[u]$, that  are the  Gel'fand--Dikii polynomials~\cite{Gelfand:1975rn} in a function $u(x)$ and its derivatives,normalized so that:
\begin{equation}
R_k[u]=u^k+\hbox{derivative terms}\ .
\label{eq:GD-norm}
\end{equation}
The first two are $R_1[u]{=}u$ and $R_2[u]{=}u^2{-}\frac{\hbar^2}{3} u^{\prime\prime}$, but since we won't explicitly use anything beyond the leading purely polynomial piece, the normalization~(\ref{eq:GD-norm}) is really all we will need. Specific  models will satisfy a non-linear differential equation (called a ``string equation'') that can be neatly expressed in terms of the object:
\begin{equation}
\Rcal[u]\equiv \sum_{k=1}^\infty t_kR_k[u]+x\ ,
 \label{eq:string-equation-object}
\end{equation}
where the  $t_k$ parameterize particular ``multicritical'' couplings in the parent random matrix model potential~\cite{Brezin:1990rb,Gross:1990vs,Gross:1990aw,Douglas:1990ve,Douglas:1990dd}.  For example, the bosonic ordinary string equation is simply $\Rcal[u]{=}0$. With the above normalization, the function $u(x,\{t_k\})$ evolves under the KdV flows:\footnote{The factor $(k+1)$ corrects an error in versions 1 and 2 of ref.~\cite{Johnson:2026twg}.}
\begin{equation}
 (k+1)\frac{\partial u}{\partial t_k}=R_{k+1}'[u]\ ,
 \label{eq:kdv-flow}
\end{equation}
where a prime denotes an $x$-derivative. In the ``double-scaling'' limit of the random matrix model where all this structure emerges, the $1/N$ topological expansion parameter becomes a parameter denoted $\hbar$. Each $x$-derivative in the $R_k[u]$ comes with a power of $\hbar$, and $u(x)$ has an expansion in powers of $\hbar$:
\begin{equation}   u(x)=u_0(x)+\sum_{g=1}^\infty u_{2g}(x)\hbar^{2g}+\cdots
    \label{eq:u-expansion}
\end{equation} The KdV equation for the leading $u_0(x)$ ({\it i.e.,} ``dispersionless'') piece\footnote{``Dispersionless' refers to  the point of view where the KdV equations are generalized wave equations, with  $t_k$ being times.} is therefore:
\begin{equation}
    \frac{\partial u_0}{\partial t_k}=u_0^{k}u_0^\prime\ .
    \label{eq:kdv-dispersionless}
\end{equation}
The central physical quantity is the free energy of the model which is related to $u(x)$ according to $u(x){=}2\hbar^2 F''$, and $F$ is understood to be evaluated at the endpoint value $x{=}\mu$. So $F$ can be expressed as a genus expansion, and is useful to write it here in  ``off-shell'' form:
\begin{equation}
    F(x)=\sum_{g=0}^\infty F_g(x) \hbar^{2g-2}
\end{equation}
where each term can be written in terms of $u_0(x)$ and its derivatives (after recursively solving the defining string equation for $u(x)$). For example at genus one:
\begin{equation}
F_1(x) = -\frac{1}{24}\log u_0^\prime\ .
    \label{eq:F1-formula}
\end{equation}

The trick is to write everything in terms of $u_0(x)$ and its derivatives. Then one can take the loop operator technology of ref.~\cite{Gross:1990aw,Banks:1990df,Dijkgraaf:1991rs} and write it as an action on $u_0(x)$ and functions of it and its derivatives. The  result of ref.~\cite{Johnson:2026twg} is that  the  resulting loop operator  is quite simple:
\begin{equation}
 \deltacal_Eu_0=\frac12\frac{u_0'}{(u_0-E)^{\frac32}}\ .
 \label{eq:loop-u0}
\end{equation}
Its natural action defines the $n$-legged loop object ${\widetilde W}(E_1,\hdots,E_n)$, which in turn has a loop expansion, with coefficients:
\begin{equation}
    {\widetilde W}_{g,n}(E_1,\hdots,E_n)=(-1)^{n-1}\delta_{E_1}\cdots\delta_{E_n}\cdot F_g(u_0,u_0^\prime,\ldots)\ .
    \label{eq:loop-object}
\end{equation}
For example acting with~(\ref{eq:loop-u0}) on~(\ref{eq:F1-formula}) gives:
\begin{equation}
\widetilde W_{1,1}(E_1)
=
\frac{u_0'}{32(u_0-E_1)^{\frac52}}
-
\frac{u_0''}
{48\,u_0'(u_0-E_1)^{\frac32}} \ .
\label{eq:W11-tilde-off}
\end{equation}
Finally, the natural variables $z_i$ are defined {\it via:}
\begin{equation}
 z^2=u_0-E,
 \qquad
 z_i^2=u_0-E_i,
 \qquad i=1,\ldots,n, 
 \label{eq:z-defs}
\end{equation}
resulting in the natural objects:\footnote{The objects $W_{g,n}$ and the natural spectral variables $z_i$ are central in the topological recursion framework~\cite{Chekhov:2006vd,Eynard:2007kz}.}
\begin{equation}
 W_{g,n}(\{ z_i \},x)=\prod_{i=1}^n(-2z_i) {\widetilde W}_{g,n}(\{ E_i\},x)\ ,
 \label{eq:the-Ws}
 \end{equation}
where the prefactor is the Jacobian, and the $x$ dependence is made explicit to show that things remain off-shell until $u_0(x)$ and its derivatives are evaluated at the endpoint $x{=}\mu$.
General universal formulae for the  $W_{g,n}$, constructed in this simple way from successive action of~$\delta_E$, are presented and explored in ref.~\cite{Johnson:2026twg}. When $u_0(x)$ defines the data for Weil-Petersson volumes, they become the volumes $V_{g,n}(\{b_i\})$ after Laplace transform (to be recalled shortly). For example, $\Vcal_{1,1}$ of equation~(\ref{eq:intro-V03-V11-off}) came from converting~(\ref{eq:W11-tilde-off}) and Laplace transforming.

Now that the review of the key players of ref.~\cite{Johnson:2026twg} is over, we should prepare to go to length space. What we need to do is first translate the action of the boundary operator~(\ref{eq:loop-u0}) on  objects depending on $u_0(x)$ and $E$ to objects depending on $u_0(x)$ and $z_i$.
 Two effects must be   distinguished: The action on  $u_0(x)$ and its derivatives on the one hand,  and on the other, the action on the spectral coordinates $z_i$ (which contain $u_0(x)$)   associated to the boundaries already present.

At fixed old energies $E_i$, we get a change:
\begin{equation}
 \deltacal_Ez_i
 =\frac{1}{2z_i}\deltacal_Eu_0
 =\frac{u_0'}{4z^3z_i}.
 \label{eq:zi-change}
\end{equation}
In fact it is this basic coordinate action that is the origin of the boundary-integral term in the length space identity!
Now, recall that the correlator with an additional boundary  is generated by
\begin{equation}
 \widetilde W_{g,n+1}=-\deltacal_E\widetilde W_{g,n}\ .
\end{equation}
Separating the fixed-$z_i$  action on $u_0(x)$ from the coordinate change \eqref{eq:zi-change} gives:
\begin{equation}
W_{g,n+1}
 =2z\,\deltacal_E^{(u_0)}W_{g,n}
 +\frac{u_0'}{2z^2}\sum_{i=1}^{n}
 \left(\frac1{z_i}\frac{\partial}{\partial z_i}-\frac1{z_i^2}\right)W_{g,n}\ .
 \label{eq:W-boundary-addition}
\end{equation}
The second term is what results from the existing coordinates $z_i$ changing (because of their secret $u_0(x)$ content). Let us see what becomes of it when we go to length variables.

The Laplace transform defining the off-shell length-space object $\Vcal_{g,n}$ is:
\begin{equation}
 W_{g,n}(\bm z;x)
 =\int_0^\infty\prod_{i=1}^{n}
 \left[b_i\dd b_i\,e^{-b_i z_i}\right]
 \Vcal_{g,n}(\bm b;x)\ ,
 \label{eq:laplace-def}
\end{equation}
where ${\bm z}$ means the set of $z_i$ and ${\bm b}$ means the set of $b_i$, $i{=}1,\hdots,n$.
Notice now that if we define the transform:
\begin{equation}
 {\cal F}(z)=\int_0^\infty b\dd b\,e^{-bz}f(b)\ ,
 \label{eq:our-laplace}
\end{equation}
then:
\begin{equation}
 \mathcal L^{-1}\!\left[
 \left(\frac1z\frac{\partial}{\partial z}-\frac1{z^2}\right){\cal F}(z)
 \right](b)
 =-\int_0^b y f(y)\dd y\ ,
 \label{eq:laplace-elementary}
\end{equation}
and so our coordinate-changing piece of the boundary adding action is just a simple integral in length space.

To complete the story, if we  define the induced jet action on functions of $u_0(x)$ and derivatives as:
\begin{equation}
 \Jcal_b
 \equiv\mathcal L^{-1}_{z\to b}\left[2z\,\deltacal_E^{(u_0)}\right]\ ,
 \label{eq:Jb-def}
\end{equation}
applying equation~\eqref{eq:laplace-elementary} boundary by boundary to equation~\eqref{eq:W-boundary-addition} yields the master (off-shell) result:
\begin{widetext}
\begin{equation}
\begin{aligned}
 \Vcal_{g,n+1}(\bm b,b;x)
 =
 -\frac{u_0'(x)}{2}\sum_{i=1}^{n}
 \int_0^{b_i}y\,\Vcal_{g,n}(b_1,\ldots,y,\ldots,b_n;x)\dd y +{}\Jcal_b\Vcal_{g,n}(\bm b;x)\ .
\end{aligned}
\label{eq:master-offshell}
\end{equation}
This is the general~$b$ boundary-addition identity.  The integral is the Laplace image of the changes induced in the  spectral coordinates $z_i$ ($i=1,\hdots,n$), while the $\Jcal_b$ term is the part that acts non-trivially on the    KdV content. We will discuss some useful forms for it shortly.

First, let's note that in moving among models (ordinary, ${\cal N}{=}1$ supersymmetric, ${\cal N}{=}2$ supersymmetric,  {\it etc.,}) there are different conventions in the literature that amount to a renormalization of the $\hbar$s between models. It is prudent to write how equation~\eqref{eq:master-offshell}, in going on-shell, converts to the physical on-shell conventions for a given ``background'' or model.  Suppose a physical background obeys:
\begin{equation}
\left.\Vcal_{g,n}\right|_{\rm bg}=K_{g,n}V^{\rm phys}_{g,n}\ ,
 \label{eq:K-normalization}
\end{equation}
for numbers $K_{g,n}$ that will be specified when the need arises.
 Then equation~\eqref{eq:master-offshell} gives:
\begin{equation}
\begin{aligned}
 V^{\rm phys}_{g,n+1}
 ={}&-\frac{K_{g,n}}{K_{g,n+1}}\frac{u_0'(\mu)}2
 \sum_i\int_0^{b_i}yV^{\rm phys}_{g,n}(\ldots,y,\ldots)\dd y
 +\frac1{K_{g,n+1}}
 \left.\Jcal_b\Vcal_{g,n}\right|_{\rm bg}\ .
\end{aligned}
\label{eq:master-physical}
\end{equation}
The same off-shell  $\Vcal_{g,n}$ remains in the second (non-integral) term.  The model-dependent normalization factors enter only when the result is specialized to a physical on-shell volume. We will see how this works in examples to come.

\end{widetext}

There is a compact way of describing the action of the length space boundary
operator~(\ref{eq:Jb-def}) on the \emph{entire} $u_0(x)$  derivative tower at the endpoint, which will turn out to be very useful. Write $u_0(\mu)=E_0$ (it is the threshold energy of the leading spectral density in the random matrix model description) and 
define the function:
\begin{equation}
 U(s)=u_0(\mu+s)\ ,
 \qquad
 U(0)=E_0\ .
 \label{eq:softedge-U}
\end{equation}
Since the $\delta_E$ operator commutes with $x$-differentiation,
$\delta_E U(s)$  generates the action of $\delta_E$ on the
entire tower of endpoint derivatives:
\begin{equation}
 \delta_E U(s)
 =
 \sum_{m=0}^{\infty}\frac{s^m}{m!}
 \delta_E u_0^{(m)}(\mu)\ ,
\end{equation} {\it i.e.,} the loop action~(\ref{eq:loop-u0}) for $u_0(x)$ gives:
\begin{equation}
 \delta_EU(s)
 =
 \frac12\frac{U'(s)}{[U(s)-E]^{3/2}}.
 \label{eq:softedge-delta-U}
\end{equation}
and since $E=E_0-z^2$ we can convert this to
\begin{equation}
 2z\,\delta_EU(s)
 =
 \frac{zU'(s)}
 {[z^2+U(s)-E_0]^{3/2}}\ .
 \label{eq:softedge-zdelta-U}
\end{equation}
Since:
\begin{equation}
 \int_0^\infty b\,\dd b\,e^{-bz}J_0(ab)
 =
 \frac{z}{(z^2+a^2)^{3/2}}
 \label{eq:softedge-Bessel-Laplace}
\end{equation}
we get,  using~(\ref{eq:Jb-def}), the result:
\begin{equation}
 \sum_{m=0}^{\infty}\frac{s^m}{m!}\,
 \Jcal_bu_0^{(m)}(\mu)
 =
 U'(s)
 J_0\!\left(
 b\sqrt{U(s)-E_0}
 \right)\ .
 \label{eq:softedge-generating}
\end{equation}
This is the exact  action of
the boundary operator on all endpoint derivatives.  Using:
\begin{equation}
J_0(y)
=
\sum_{m=0}^{\infty}
\frac{(-1)^m}{(m!)^2}
\left(\frac{y^2}{4}\right)^m
=
1-\frac{y^2}{4}
+\frac{y^4}{64}
+\cdots ,
\end{equation}
expanding in $s$ gives
\begin{align}
 \Jcal_bu_0
 &=u_0'\ ,
 \nonumber\\
 \Jcal_bu_0'
 &=u_0''
 -\frac{b^2}{4}(u_0')^2\ ,
 \nonumber\\
 \Jcal_bu_0''
 &=u_0'''
 -\frac{3b^2}{4}u_0'u_0''
 +\frac{b^4}{32}(u_0')^3\ ,
 \nonumber\\
 \Jcal_bu_0'''
 &=u_0''''
 -b^2u_0'u_0'''
 -\frac{3b^2}{4}(u_0'')^2
 \nonumber\\
 &\hskip2.5cm+\frac{3b^4}{16}(u_0')^2u_0''
 -\frac{b^6}{384}(u_0')^4\ ,
 \label{eq:softedge-first-jet-actions}
\end{align}
reproducing the universal actions~(\ref{eq:intro-Jupp})  used earlier and adding one for good measure.

In a model with an endpoint that has  $u_0(\mu){=}0$, $\Jcal_b$ may be written directly in terms of the KdV flows, and this is useful to record here too.
In our conventions the energy-space loop operator  is:
\begin{equation}
 \delta_E
 =
 \sum_{k=0}^\infty
 \frac{2(k+1)C^{\rm loop}_{k+1}}
 {(-E)^{k+3/2}}
 \frac{\partial}{\partial t_k}\ ,
\end{equation}
where:
\begin{equation}
 C^{\rm loop}_k
 =
 (-1)^{k+1}
 \frac{(2k-1)!!}{2^{k+1}k!}\ .
\end{equation}
Acting on just $u_0(x)$ and using the dispersionless KdV flows~(\ref{eq:kdv-dispersionless}) yields the simple form~(\ref{eq:loop-u0}) given above. 

With $u_0(\mu){=}0$, we have $z^2=-E$, and so, with an eye on~(\ref{eq:Jb-def}), we write:
\begin{equation}
 2z\,\delta_E
 =
 \sum_{k=0}^\infty
 (-1)^k
 \frac{(2k+1)!!}{2^k k!}
 \frac{1}{z^{2k+2}}
 \frac{\partial}{\partial t_k}.
\end{equation}
Using our Laplace convention~(\ref{eq:our-laplace}) we have:
\begin{equation}
 \mathcal L^{-1}_{z\to b}
 \left[
 \frac{1}{z^{2k+2}}
 \right]
 =
 \frac{b^{2k}}{(2k+1)!}\ ,
\end{equation} and after a bit of algebra we get the simple representation in terms of the KdV
times $t_k$:
\begin{equation}
 \Jcal_b
 =\sum_{k=0}^\infty
 \frac{(-1)^k b^{2k}}{4^k(k!)^2}\frac{\partial}{\partial t_k}\ ,
 \label{eq:intro-Jb-t}
\end{equation}
which will be useful later.
Notice that while this starts life as an infinite sum, it truncates when acting on any fixed $m$th $x$-derivative of $u_0$, as we already saw in the examples~(\ref{eq:softedge-first-jet-actions}).
Indeed, using the dispersionless flow~(\ref{eq:kdv-dispersionless})
and  commutating $t_k$ and $x$ derivatives, we have:
\begin{equation}
 \frac{\partial}{\partial t_k}
 \left(
 \frac{\partial^m u_0}{\partial x^m}
 \right)
 =
 \frac{\partial^m}{\partial x^m}
 \left(u_0^k u_0'\right).
\end{equation}
After evaluating at $u_0{=}0$, this vanishes for $k{>}m$ (since after 
$m$ derivatives there'll be an explicit power of
$u_0$ remaining). 
So $\Jcal_b$ becomes a finite polynomial in $b^2$. This makes sense since the left hand side of the identity,  a Weil-Petersson volume, is itself a finite polynomial in~$b^2$.

\section{Ordinary Weil-Petersson Volumes}
\label{sec:boundary-addition-WP}

\subsection{Specialization to WP}
\label{sec:WP-specialization}

For the ordinary JT background, as already mentioned the leading string equation defining $u_0(x)$ is
\begin{equation}
 \frac{\sqrt{u_0}}{2\pi}\,
 I_1\!\left(2\pi\sqrt{u_0}\right)+x=0\ ,
 \label{eq:JT-leading-string}
\end{equation}
which amounts to the particular multicritical background
\begin{equation}
 t_k^{\WP}=\frac{\pi^{2k-2}}{2k!(k-1)!},
 \qquad k\geq1,
 \label{eq:JT-couplings}
\end{equation}
and the endpoint is at $x=\mu=0$, where $u_0(0)=0$.

 Differentiating~\eqref{eq:JT-leading-string}  gives
\begin{equation}
 u_0'(x)=-\frac{2}{I_0(2\pi\sqrt{u_0})}\ ,
 \label{eq:JT-uprime}
\end{equation}
which gives the endpoint value $u_0'(0)=-2$. Further differentiation  generates the additional endpoint derivatives already displayed  in equation~\eqref{eq:intro-JT-low-jet}.

There is no additional normalization factor between $\Vcal_{g,n}|_{\WP}$ and the ordinary WP volume, so we can set all the $K_{g,n}$ factors to unity, and for convenience, repeat the identity here:
\begin{equation}
\begin{aligned}
 V_{g,n+1}(\bm b,b)
 ={}&\sum_i\int_0^{b_i}yV_{g,n}(\ldots,y,\ldots)\dd y\\
 &\hskip1.5cm
 +\left.\Jcal_b\Vcal_{g,n}(\bm b;x)\right|_{\WP,x=0}\ .
\end{aligned}
\label{eq:WP-full-identity}
\end{equation}

As a first simple consequence, consider the genus zero cusp
specialization {\it i.e.} having all boundary lengths be zero. This is the setting for one of the earlier recursive relationships among Weil-Petersson volumes, due to Zograf~\cite{Zograf1993WeilPetersson}, who showed that  a generating function for them can be written, non-linearly, in terms of a Bessel function. That this follows from the Bessel form of the leading JT/WP string equation~(\ref{eq:JT-leading-string}) was noticed recently by Mertens and Turiaci~\cite{Mertens:2020hbs}.
Our boundary-addition identity gives a particularly direct way of seeing the relation, and it will set us up for later derivations of many more general new formulae in Sections~\ref{sec:WP-all-cusps-genus-zero} and~\ref{sec:WP-higher-genus-cusps}.

Set all of the existing boundary lengths $b_i$ to zero, and take the added
boundary to be a cusp as well, so $b{=}0$.  The integral term in
equation~\eqref{eq:WP-full-identity} then vanishes, while at the endpoint the
boundary operator acts particularly simply, as only the $k{=}0$ term survives in~(\ref{eq:intro-Jb-t}), and~$\partial_{t_0}{\equiv}\partial_x$, hence:
\begin{equation}
 \left.\Jcal_0 u_0^{(m)}\right|_{\WP}
 =
 u_0^{(m+1)}(0)\ .
 \label{eq:WP-cusp-J-action}
\end{equation}
Starting with $\Vcal_{0,3}{=}{-}u_0'/2$ from~(\ref{eq:intro-V03-V11-off}), repeated boundary addition therefore
gives:
\begin{equation}
 V_{0,n}(0,\ldots,0)
 =
 -\frac12 u_0^{(n-2)}(0)\ .
 \label{eq:Zograf-cusp-from-u0}
\end{equation}
So the complete sequence of genus zero cusp volumes is simply contained within the
Taylor expansion of the same off-shell function $u_0(x)$ appearing in
the leading string equation~(\ref{eq:JT-leading-string}). (This will have powerful consequences later.) Indeed, using the values given in~(\ref{eq:intro-JT-low-jet}), we get the first few of the sequence as:
\begin{equation}
V_{0,3}({\bf 0})=1\ ,\quad 
V_{0,4}({\bf 0})=2\pi^2\ ,\quad 
V_{0,5}({\bf 0})=10\pi^4\ ,\cdots\ .
    \label{eq:low-cusp-volumes}
\end{equation}

It is useful to write $u_0(x)$ as a generating function for the cusp volumes. Since $u_0(0){=}0$,
we have directly:
\begin{equation}
 -\frac12u_0(x)
 =
 \sum_{n=3}^{\infty}
 V_{0,n}(0,\ldots,0)\frac{x^{n-2}}{(n-2)!}\ .
 \label{eq:Zograf-u0-generating}
\end{equation}
The  string equation~\eqref{eq:JT-leading-string} can then be written
as:
\begin{equation}
 2\pi^2x
 =
 \sqrt{-\pi^2u_0(x)}\,
 J_1\!\left(2\sqrt{-\pi^2u_0(x)}\right),
 \label{eq:Zograf-Bessel}
\end{equation}
where we have used the continuation from $I_1$ to $J_1$, and  together
with equation~\eqref{eq:Zograf-u0-generating}, this is precisely the
Bessel-function inversion form of  Zograf's genus zero generating
function~\cite{Zograf1993WeilPetersson}. To make contact with the notation there,  define the variable\footnote{When used in this context of generating functions through this paper, the variable $z$ has nothing to do with the spectral coordinate $z$ introduced earlier when working in energy space.} $z{=}2\pi^2x$ and the generating function~$h(z)$ so that: \begin{equation}
 h'(z)
 =
 -\pi^2
 u_0\!\left(\frac{z}{2\pi^2}\right)\ ,
 \label{eq:all-cusp-hprime-u0}
\end{equation}
giving (a prime will denote a $z$-derivative for the next several steps where $z$ dependence is clear):
\begin{equation}
    h(z)
 =
\sum_{n=3}^{\infty}
 \frac{
 V_{0,n}(0,\ldots,0)
 }{
 (2\pi^2)^{n-3}(n-1)!
 }
 z^{n-1}\ ,
 \label{eq:all-cusp-h}
\end{equation}
and then the string equation is:
\begin{equation}
    z=\sqrt{h'}J_1(2\sqrt{h'})\ .
    \label{eq:zograf-inversion-form}
\end{equation}
With this way of writing things, the cuspy volumes obey a very non-linear recurrence relation. A direct way to see it is to rewrite~(\ref{eq:zograf-inversion-form})  as a differential equation for $h$, following ref.~\cite{Matone:1994pz}:\footnote{Differentiate~(\ref{eq:zograf-inversion-form})  with respect to $y{=}h'$ to discover, with the help of Bessel identities, that $\frac{\dd z}{\dd y}{=}J_0(2\sqrt y)$ and $y\frac{\dd^2z}{\dd y^2}{+}z{=}0$. Then the $y$-derivative of the  combination on the LHS of \eqref{eq:all-cusp-Zograf-differential} vanishes. So it is a conserved quantity. The initial condition shows that it vanishes at $y{=}0$, and hence for all~$y$.}
\begin{equation}
 z h''-h'
 -
 h''(zh'-h)=0\ .
 \label{eq:all-cusp-Zograf-differential}
\end{equation}
 Now simply substitute into this the expansion of $h(z)$ from \eqref{eq:all-cusp-h} and collect powers of $z$. Reading off the coefficient of $z^{n-2}$ gives:
\begin{equation}
 V_{0,n}
 =
 \frac{2\pi^2}{n-3}
 \sum_{\substack{r,s\geq3\\ r+s=n+2}}
 \binom{n-2}{r-3}
 (s-2)\,
 V_{0,r}V_{0,s}\ ,
 \quad n\geq4\ ,
 \label{eq:cusp-nonlinear}
\end{equation}
seeded by $V_{0,3}{=}1$.
For $n{=}4$,  $(r,s){=}(3,3)$ only, so:
\begin{align}
 V_{0,4}
 =
 2\pi^2
 \binom{2}{0}
 (3-2)\,
 V_{0,3}V_{0,3}
 =
 2\pi^2\ .
\end{align}
For $n{=}5$,   the only cases are
$(r,s){=}(3,4)$ and $(4,3)$ giving:
\begin{align}
 V_{0,5}
 &=
 \pi^2
 \left[
 \binom{3}{0}(4-2)\,
 V_{0,3}V_{0,4}
 +
 \binom{3}{1}(3-2)\,
 V_{0,4}V_{0,3}
 \right]
 \nonumber\\&=
 \pi^2
 \left[
 2(2\pi^2)+3(2\pi^2)
 \right]
 =
 10\pi^4\ .
\end{align}
 This is a fun way of recovering the volumes we got in equation~\eqref{eq:low-cusp-volumes}, but it hides the clear simplicity of the underlying structure!  The key point is that the  boundary-addition
identity~(\ref{eq:WP-full-identity}) itself is  simple and linear: Adding another cusp just produces one more
$x$-derivative on~$u_0(x)$.  The
 nonlinear recursion among the cusp volumes~(\ref{eq:cusp-nonlinear})  emerges only after
using the string equation for $u_0(x)$. In other words, putting the system on-shell can sometimes obscure the simplicity of the underlying organizational structure. This is really the  overarching theme  playing, sometimes loudly,  throughout  most of the work presented here and in ref.~\cite{Johnson:2026twg}.  

Now let us move on to study the full identity~(\ref{eq:WP-full-identity}). As explained in the preceding derivation, although $\Jcal_b$ starts
life as an infinite sum over KdV times, its action on any fixed
$x$-derivative of $u_0$ truncates at the endpoint $u_0{=}0$.  It is
useful to record what this looks like explicitly in the first few
cases.  Using the endpoint derivatives listed in
equations~\eqref{eq:softedge-first-jet-actions} with equation~(\ref{eq:intro-JT-low-jet}), we  get:
\begin{align}
 \left.\Jcal_bu_0\right|_{\WP}
 &=-2,\nonumber\\
 \left.\Jcal_bu_0'\right|_{\WP}
 &=-(b^2+4\pi^2),\nonumber\\
 \left.\Jcal_bu_0''\right|_{\WP}
 &=-\frac14(b^2+4\pi^2)(b^2+20\pi^2),\nonumber\\
 \left.\Jcal_bu_0'''\right|_{\WP}
 &=-\frac1{24}(b^2+4\pi^2)
 \left(b^4+68\pi^2b^2+976\pi^4\right).
 \label{eq:WP-first-finite-actions}
\end{align}
So even before making any special choice of $b$, the arbitrary-$b$
insertion has become a finite polynomial for each of the endpoint
derivatives entering a fixed-genus volume.  The first factor appearing
in the last three expressions will acquire a special interpretation
below.

This is also consistent with the familiar polynomial structure of the
ordinary Weil--Petersson volumes.  At fixed $(g,n)$, $V_{g,n}$ has
total degree $3g-3+n$ when each $b_i^2$ and $\pi^2$ is assigned
degree one.  The boundary-addition identity therefore maps the finite
endpoint data contained in $\Vcal_{g,n}$ (built from  $u_0$ derivatives) into a finite polynomial in
the new boundary length.  We will make more detailed use of this
grading below.

We studied some examples already in the Introduction, and we will do some more in due course.  Before doing so, it is useful to reorganize the full boundary-addition identity around the removable cone point $b{=}2\pi\ii$.  Introducing $T{=}b^2{+}4\pi^2$ gives an exact rewriting of the arbitrary-$b$ insertion as a finite polynomial in $T$ at fixed $(g,n)$.  The removable cone corresponds to the expansion's natural origin $T{=}0$, while the successive powers of $T$ organize the complete hierarchy away from that point.  The three identities of Do and Norbury~\cite{do:2009weil} will appear as the zeroth and next two orders in this exact organization, and higher orders will contain interesting extra information.  This organization will then allow us to compare how the geometrical information is organized when we return to further examples in Section~\ref{sec:WP-more-examples}.

\subsection{Exact $T$-expansion at the removable cone point}
\label{sec:removable-cone-point}
Let's now fully explore the special removable cone point $b{=}2\pi \ii$, the geometric significance of which was briefly recalled in footnote~\ref{fn:conical}.  
In the framework we're using, there's also a simplification occurring at this special point. At $b{=}2\pi \ii$, since:
\begin{equation}
 \frac{(-1)^k(2\pi \ii)^{2k}}{4^k(k!)^2}
 =\frac{\pi^{2k}}{(k!)^2}
 =2(k+1)t^{\WP}_{k+1}\ ,
\end{equation}
we see from  equation~\eqref{eq:intro-Jb-t} that our $\Jcal_b$ operator becomes:
\begin{equation}
 \Jcal_{2\pi\ii}=2\Dcal_{-1}\ ,
 \quad
 \text{where}\quad
 \Dcal_{-1}=\sum_{k=1}^{\infty}k t_k^{\WP}
 \frac{\partial}{\partial t_{k-1}}\ .
 \label{eq:Jstar-string}
\end{equation}

\noindent The leading string equation in this case is:\footnote{In our  normalization, the $L_{-1}$ Virasoro generator is
\begin{equation*}
 L_{-1}
 =\sum_{k=1}^{\infty}k t_k\frac{\partial}{\partial t_{k-1}}
 -\frac{x^2}{4\hbar^2}\ .
\end{equation*}
So  we see that $\Dcal_{-1}$  is its specialization to the WP/JT background $t_k=t_k^{\WP}$, with $x=0$.}
\begin{equation}
\Rcal_0[u_0]\equiv\sum_{k=1}^{\infty}t_ku_0^k+x=0\ ,
 \label{eq:string-equation}
 \end{equation}
Differentiating it and using equation~\eqref{eq:kdv-flow} gives
$  \Dcal_{-1}u_0{=}{-}1$.
So we can write:
\begin{equation}
 \Jcal_{2\pi\ii}u_0=-2,
 \qquad
 \Jcal_{2\pi\ii}u_0^{(m)}=0
 \quad(m\geq1).
 \label{eq:Jstar-jets-JT}
\end{equation}
The (stable) off-shell volumes $\Vcal_{g,n}$ depend entirely on the nontrivial derivatives $u_0^{(m)}$, and so the second   term in the master identity vanishes~\eqref{eq:WP-full-identity}, and we recover the  Do--Norbury  identity \eqref{eq:intro-DN-value}.

We can now reorganize the full arbitrary $b$ identity exactly about this point by 
substituting $b^2{=}T{-}4\pi^2$ into the expression (\ref{eq:intro-Jb-t}) for $\Jcal_b$ and resumming the resulting binomial expansion, giving:
\begin{align}
 \Jcal_b
 &=\sum_{k=0}^{\infty}
 \frac{(\pi^2-T/4)^k}{(k!)^2}
 \frac{\partial}{\partial t_k}\nonumber\\
 &=\sum_{m=0}^{\infty}\frac{(-T/4)^m}{m!}
 \sum_{j=0}^{\infty}
 \frac{\pi^{2j}}{j!(j+m)!}
 \frac{\partial}{\partial t_{j+m}}
 \nonumber\\
 &=\sum_{m=0}^{\infty}\frac{(-T/4)^m}{m!} {\cal B}_m
 \ .
 \label{eq:J-T-expansion}
\end{align}
It is important to note that this is an exact expansion of the boundary operator, rather than a small $T$
approximation.  When it acts on a  volume at some given genus, the series truncates
at finite order, so it gives the complete dependence on the new boundary
length $b$.
The coefficients are:
\begin{align}
 {\cal B}_m
 & =\sum_{j=0}^{\infty}\frac{\pi^{2j}}{j!(j+m)!}
 \frac{\partial}{\partial t_{j+m}},\nonumber\\
 & =2\sum_{k=1}^{\infty}
 \frac{k!}{(k+m-1)!}t_k^{\WP}
 \frac{\partial}{\partial t_{k+m-1}}\ .
 \label{eq:Bm}
\end{align}
The first two ${\cal B}_m$ are:
\begin{equation}
 {\cal B}_0=2\Dcal_{-1}\ ,
 \qquad
 {\cal B}_1=2E_t\big|_{\WP}\ ,
 \qquad
 E_t=\sum_{k=1}^{\infty}t_k\frac{\partial}{\partial t_k}\ .
 \label{eq:B01}
\end{equation}
Let us see what $E_t$ does.
Using the classical string equation~\eqref{eq:string-equation} and the KdV flows \eqref{eq:kdv-flow}, we have:
\begin{equation}
 E_tu_0=\sum_{k=1}^{\infty}t_k\frac{\partial u_0}{\partial t_k}
 =\sum_{k=1}^{\infty}t_ku_0^ku_0'
 =-xu_0'\ ,
\end{equation}
and so:
\begin{equation}
 E_tu_0^{(m)}
 =-xu_0^{(m+1)}-m u_0^{(m)}\ ,
 \label{eq:Eetee-weight}
\end{equation}
which means that
 at the endpoint $x{=}0$ ({\it i.e.,} on-shell)~$E_t$  assigns weight $-m$ to $u_0^{(m)}$. We can work out what the derivative weight of a (stable) $\Vcal_{g,n}$ volume must be from how they are constructed in ref.~\cite{Johnson:2026twg}.
The  genus expansion of the full string equation, which involves the Gel'fand-Dikii polynomials, gives $u_{2g}(x)$ differential weight $2g$.  So $F_g$, given that  $u_{2g}=2\partial_x^2F_g$, has weight $2g-2$.
 Each loop insertion by an action with $\delta_E$ in (\ref{eq:loop-u0}) raises the differential weight by one, and so the  differential weight of $\Vcal_{g,n}$ is consequently $2g-2+n$.
So multiplying by $-2$, as instructed by~(\ref{eq:B01}) and~(\ref{eq:Eetee-weight}),  gives us that the first nonzero $T$ term is $T(2g-2+n)V_{g,n}/2$. Since $\partial/\partial b=2b\partial/\partial T$, setting $b{=}2\pi\ii$ yields the cone-point identity~\eqref{eq:intro-DN-first}.

The next $T$ order is even more interesting. We must identify what the  $m=2$ operator in~\eqref{eq:Bm} measures about a volume.  We have:
\begin{equation}
 {\cal B}_2
 =2\sum_{k=1}^{\infty}\frac{t_k^{\WP}}{k+1}
 \frac{\partial}{\partial t_{k+1}}\ .
 \label{eq:B2-first}
\end{equation}
Note that the WP couplings obey:
\begin{equation}
 t_{k+1}^{\WP}
 =\frac{\pi^2}{k(k+1)}t_k^{\WP},
\end{equation}
so, after shifting the summation index we can write:
\begin{equation}
 {\cal B}_2
 =\frac{2}{\pi^2}
 \sum_{k=1}^{\infty}(k-1)t_k^{\WP}
 \frac{\partial}{\partial t_k}.
 \label{eq:B2-Euler}
\end{equation}
Aside from the overall factor, the  operator measures how the volume changes under scaling of the couplings according to $t_k^{\WP}\longrightarrow s^{k-1}t_k^{\WP}$,
since the infinitesimal generator of this scaling at $s{=}1$ is exactly what appears in the sum.  
This can be readily kept track of since the $t_k$ introduce $\pi^{2k-2}$ into the volumes, and hence we get a power $s^{k-1}$ for every such occurrence.  On the other hand, the powers of $\pi^2$ can change from term to term. But since the powers of the boundary lengths adjust to compensate, every term in the volume polynomial  can be given the same total overall grading  if the contributions of the $b_i$ polynomials are taken into account, counting powers appropriately. That grading is simply $3g-3+n$, the complex dimension of
$\mathcal M_{g,n}$.  This follows because geometrically, the Weil--Petersson volume is obtained
from the degree $d=3g-3+n$ part of the Weil--Petersson
 class, which contains~\cite{Wolpert:1983,Mirzakhani2007WPVolumesIntersection} a term proportional to
$2\pi^2\kappa_1$ and one proportional to
$\frac12 b_i^2\psi_i$.  Every term in $V_{g,n}$ therefore has total
degree $d$ when one counts each power of $\pi^2$ and each power of $b_i^2$
as degree one. The natural operator to add to get the grading contribution from the $b_i$  is $\frac12\sum_i^n b_i\frac{\partial}{\partial b_i}$, and so together we have:
\begin{widetext}
\begin{align}
 &\left.
 \sum_{k=1}^{\infty}(k-1)t_k^{\WP}
 \frac{\partial}{\partial t_k}
 \Vcal_{g,n}
 \right|_{\WP}\!\!+\frac12\sum_i^n b_i\frac{\partial}{\partial b_i}V_{g,n}\label{eq:weighted-Euler-WP}
=
 \left(3g-3+n\right)V_{g,n}\ .
\end{align}
Equations~\eqref{eq:B2-Euler} and \eqref{eq:weighted-Euler-WP} therefore give:
\begin{align}
 \left.{\cal B}_2\Vcal_{g,n}\right|_{\WP}
 &=\frac{2}{\pi^2}
 \left(3g-3+n-\frac12\sum_i^n b_i\frac{\partial}{\partial b_i}\right)V_{g,n}
\ .
 \label{eq:B2-DN}
\end{align}
So now we see that the 
the first two nonvanishing levels of the exact $T$-expansion are:
\begin{equation}
 \left.\Jcal_b\Vcal_{g,n}\right|_{\WP}
 =(2g-2+n)\frac{T}{2}V_{g,n}
 +\frac{T^2}{32}
 \left.{\cal B}_2\Vcal_{g,n}\right|_{\WP}
 +O(T^3)\ ,
 \label{eq:J-T-firstterms}
\end{equation}
where the ${\cal B}_0$ term vanishes on a stable volume and
${\cal B}_1{=}2E_t|_{\WP}$.  Since $T=b^2+4\pi^2$, the second $b$ derivative is:
\begin{equation}
 \left.\frac{\partial^2}{\partial b^2}\right|_{b=2\pi\ii}
 =2\frac{\partial}{\partial T}
 -16\pi^2\frac{\partial^2}{\partial T^2}.
\end{equation}
Applying this to equation~\eqref{eq:J-T-firstterms}, and using
equation~\eqref{eq:B2-DN}, gives:
\begin{align}
 \left.
 \frac{\partial^2}{\partial b^2}
 \Jcal_b\Vcal_{g,n}
 \right|_{b=2\pi\ii,\WP}
 &=
 (2g-2+n)V_{g,n}
 -\pi^2
 \left.{\cal B}_2\Vcal_{g,n}\right|_{\WP}
 =
 \left[\sum_i^n b_i\frac{\partial}{\partial b_i}-(4g-4+n)\right]V_{g,n}\ .
\end{align}
Again, the integral term in equation~\eqref{eq:WP-full-identity} is independent of  $b$, so we can equate  this  also to the second derivative of~$V_{g,n+1}$.  We therefore recover equation~\eqref{eq:intro-DN-second}, the second derivative cone-point identity of  Do and Norbury~\cite{do:2009weil}.\footnote{Meaning the  published identity. There's an extra factor in the ArXiv version.}

There is another exact way to see the same $T$-organization, which is a little less economical for ordinary WP than the coupling-space ${\cal B}_m$ construction above, but will be useful when  $u_0(\mu)$ is nonzero in the ${\cal N}=2$ and small ${\cal N}=4$ cases. Starting directly from the general soft-edge generating relation~\eqref{eq:softedge-generating}, define:
\begin{equation}
 \Delta(s)=U(s)-U(0)\ ,
 \qquad
 T=b^2+4\pi^2\ .
 \label{eq:softedge-T-Delta}
\end{equation}
For ordinary WP, $U(0)=0$, but it is useful to keep the endpoint-subtracted notation. Reorganizing the Bessel series gives:
\begin{equation}
 \sum_{m=0}^{\infty}\frac{s^m}{m!}\,
 \Jcal_bu_0^{(m)}(\mu)
 =
 \sum_{r=0}^{\infty}
 \frac{(-T/4)^r}{r!}\,
 U'(s)
 \frac{\Delta(s)^{r/2}}{\pi^r}
 I_r\!\left(2\pi\sqrt{\Delta(s)}\right)
 \ ,
 \label{eq:softedge-exact-T}
\end{equation}
\end{widetext}
where $I_r$ is the modified Bessel function of order $r$.  Since $\Delta(s)$ is of order $s$ and the $r$th term begins at order $\Delta(s)^r$, the coefficient of a fixed $s^m$ receives contributions only from $r\leq m$. So the action on each endpoint derivative, and hence on a fixed-genus volume, contains only finitely many powers of $T$.

For the ordinary WP background this representation also reproduces the first ${\cal B}_m$ actions very directly. The $r=0$ term gives:
\begin{equation}
 U'(s)I_0\!\left(2\pi\sqrt{U(s)}\right)=-2\ ,
 \label{eq:WP-Bessel-r0}
\end{equation}
using equation~\eqref{eq:JT-uprime}, and hence reproduces equation~\eqref{eq:Jstar-jets-JT}. For the next term, the string equation~\eqref{eq:JT-leading-string} gives
\begin{equation}
 \frac{\sqrt{U(s)}}{\pi}
 I_1\!\left(2\pi\sqrt{U(s)}\right)=-2s\ ,
 \label{eq:WP-Bessel-I1}
\end{equation}
so the $r=1$ contribution to equation~\eqref{eq:softedge-exact-T} is
\begin{equation}
 \left.
 \sum_{m=0}^{\infty}\frac{s^m}{m!}\,
 \Jcal_bu_0^{(m)}(\mu)
 \right|_{T^1}
 =
 \frac{T}{2}sU'(s)\ .
 \label{eq:WP-Bessel-r1}
\end{equation}
This is indeed the ${\cal B}_1=2E_t|_{\WP}$ result encoded in equation~\eqref{eq:Eetee-weight}. Similarly, using $I_2(x)=I_0(x)-2I_1(x)/x$, the next term is:
\begin{equation}
 \left.
 \sum_{m=0}^{\infty}\frac{s^m}{m!}\,
 \Jcal_bu_0^{(m)}(\mu)
 \right|_{T^2}
 =
 \frac{T^2}{16\pi^2}
 \left[sU'(s)-U(s)\right]\ ,
 \label{eq:WP-Bessel-r2}
\end{equation}
an alternative form of equation~\eqref{eq:B2-Euler}. The ${\cal B}_m$ description is a particularly simple explicit realization in terms of $t_k$ that  emerges when $u_0(\mu)=0$.  The more general exact Bessel $T$-expansion~\eqref{eq:softedge-exact-T} form will be the useful one when we later come to the non-zero $u_0(\mu)$ soft-edge cases.

\subsection{Intersection Theory Perspective}
\label{sec:intersection-theory-WP}
Our boundary-addition identity compares an $(n+1)$-boundary
volume with the corresponding $n$-boundary data.  This immediately
suggests the ``forgetful map'' apparatus, (see {\it e.g.,}
refs.~\cite{KockPsiClasses,Zvonkine2012} for reviews) which is centered around
the  map:
\begin{equation}
 \pi:\Mbar_{g,n+1}\longrightarrow\Mbar_{g,n},
\end{equation}
which forgets the $(n+1)$st marked point in going from the larger space to the smaller. Here, $\Mbar_{g,n}$ is the Deligne--Mumford compactified moduli space of stable genus $g$ Riemann surfaces with~$n$ marked points, where ``stable'' means $2g{-}2{+}n{>}0$.\footnote{This is different from the object ${\cal M}_{g,n}$ pertaining to surfaces with geodesic boundaries with finite lengths. The geodesic boundary-length dependence in the intersection theory treatment is encoded by the $\psi_i$-class insertions in the volume formula~\eqref{eq:intersection-volume}.} In dealing with forgetful maps on  this  space, there needs to be a stabilization process: If forgetting the point leaves an unstable  component ({\it i.e.} if it has ($g',n'$) such that $2g'{-}2{+}n'{\leq}0$), then that component is contracted.

Weil--Petersson volumes have an intersection-theoretic description in
terms of the  $\kappa$- and $\psi$-classes~\cite{Wolpert:1983,Mirzakhani2007WPVolumesIntersection},
so the standard rules  for how these classes behave (and relate to each other) under the
forgetful map provide a natural geometric language for tracing the
effects of the new boundary.  Wolpert and Mirzakhani's work give us:
\begin{equation}
 V_{g,n}(\bm b)
 =\int_{\Mbar_{g,n}}
 \exp\!\left(
 2\pi^2\kappa_1+\frac12\sum_i b_i^2\psi_i
 \right),
 \label{eq:intersection-volume}
\end{equation}
where it is understood that the top-degree component  is picked out, and recall that $\Mbar_{g,n}$ has complex dimension $3g{-}3{+}n$.  Let's additionally define $\kappa$-class insertions generalizing the volumes (see refs.~\cite{Kaufmann:1996aa,Liu:2007au}) {\it via}:
\begin{equation}
 \Kcal_{g,n}^{(j)}(\bm b)
 =\int_{\Mbar_{g,n}}\kappa_j
 \exp\!\left(
 2\pi^2\kappa_1+\frac12\sum_i b_i^2\psi_i
 \right).
 \label{eq:kappa-decorated}
\end{equation}
The standard forgetful-map relations are:
\begin{equation}
 \psi_i^{(n+1)}
 =
 \pi^*\psi_i^{(n)}+D_{i,n+1},
 \qquad
 \kappa_1^{(n+1)}
 =
 \pi^*\kappa_1^{(n)}+\psi_{n+1},
 \label{eq:forgetful-class-relations}
\end{equation}
where $\pi^*$ denotes the pullback.  Here $D_{i,n+1}$ represents the boundary
component  (or ``divisor'') on which the $i$th and $(n+1)$st marked points lie together
on a rational ``tail'' attached to the rest of the curve at a node.  This tail is basically a sphere with three points, $i$, $n+1$, and the node. (As is standard, we will use the same symbol $D_{i,n+1}$  for the boundary divisor and for its associated cohomology class.) After
the $(n+1)$st point is forgotten, this thrice-punctured component
becomes unstable (since it now has two points so $2g'-2+n'{=}0$), so it must be  contracted in the
stabilization step mentioned above.

Substituting~\eqref{eq:forgetful-class-relations} into the
intersection definition of~$V_{g,n+1}$ gives:
\begin{widetext}
\begin{equation}
\begin{aligned}
 V_{g,n+1}(\bm b,b)
 =\int_{\Mbar_{g,n+1}}
 &\pi^*\!\left[
 \exp\!\left(
 2\pi^2\kappa_1+\frac12\sum_i b_i^2\psi_i
 \right)\right]
 \exp\!\left[
 \frac{b^2+4\pi^2}{2}\psi_{n+1}
 \right]
 \prod_i
 \exp\!\left(\frac{b_i^2}{2}D_{i,n+1}\right).
 \label{eq:forgetful-before-resum}
\end{aligned}
\end{equation}
The other ingredient needed is the basic relation between $\kappa$-classes on $\Mbar_{g,n}$ and $\psi$-classes on $\Mbar_{g,n+1}$:
\begin{equation}
\pi_*\!\left(\psi_{n+1}^m\right)=\kappa_{m-1},
 \qquad m\geq1\ .
 \label{eq:psi-push-kappa}
\end{equation}
Expanding the middle exponential and using the relation gives a sum over the $\kappa$-class insertions we defined earlier:
\begin{equation}
 \sum_{m=1}^{\infty}
 \frac{(b^2+4\pi^2)^m}{2^m m!}
 \Kcal_{g,n}^{(m-1)}(\bm b)\ .
 \label{eq:forgetful-kappa-piece}
\end{equation}

Now turn to the terms involving the divisor $D_{i,n+1}$.  There are several simplification steps. First, note that on $D_{i,n+1}$, the $i$th and $n+1$st points lie on the aforementioned rational tail, {\it i.e.,} just a sphere with three punctures, which has no moduli. This means that  
$\pi$, when restricted to $D_{i,n+1}$ just gives back $\Mbar_{g,n}$, and moreover,  $\psi_{n+1}$ restricted there must vanish (the first Chern class can't give anything interesting there).  There's also a standard self-intersection result~\cite{KockPsiClasses} that is very useful: $D^2_{i,n+1}=-\pi^*\psi_i D_{i,n+1}$, and so in simplifying  we use:
\begin{equation}
 \left.\psi_{n+1}\right|_{D_{i,n+1}}=0\ ,
 \qquad
 D_{i,n+1}^{\,k}
 =
 (-1)^{k-1}
 (\pi^*\psi_i)^{k-1}D_{i,n+1}\ .
 \qquad k\geq1.
 \label{eq:forgetful-divisor-powers}
\end{equation}
Different $D_{i,n+1}$ do not contribute mixed products (since the forgettable point can't be in two places at once), so for each $i$,
the divisor series therefore reduces, after pushing forward and using $\pi_*(D_{i,n+1})=1$, to:
\begin{align}
 &e^{b_i^2\psi_i/2}
 \sum_{k=1}^{\infty}
 \frac{(b_i^2/2)^k}{k!}(-\psi_i)^{k-1}
 =
 \frac{e^{b_i^2\psi_i/2}-1}{\psi_i}
 =
 \int_0^{b_i}y\,e^{y^2\psi_i/2}\dd y\ ,
 \label{eq:forgetful-divisor-resum}
\end{align}
where in the first step the $k{=}0$ term is omitted since it does not contain the divisor. Inserting the last result into the remaining intersection integral we get:
\begin{equation}
 \int_0^{b_i}
 y\,V_{g,n}(b_1,\ldots,y,\ldots,b_n)\dd y\ .
\end{equation}
Combining the two contributions gives:
\begin{equation}
\begin{aligned}
 V_{g,n+1}(\bm b,b)
 ={}&\sum_i\int_0^{b_i}
 y\,V_{g,n}(b_1,\ldots,y,\ldots,b_n)\dd y
+\sum_{m=1}^{\infty}
 \frac{(b^2+4\pi^2)^m}{2^m m!}
 \Kcal_{g,n}^{(m-1)}(\bm b)\ .
\end{aligned}
\label{eq:forgetful-resummed}
\end{equation}
\end{widetext}
This has precisely the structure of our boundary-addition identity~(\ref{eq:WP-full-identity}) for Weil-Petersson volumes! The geometric origin of the two terms is now explicit in this language.  The boundary-divisor corrections induced by the need to  stabilize (those rational tails) produces
the integral term, while powers of
the new $\psi$-class generate the hierarchy of $\kappa$-class insertions.

Since $T=b^2+4\pi^2$, the removable cone is exactly the point where
all of the $\kappa$-insertion terms vanish, leaving the rational-tail
term.  
Comparing equations~\eqref{eq:J-T-expansion} and
\eqref{eq:forgetful-resummed} yields the  dictionary between geometry and KdV data:
\begin{equation}
 \Kcal_{g,n}^{(m-1)}(\bm b)
 =\frac{(-1)^m}{2^m}
 \left.{\cal B}_m\Vcal_{g,n}(\bm b)\right|_{\WP},
 \qquad m\geq1\ .
 \label{eq:kappa-KdV-dictionary}
\end{equation}
Put differently, our off-shell KdV formalism is naturally computing the full
$\kappa$-decorated volume series directly, examples of what are known as ``higher Weil-Petersson volumes'' in the literature~\cite{Kaufmann:1996aa,Liu:2007au}.\footnote{Connections of such objects to KdV are of course known, starting with the work of ref.~\cite{Manin:1999wj}, but this off-shell KdV approach, encapsulated in $\Jcal_b$, seems simple and natural. It would be interesting to explore the whole family using this language.}
Their appearance here will also make precise why the on-shell ({\it i.e.,} geometric) volume~$V_{g,n}$ is not
always enough to construct~$V_{g,n+1}$, as we will see  directly in the next Section.

It is worth making a last comment on the integral term. It seems to have two very different origins, but they are morally the same mechanism. The rational tail origin in intersection theory results from the newly added point coming near to one of the existing $n$ points, and we had to  handle that carefully in the pin-the-tail-on-the-donkey procedure that resulted in~(\ref{eq:forgetful-divisor-resum}). The boundary operator origin is because, again, adding the  point (through the action of $\delta_E^{(u_0)}$) affected the $z_i$ coordinates (see~(\ref{eq:zi-change})), which depend on $u_0(x)$. The common language here is that contact terms were generated when the point was added.

\subsection{More examples and the need for higher terms}
\label{sec:WP-more-examples}
We have already explored some simple examples in the Introduction, constructing $V_{0,4}$ and $V_{1,2}$. It is instructive to explore some more. But first note that the  genus one example in equation~\eqref{eq:intro-11example} is quite informative because the $\Jcal_b$ term contains two orders in powers of $T=b^2+4\pi^2$:
\begin{equation}
 \left.\Jcal_b\Vcal_{1,1}(b_1)\right|_{\WP}
 =\frac{T}{2}V_{1,1}(b_1)+\frac{T^2}{192}\ ,
 \label{eq:J11}
\end{equation}
so even though $V_{1,1}$ is only linear in $b_1^2$, the identity supplies the additional $b_1^4$ information needed to reconstruct the complete $V_{1,2}$. 
Our intersection theory explorations give us another perspective on these terms, looking at~(\ref{eq:kappa-KdV-dictionary}). We  use some standard intersection theory results (some of which are collected in Appendix~\ref{app:intersection-integrals} for convenience) in what follows. The ${\cal K}_{1,1}^{(0)}$ is just an insertion of $\kappa_0$ into the volume integral, giving a factor $(2g-2+n)V_{1,1}=V_{1,1}$, times the $T/2$, confirming the linear result. Meanwhile since the degree of $\kappa_1$ already saturates the dimension of the moduli space for this example, we have
 ${\cal K}_{1,1}^{(1)}
 =
 \int_{\overline{\mathcal M}_{1,1}}\kappa_1
 =
 \langle\tau_1\rangle_1
 =
 {1}/{24}$.
The corresponding $T^2$ contribution is then $ \frac{T^2}{2^2\,2!}\,{\cal K}_{1,1}^{(1)}
 {=}
 {T^2}/{192}$, 
in  agreement with equation~\eqref{eq:J11}.

In this low genus example, the additional $\kappa$-decorated information is still determined by the on-shell $V_{1,1}$ data through the alternative interpretation of equation~\eqref{eq:B2-DN}. So when the result is viewed as an expansion about the removable-cone point, it supplies the corresponding derivative data of $V_{1,2}$. Here the order of these polynomials is low enough for this information to reconstruct the complete $V_{1,2}$ from purely $V_{1,1}$ data. A natural question is whether this can persist indefinitely.

A useful illustration starts with the known genus two volume:
\begin{widetext}
\begin{align}
 V_{2,1}(b)={}&\frac{(b^2+4\pi^2)(b^2+12\pi^2)}{2211840}
 (5b^4+384\pi^2b^2+6960\pi^4)\ ,
 \label{eq:V21}
\end{align}
and our goal is to add a point to make $V_{2,2}$. With
$T=b_2^2+4\pi^2$, the identity gives:
\begin{equation}
 V_{2,2}=\int_0^{b_1}yV_{2,1}(y)\dd y+\frac32TV_{2,1}
 +T^2C_2+T^3C_3+T^4C_4+T^5C_5\ ,
 \label{eq:V22-expansion}
\end{equation}
where:
\begin{align}
 C_2&=\frac{8-b_1\partial_{b_1}}{32\pi^2}V_{2,1}
 =\frac{29b_1^6+1668\pi^2b_1^4+24336\pi^4b_1^2+83520\pi^6}{2211840},
 \nonumber\\
 C_3&=\frac{29b_1^4+808\pi^2b_1^2+4144\pi^4}{2211840},
 \qquad
 C_4=\frac{5b_1^2+52\pi^2}{1474560},
 \qquad
 C_5=\frac1{4423680}.
 \label{eq:V22-Cs}
\end{align}
However, in this  example the information in the terms $T^3$,
$T^4$, and $T^5$ is not actually needed in order to reconstruct the
complete answer. The first three orders ($T^0, T^1,T^2$) already suffice once
the symmetry of $V_{2,2}$ is imposed.
\end{widetext}

To see this, we can extend an argument  used in ref.~\cite{do:2009weil} at genus zero and one. Suppose that $V_{2,2}(b_1,b_2)$ and
$\widetilde V_{2,2}(b_1,b_2)$ are two symmetric candidate polynomials
in~$b_1^2$ and $b_2^2$ which have the same value and the same first two
derivatives at the cone point. Their difference,
\begin{equation}
 D(b_1,b_2)
 =
 V_{2,2}(b_1,b_2)-\widetilde V_{2,2}(b_1,b_2)\ ,
\end{equation}
therefore has a third-order zero at $b_2^2=-4\pi^2$. Since $V_{2,2}$ is polynomial in $b_2^2$, we must have:
\begin{equation}
 D(b_1,b_2)
 =
 (b_2^2+4\pi^2)^3Q(b_1^2,b_2^2)\ .
\end{equation}
But $D$ is symmetric under interchange of $b_1$ and $b_2$, so  also must have a third-order zero at $b_1^2=-4\pi^2$, so:
\begin{equation}
 D(b_1,b_2)
 =
 (b_1^2+4\pi^2)^3(b_2^2+4\pi^2)^3
 R(b_1^2,b_2^2)\ .
\end{equation}
On the other hand, the total degree of $V_{2,2}$ in the squared boundary lengths is $3g-3+n=5$. But $D$'s  prefactor of $(b_1^2+4\pi^2)^3(b_2^2+4\pi^2)^3$
already has total degree six in the squared boundary lengths, so  $D$  cannot exist Therefore the first three orders in $T$ of the boundary-addition identity (equivalently the three
cone-point identities) together with symmetry uniquely determine the
full $V_{2,2}$.

So, although the full boundary-addition identity directly supplies the
higher coefficients $C_3$, $C_4$, and $C_5$, which we know are beyond ordinary $V_{2,1}$ volume data, in this example they were
already fixed by the lower cone-point data and symmetry.

The preceding argument extends immediately to arbitrary genus and number of boundaries. Consider the problem of reconstructing $V_{g,n+1}$ from $V_{g,n}$ using the three cone-point identities. Suppose that two symmetric candidate polynomials in the squared boundary lengths have the same value and the same first two derivatives when the $(n+1)$th boundary is placed at the removable-cone point $
 b_{n+1}=2\pi \ii.$ 
Their difference $D(b_1,\ldots,b_{n+1})$ must then be divisible by
$ (b_{n+1}^2+4\pi^2)^3$, but 
since $D$ is symmetric in all of the boundary variables, we must have:
\begin{equation}
 D(b_1,\ldots,b_{n+1})
 =
 \prod_{i=1}^{n+1}(b_i^2+4\pi^2)^3\,
 R(b_1^2,\ldots,b_{n+1}^2)\  .
 \label{eq:cone-ambiguity-factor}
\end{equation}

On the other hand, the total degree of  $V_{g,n+1}$ in the squared boundary lengths is $3g{-}3{+}(n{+}1){=}3g{+}n{-}2$.
The prefactor in equation~\eqref{eq:cone-ambiguity-factor} has total degree of $3(n{+}1)$ and so a nonzero $D$ polynomial ambiguity can exist only when:
\begin{equation}
 3(n+1)\leq 3g+n-2\ ,\quad \text{\it i.e., }\quad 2n+5\leq 3g \ .
 \label{eq:cone-ambiguity-condition}
\end{equation}
In other words, whenever
$ 2n+5>3g$, the cone-point value and first two derivatives, together with symmetry, determine the complete $(n+1)$-boundary volume uniquely.

So for  our $g{=}2$ example, condition~\eqref{eq:cone-ambiguity-condition}  requires
$2n+5\leq6$,
which has no solution for $n\geq1$. So at genus two the three cone-point identities and symmetry determine $V_{2,n+1}$ uniquely for any number of starting boundaries.

It is at genus three that we get something new. If  we have $n$ such that $2n+5\leq9$, we need data beyond the cone-point identities. This begins with the simplest case  of  $V_{3,1}\to V_{3,2}$,  and also the case of $V_{3,2}\to V_{3,3}$, where there is a constant in $R$ that can't be fixed.
 By $V_{3,3}\to V_{3,4}$, however, there are enough boundaries that symmetry forces the right answer without appeal to the higher data.

So overall   the issue is controlled by the balance between genus and the number of boundaries. Increasing the genus makes an ambiguity more likely, while increasing the number of boundaries makes the cone-point identities, together with symmetry, more constraining. The full-$b$ identity becomes genuinely stronger than the three cone-point identities precisely in those cases satisfying equation~\eqref{eq:cone-ambiguity-condition}. The higher powers in $T=b^2+4\pi^2$ contain information which is not fixed by the  data obtainable from the ordinary   $V_{g,n}$  and symmetry alone.

\subsection{Some genus zero finite-boundary applications}
\label{sec:WP-all-cusps-genus-zero}

Let us apply  our boundary-addition identity to  some very recent work, perhaps as further illustration of the usefulness of the whole framework.  Ancona and Gayet~\cite{ancona2026recursionvolumemodulispace} have derived a nonlinear recursion for genus zero Weil--Petersson volumes with a mixture of finite geodesic boundaries and cusps,
generalizing the all-cusp recursion of Zograf~\cite{Zograf1993WeilPetersson} that we already studied in Section~\ref{sec:WP-specialization}. For illustration, we will focus initially on the case in which
one boundary has finite length $b$ and all the others are cusps.
The reason this case is especially simple from our point of view is that
the finite boundary can be added last.  Starting with the $(n-1)$-boundary
all-cusp volume we already computed in~(\ref{eq:Zograf-cusp-from-u0}), every integral term in equation~\eqref{eq:WP-full-identity}
vanishes, and  so we get:
\begin{equation}
 V_{0,n}(b,0,\ldots,0)
 =
 -\frac12
 \left.
 \Jcal_b u_0^{(n-3)}(x)
 \right|_{x=0}\ .
 \label{eq:all-cusp-q1-volume}
\end{equation}
Happily, we have worked out the soft-edge generating relation~\eqref{eq:softedge-generating}  for how $\Jcal_b$ acts on all derivatives of $u_0(x)$, and it immediately gives:
\begin{equation}
 \sum_{n=3}^{\infty}
 V_{0,n}(b,0,\ldots,0)
 \frac{s^{n-3}}{(n-3)!}
 =
 -\frac12u_0'(s)
 J_0\!\left(b\sqrt{u_0(s)}\right)\ .
 \label{eq:all-cusp-q1-u0}
\end{equation}
Equation~\eqref{eq:all-cusp-q1-u0} is  the  finite-boundary
generalization  of the all-cusp generating relation~(\ref{eq:Zograf-u0-generating}).\footnote{\label{fn:lowenstein}Up to an overall normalization, the appropriate specialization of
the general genus zero formulae presented in work of Lowenstein~\cite{Lowenstein:2024fji} also yields
this compact form in which $b$ enters {\it via} a Bessel function.  Building
on the earlier open--closed KdV boundary technology of
refs.~\cite{Gross:1990aw,Banks:1990df, Johnson:1992wr,Johnson:1993vk}, the method used there directly represents
a boundary insertion as an infinite combination of KdV flow insertions,
heroically resumming their contributions to obtain the closed form result.}

To compare with the generating functions used in refs.~\cite{ancona2026recursionvolumemodulispace,Zograf1993WeilPetersson}, we 
write their generating variable as~$z$, so as not to confuse with our $x$.  In terms of our physical volume normalization, the two
generating functions needed here are the original $h(z)$ of Zograf, that we have already introduced in~\eqref{eq:all-cusp-h} and the one-finite-boundary one $h_1(z)$, which is: 
\begin{align}
 h_1(z;b)
 &=
 \sum_{n=3}^{\infty}
 \frac{
 V_{0,n}(b,0,\ldots,0)
 }{
 (2\pi^2)^{n-3}(n-1)!
 }
 z^{n-1}.
 \label{eq:all-cusp-h-h1}
\end{align}
Recalling from~\eqref{eq:all-cusp-hprime-u0} that 
$h'(z)
 =
 -\pi^2
u_0\!\left(\frac{z}{2\pi^2}\right)$, we see that equation~\eqref{eq:all-cusp-q1-u0}  gives:
\begin{equation}
h_1''(z;b)
 =
 h''(z)\,
 I_0\!\left(
 \frac{b}{\pi}\sqrt{h'(z)}
 \right)\ ,
 \label{eq:all-cusp-h1-second}
\end{equation}
where we've done the continuation
$J_0(iy)=I_0(y)$, since 
$u_0{=}{-}h'/\pi^2$.
Equation~\eqref{eq:all-cusp-h1-second} can be integrated once, and since
$h'(0)=h_1'(0;b)=0$, the result is:
\begin{equation}
 h_1'(z;b)
 =
 \frac{2\pi}{b}
 \sqrt{h'(z)}\,
 I_1\!\left(
 \frac{b}{\pi}\sqrt{h'(z)}
 \right)\ .
 \label{eq:all-cusp-h1-first}
\end{equation}
The $b{\to}0$ limit of this  yields  $h_1'{=}h'$, as it should.

The simple form~\eqref{eq:all-cusp-h1-first} is an especially compact encapsulation of all the volumes, a companion to the non-linear recurrence approach of ref.~\cite{ancona2026recursionvolumemodulispace}. We can directly check the contents by expanding.
The already computed
$ V_{0,3}({\bf 0}){=}1$,
 $V_{0,4}({\bf 0}){=}2\pi^2$, and
$ V_{0,5}({\bf 0}){=}10\pi^4$
mean that:
\begin{equation}
 h(z)
 =
 \frac{z^2}{2}
 +\frac{z^3}{6}
 +\frac{5z^4}{48}
 +\cdots
 \label{eq:h-expansion}
\end{equation}
 Next, expanding equation~\eqref{eq:all-cusp-h1-first} gives
\begin{align}
 h_1'(z;b)
 &=
 h'(z)
 +\frac{b^2}{8\pi^2}\bigl(h'(z)\bigr)^2
 +\frac{b^4}{192\pi^4}\bigl(h'(z)\bigr)^3
 +\cdots
 \nonumber\\
 &=
 z
 +
 \left(
 \frac12+\frac{b^2}{8\pi^2}
 \right)z^2
 \nonumber\\
 &\hskip0.7cm+
 \left(
 \frac{5}{12}
 +\frac{b^2}{8\pi^2}
 +\frac{b^4}{192\pi^4}
 \right)z^3
 +\cdots.
 \label{eq:all-cusp-h1-z-expansion}
\end{align}
Finally, comparing to the derivative of~(\ref{eq:all-cusp-h-h1}) gives:
\begin{align}
 &V_{0,3}(b,0,0)
 =1\ ,
 \nonumber\\
 &V_{0,4}(b,0,0,0)
 =2\pi^2+\frac{b^2}{2}\ ,
 \nonumber\\
 &V_{0,5}(b,0,0,0,0)
 =10\pi^4+3\pi^2b^2+\frac{b^4}{8}\ ,
\end{align}
which match standard results.

Ref.~\cite{ancona2026recursionvolumemodulispace} also gives a nonlinear differential equation for~$h_1(z)$ in their Theorem 1.5.  A direct way to the equation for $h_1(z)$ is to recall that  $h(z)$ obeys a differential equation, given in equation~\eqref{eq:all-cusp-Zograf-differential}, where the non-linearity enters through a $(zh'{-}h)$ factor. So we can simply ask how much $h_1(z)$ fails to satisfy an equation of the same form.
So write:
\begin{align}
 &z h_1''-h_1'
 -h_1''(zh'-h)
 =
 h'\frac{h_1''}{h''}-h_1'
 \nonumber\\
 &\qquad\qquad=
 h'\,
 I_0\!\left(
 \frac{b}{\pi}\sqrt{h'}
 \right)
 -
 \frac{2\pi}{b}\sqrt{h'}\,
 I_1\!\left(
 \frac{b}{\pi}\sqrt{h'}
 \right).
 \label{eq:all-cusp-nonlinear-step}
\end{align}
where the first equality used~(\ref{eq:all-cusp-Zograf-differential}), and the last line used ~\eqref{eq:all-cusp-h1-second} and~(\ref{eq:all-cusp-h1-first}).
Finally,
using   Bessel recurrence we get the compact nonlinear
differential equation:\footnote{
This does not quite match the ODE presented in version~1 of ref.~\cite{ancona2026recursionvolumemodulispace}, where their $L$ is our $b$.  In the normalization used there, their ODE gives at its first nontrivial order
$V_4^1(L){=}1+3L^2/(4\pi^2)$, whereas it should be
$V_4^1(L){=}1+L^2/(4\pi^2)$.
At the next order it gives
$V_5^1(L){=}5+19L^2/(4\pi^2)+9L^4/(8\pi^4)$,
whereas the correct result is
$V_5^1(L){=}5+3L^2/(2\pi^2)+L^4/(16\pi^4)$.
So the discrepancy cannot be repaired by a rescaling.  Presumably, there is an error in the (more difficult) computation they must perform.}
\begin{equation}
 z h_1''-h_1'
 =
 h_1''(zh'-h)
 +
 h'\,
 I_2\!\left(
 \frac{b}{\pi}\sqrt{h'}
 \right)\ ,
 \label{eq:all-cusp-nonlinear}
\end{equation} and the $b$-dependence arrives neatly packaged inside a Bessel function again.
For $b{=}0$, that $I_2$ term vanishes and $h_1{=}h$, so the equation of course reduces
immediately to the Zograf equation~\eqref{eq:all-cusp-Zograf-differential}.

 As remarked before, the boundary-addition identity gives the linear
off-shell relation~\eqref{eq:all-cusp-q1-u0}, while elimination of $u_0(x)$
using the genus zero string equation reorganizes the same information
into the nonlinear equation~\eqref{eq:all-cusp-nonlinear}.

Next let's turn to  a  connection between the results
above and recent work of Hide and Thomas~\cite{HideThomas2025}. 
They study the large-$n$
behaviour of Weil--Petersson volumes with a fixed number, $k$ of finite
geodesic boundaries with all the remaining boundaries cusps.  For fixed~$g$ and~$k$, they prove that, asymptotically as $n$ gets large:
\begin{equation}
 \frac{
 V_{g,n}(b_1,\ldots,b_k,0,\ldots,0)
 }{
 V_{g,n}(0,\ldots,0)
 }
 =
 \prod_{i=1}^k
 I_0\!\left(\frac{j_0b_i}{2\pi}\right)
 +\cdots 
 \ ,
 \label{eq:HT-large-n}
\end{equation}
computing also  subleading behaviour we won't need here. In this expression,  $j_0$ is the first positive zero of $J_0$. From the perspective of our computations so far,  the origin of this appearance of Bessel data has a natural explanation. Notice that the result is $g$ independent, and moreover each boundary brings in the same factor, so we should learn something even at $g{=}0$ and with a single boundary, which is exactly what we have for our results above. (We will study higher genus in the next subsection.) Let's first recognize the origin of the form of the large $n$  behaviour.

Our exact result for $h_1(z;b)$, the generating function of the one-finite boundary volumes is given in~(\ref{eq:all-cusp-h1-second}). It is in turn given in terms of $h(z)$, the corresponding all-cusp generating function. Kaufmann {\it et. al.}~\cite{Kaufmann:1996aa} show that the
 large $n$ behaviour of the cusp volumes contained in $h(z)$ can be extracted from the first critical
point of  Zograf's relation~(\ref{eq:Zograf-Bessel}), rewritten as:
\begin{equation}
 z=\sqrt{h'}\,J_1\!\left(2\sqrt{h'}\right)\ .
 \label{eq:Zograf-Bessel-z-h}
\end{equation}
Differentiating:
\begin{equation}
 \frac{\dd z}{\dd h'}
 =
 J_0\!\left(2\sqrt{h'}\right)\ ,
\end{equation}
so  there is a singularity in inverting to get $h'(z)$ when this first vanishes. This  first branch point is controlled by $j_0$ the first zero of $J_0$, and there we have:
\begin{equation}
 h'=\frac{j_0^2}{4}\ ,\quad\text{at which}\quad z=\frac{j_0}{2}J_1(j_0)\ .
 \label{eq:HT-Zograf-branch}
\end{equation}
In fact Manin and Zograf~\cite{Manin:1999wj} then showed that  this singularity applies to the  genus $g$ all-cusp situation as well. (We will see that in Section~\ref{sec:WP-higher-genus-cusps}.) This says that the leading physics is dominated by the large number of punctures, and some finite number of handles is a sub-leading effect.

But now we have an exact result~\eqref{eq:all-cusp-h1-second} for the generating function of volumes with one finite boundary~$b$, where the~$b$ dependence enters through the $I_0$ Bessel factor. At the singularity that  factor in our exact solution becomes:
\begin{equation}
 I_0\!\left(
 \frac{b}{\pi}\sqrt{h'}
 \right)
 \longrightarrow
 I_0\!\left(\frac{j_0b}{2\pi}\right)\ .
 \label{eq:HT-asymptotic-factor}
\end{equation}
precisely the per-boundary factor appearing in ref.~\cite{HideThomas2025}'s
large-$n$ result! In our language and approach, its precise form simply follows from the first critical point of the genus zero (off-shell) string equation. 

A natural interpretation is that the singularity is really a saddle-point of an effective action $S_{\rm eff}{\sim} \log(\sqrt{h'}J_1(2\sqrt{h'}))$ that controls the ``gas'' of puncture insertions.\footnote{There is of course an analogy to be made to the instanton of the genus expansion, which (interestingly) is also being controlled by the form of the string equation, since $u_0(x)$ builds the random matrix spectral density $\rho_0(E)$ through the Abel transform.}  The result of introducing some finite number of loops (by simply acting with operator $\Jcal_{b}$ in our language), shows that they really act as probes of the large~$n$ system. The simple Bessel factor in our exact solution suggests this already, but the fact that it factorizes so nicely in the full result~(\ref{eq:HT-large-n}) strongly supports that picture. It would be interesting to explore this more.

Finally, it is worth noting that there is another, even more detailed point of contact with ref.~\cite{HideThomas2025}'s work.  In proving their
general asymptotic result, they needed to determine the
large-$n$ behaviour of the successive coefficients in the finite
boundary lengths within the volumes.  In the genus zero one finite-boundary case, we can write their normalization of these coefficients as
\begin{equation}
 V_{0,n}(b,0,\ldots,0)
 =
 \sum_{\ell=0}^{n-3}
 \frac{b^{2\ell}}{4^\ell(2\ell+1)!}
 \left[\tau_0^{n-1}\tau_\ell\right]_{0,n}\ ,
 \label{eq:HT-volume-expansion}
\end{equation}
where the square-bracket quantities  here are 
{\it normalized} coefficients of the successive powers $b^{2\ell}$ in the
volume (so not exactly the $\langle\tau_0^{n-1}\tau_\ell\rangle_{0,n}$ objects of the usual intersection theory notation).  Each such coefficient, as $n$ varies, gets packaged into a generating function:
\begin{equation}
 \Phi_\ell(z)
 =
 \sum_{n=3}^{\infty}
 \frac{
 \left[\tau_0^{n-1}\tau_\ell\right]_{0,n}
 }{
 (2\pi^2)^{n-2}n!
 }
 z^n\ .
 \label{eq:HT-Phi}
\end{equation}
So for $\ell=0$ this is the corresponding all-cusp generating function,
while increasing $\ell$ picks out successively higher powers of the
single finite boundary length $b$.

Comparing our equations~\eqref{eq:all-cusp-h-h1} to 
\eqref{eq:HT-volume-expansion} reveals that:
\begin{equation}
 h_1'(z;b)
 =
 2\pi^2
 \sum_{\ell=0}^{\infty}
 \frac{b^{2\ell}}{4^\ell(2\ell+1)!}
 \Phi_\ell''(z)\ ,
 \label{eq:HT-h1-Phi}
\end{equation}
but we can expanding our exact Bessel relation~\eqref{eq:all-cusp-h1-first}:
\begin{equation}
 h_1'(z;b)
 =
 \sum_{\ell=0}^{\infty}
 \frac{
 b^{2\ell}
 }{
 4^\ell\pi^{2\ell}\ell!(\ell+1)!
 }
 \bigl(h'(z)\bigr)^{\ell+1}.
 \label{eq:HT-h1-Bessel-expansion}
\end{equation}
Picking out the  $\ell=0$ terms (no boundary, hence just coming from all-cusps) sets:
\begin{equation}
 h'(z)=2\pi^2\Phi_0''(z),
 \label{eq:HT-h-Phi0}
\end{equation}
and comparing orders in $b^2$ yields:
\begin{align}
 \Phi_\ell''(z)
 &=
 2^\ell
 \frac{(2\ell+1)!}{\ell!(\ell+1)!}
 \bigl(\Phi_0''(z)\bigr)^{\ell+1}
 \nonumber\\&=
 \frac{2^{3\ell+1}}{\sqrt{\pi}}\,
 \frac{\Gamma(\ell+\frac32)}{\Gamma(\ell+2)}
 \bigl(\Phi_0''(z)\bigr)^{\ell+1}\ .
 \label{eq:HT-Phi-closed}
\end{align}
This relation was obtained in ref.~\cite{HideThomas2025} during  their large~$n$ analysis, emerging
from a recursively derived family of differential equations.  Here
we see it as  the coefficient expansion of the single finite-boundary identity~\eqref{eq:all-cusp-h1-first}.

We can also see the origin of their differential
recursion in our approach.  Using the
modified Bessel equation for $I_0$ on
equation~\eqref{eq:all-cusp-h1-second}, and combining the result with
equation~\eqref{eq:all-cusp-h1-first}, gives:
\begin{equation}
 4\left(b\frac{\partial}{\partial b}\right)
 \left(b\frac{\partial}{\partial b}+1\right)h_1''
 =
 \frac{2b^2}{\pi^2}
 \left(
 2h'h_1''+h''h_1'
 \right)\ .
 \label{eq:HT-Bessel-ODE}
\end{equation}
Then using expansion~(\ref{eq:HT-h1-Bessel-expansion}) and relation~(\ref{eq:HT-h-Phi0})
and equating coefficients of $b^{2\ell}$ immediately gives:
\begin{equation}
 \Phi_\ell'''
 =
 8\Phi_0''\Phi_{\ell-1}'''
 +
 4\Phi_0'''\Phi_{\ell-1}''\ ,\quad \ell\ge1\ ,
 \label{eq:HT-Phi-ODE}
\end{equation}
which is precisely what was obtained in
ref.~\cite{HideThomas2025}.

\subsection{Cusps and finite boundaries at higher genus}
\label{sec:WP-higher-genus-cusps}

The genus zero results above suggest that we look to see if some of the 
simplifications can also occur at higher genus. We will first do the case of  all cusps at general~$g$, and indeed recover some more classic results  of Zograf~\cite{Zograf:1998ur,ZografP.G.1998Wvol}. Then we will add one boundary of finite length, obtaining a useful compact generalization applicable to all genus, which is new. Finally, we'll spot the  pattern that invites an immediate generalization, and present a compact formula for the generating function of the volumes $V_{g,n}({\bm b},{\bm 0})$ where ${\bm b}$ is the set of $k$ finite lengths~$b_i$, $i{=}1,\ldots,k$, and~${\bm 0}$  are the $(n{-}k)$ cusps. It serves as a remarkable rewriting of {\it all} Weil-Petersson volumes entirely in terms of a nested family of Bessel-flavoured operations on solutions of the string equation, which is  worth further exploration.

\subsubsection{All-cusp volumes at arbitrary genus}
\label{sec:WP-all-cusps-arbitrary-genus}

A quick way into the problem is to note, following  ref.~\cite{Johnson:2026twg} that the (off-shell)  one-boundary loop~(\ref{eq:loop-object}) can be written as:
\begin{equation}
 \widetilde W(x,E)
 =\hbar^2\delta_E F(x)
 =\int^x \widehat R(x',E)\dd x'\ ,
 \label{eq:all-cusp-GD-start}
\end{equation}
where $\widehat R$ is the diagonal Gel'fand--Dikii resolvent that solves the ODE:
\begin{equation}
 4(u-E)\widehat R^2
 -2\hbar^2\widehat R\,\widehat R''
 +\hbar^2(\widehat R')^2=1\ .
 \label{eq:GD-resolvent}
\end{equation}
Expanding
both sides in genus gives:
\begin{equation}
 \widetilde W_{g,1}(x,E)
 =\delta_E F_g(x)
 =\int^x \widehat R_g(x',E)\dd x'\ ,
 \label{eq:all-cusp-GD-genus}
\end{equation}
and recursively solving~(\ref{eq:GD-resolvent}) gives a useful expansion for ${\widehat R}(x,E)$ in terms of $u_0$ and its derivatives, where the terms are organized in inverse powers of $z{=}(u_0(x)-E)^{\frac12}$, what will ultimately become the spectral coordinate. The fact that $u{=}2\hbar^2\partial_x^2 F$ ensures that $\widehat R_g(x',E)$ is a total derivative and so the right hand side is a local function of~$u_0$ and its derivatives.
Recall that the loop operator action~\eqref{eq:loop-u0} on any $u_0$ derivative is:\footnote{Briefly, while revisiting notions from Subsection~\ref{sec:boundary-addition-derivation}, we are using $z$ here in its old role as spectral coordinate, until just  before equation~\eqref{eq:all-cusp-V21-check}. After that we will return to it being a generating function expansion parameter.}
\begin{equation}
 \delta_Eu_0^{(m)}
 =\frac{\partial^m}{\partial x^m}
 \left(\frac{u_0'}{2z^3}\right)
 =\frac{u_0^{(m+1)}}{2z^3}+O(z^{-5})\ .
 \label{eq:all-cusp-jet-leading}
\end{equation}
where we will see shortly that the terms neglected will not be needed. Now $F_g$ is a local function of $u_0$'s derivatives, so its loop
variation therefore begins as:
\begin{equation}
\delta_EF_g
 =\frac{1}{2z^3}\frac{\partial F_g}{\partial x}
 +O(z^{-5})\ ,
 \label{eq:all-cusp-Q-leading}
\end{equation}
where the polynomial in $1/z$ truncates. Following the procedure in~(\ref{eq:the-Ws})
we convert using the Jacobian $(-2z)$ to get:
\begin{equation}
 W_{g,1}(z;x)
 =-\frac{F_g'(x)}{z^2}
 +\frac{A_{g,1}(x)}{z^4}
 +\frac{A_{g,2}(x)}{z^6}
 +\cdots\ ,
 \label{eq:all-cusp-W-leading}
\end{equation}
where $A_{g,i}$ are names for functions of the $u_0(x)$ jet that  we won't need.
Anticipating the transform~\eqref{eq:laplace-def} to volume with $b$ dependence, write:
\begin{equation}
 \Vcal_{g,1}(b;x)
 =v_{g,0}(x)+v_{g,1}(x)b^2+v_{g,2}(x)b^4+\cdots\ ,
\end{equation}
and so:
\begin{equation}
 W_{g,1}(z;x)
 =
 \frac{v_{g,0}(x)}{z^{2}}+\frac{6\,v_{g,1}(x)}{z^{4}}+\dots\ .
 \label{eq:all-cusp-Laplace-powers}
\end{equation}
We are interested in the cusp limit where  $b{=}0$, and so we see that this  picks out precisely the leading $z^{-2}$ term of the loop object.
Combining equations~\eqref{eq:all-cusp-W-leading} and
\eqref{eq:all-cusp-Laplace-powers} gives the all-genus one-cusp off-shell volume:
\begin{equation}
 \Vcal_{g,1}(0;x)=-F_g'(x)\ ,
 \qquad g\geq1\ .
 \label{eq:all-cusp-one-seed}
\end{equation}
For example, at genus one,
recall from equation~\eqref{eq:F1-formula} that $F_1=-\frac{1}{24}\log u_0^\prime$, and  acting with $\delta_E$ given in~(\ref{eq:loop-u0}) gave~(\ref{eq:W11-tilde-off}), which is:
\begin{equation}
 {\widetilde W}_{1,1}
 =-\frac{u_0''}{48u_0'z^3}
 +\frac{u_0'}{32z^5}
 =\frac{F_1'}{2z^3}+O(z^{-5})\ .
\end{equation}
Hence:
\begin{equation}
 \Vcal_{1,1}(0;x)
 =\frac{u_0''}{24u_0'}=-F_1'(x)\ ,
\end{equation}
and of course putting in the JT/WP values of the derivatives~(\ref{eq:intro-JT-low-jet}) gives $\pi^2/12$, the $b{=}0$ limit of $V_{1,1}$ in~equation~(\ref{eq:intro-V03-V11-WP}).

At genus two we have:
\begin{equation}
 F_2
 =\frac{u_0^{(4)}}{576(u_0')^2}
 -\frac{7u_0''u_0^{(3)}}{960(u_0')^3}
 +\frac{(u_0'')^3}{180(u_0')^4}\ ,
 \label{eq:all-cusp-F2}
\end{equation}
and the leading large $z$ term of $\int^x\widehat R_2(x',E)\dd x'$, given in ref.~\cite{Johnson:2026twg}, is indeed seen to be $F_2^\prime/(2z^3)$, which  leads to: 
\begin{equation}
 \Vcal_{2,1}(0;x)=-F_2^\prime\ ,\quad\longrightarrow \quad V_{2,1}(0)=\frac{29\pi^8}{192}\ .
 \label{eq:all-cusp-V21-check}
\end{equation}
Now that we know the result for one cusp, adding further cusps
is immediate from our boundary addition technology.  As before, at any stage,  since all existing
boundaries have zero length, every integral term in
 equation~\eqref{eq:WP-full-identity} vanishes, and the action of $\Jcal_0$ is just an $x$-derivative, so we use
 equation~\eqref{eq:WP-cusp-J-action}.  Repeated cusp insertion  gives therefore that:
$\Vcal_{g,n}(0,\ldots,0;x)
 =-\frac{\partial^nF_g(x)}{\partial x^n}$,
and so evaluating at the $x=0$ endpoint:
\begin{equation}
 V_{g,n}(0,\ldots,0)
 =-\left.
 \frac{\partial^nF_g(x)}{\partial x^n}
 \right|_{x=0}\ , \qquad g\geq1\ .
 \label{eq:all-cusp-Fg-master}
\end{equation}
The genus zero relation~\eqref{eq:Zograf-cusp-from-u0} is the corresponding
statement with $u_0=2F_0''$.

In analogy with what we did in the genus zero case, define the all-cusp generating function of volumes, as a function of an expansion parameter~$z$:
\begin{equation}
 H_g(z)
 =\sum_{\substack{n\geq0\\2g-2+n>0}}
 \frac{V_{g,n}(0,\ldots,0)}
 {(2\pi^2)^{3g-3+n}n!}\,z^n\ .
 \label{eq:higher-genus-hg-def}
\end{equation}
For $g\geq2$, equation~\eqref{eq:all-cusp-Fg-master} tells us that:
\begin{equation}
 H_g(z)
 =-\frac{1}{(2\pi^2)^{3g-3}}
 F_g\!\left(\frac{z}{2\pi^2}\right),
 \qquad g\geq2 \ .
 \label{eq:higher-genus-hg-Fg}
\end{equation}
For genus one a constant should be  included, giving:
\begin{equation}
 H_1(z)
 =-F_1\!\left(\frac{z}{2\pi^2}\right)+F_1(0)\ .
 \label{eq:higher-genus-h1-F1}
\end{equation}
and we can use equation~\eqref{eq:all-cusp-hprime-u0} to relate $u_0$ $z$-derivatives to the genus zero generating function $h$ by writing
$u_0'(z/(2\pi^2))=-2h''(z)$. This gives:
\begin{equation}
 H_1(z)=\frac1{24}\log h''(z)\ .
 \label{eq:higher-genus-h1}
\end{equation}
We can do the same substitution in equation~\eqref{eq:all-cusp-F2}
to give a  compact genus two result:
\begin{equation}
 H_2(z)
 =\frac{h^{(5)}}{1152(h'')^2}
 -\frac{7h'''h^{(4)}}{1920(h'')^3}
 +\frac{(h''')^3}{360(h'')^4}\ ,
 \label{eq:higher-genus-h2}
\end{equation}
and it should be remembered here that  all derivatives are with respect to $z$. Also, $h^{(m)}$ denotes the $m$th $z$-derivative. Note that from  equation~\eqref{eq:all-cusp-h-h1} we have at genus zero: $H_0'(z)=h(z)$.

We've actually made direct contact  with the results of Zograf in refs.~\cite{Zograf:1998ur,ZografP.G.1998Wvol}, where generating functions for the  all-cusp
volumes at genus $g$ were observed to have an identification with 
identification with the  genus expansion for the free energy~\cite{Itzykson:1992ya}.  Here we see the result emerge readily from the  identity, naturally using off-shell data fundamentally built from KdV material, and for this all-cusp case adding more cusps is just a matter of elementary differentiations.

\subsubsection{Adding one finite boundary}
\label{sec:WP-one-boundary-arbitrary-genus}

Let's now consider $V_{g,n}$ where all boundaries are cusps except one, which has length $b$.  Again this is easy to do, by starting with the case of  there being $n-1$ cusps already present,  and then adding the finite length  boundary. So every upper limit in the integral term of identity~\eqref{eq:WP-full-identity} is again zero.  and we just need to operate once with $\Jcal_b$. So we have  the immediate result:
\begin{equation}
 V_{g,n}(b,0,\ldots,0)
 =-
 \left.
 \Jcal_b\frac{\partial^{n-1}F_g(x)}{\partial x^{n-1}}
 \right|_{x=0}\ ,
 \quad 2g-2+n>0\ .
 \label{eq:higher-genus-one-boundary-master}
\end{equation}
Again, the action needed in equation~\eqref{eq:higher-genus-one-boundary-master} is known on the entire set of $u_0(x)$ derivatives.  Setting $E_0{=}0$ in
 equation~\eqref{eq:softedge-generating} gives:
\begin{equation}
 \sum_{m=0}^{\infty}\frac{s^m}{m!}
 \left.\Jcal_bu_0^{(m)}\right|_{x=0}
 =u_0'(s)J_0\!\left(b\sqrt{u_0(s)}\right)\ ,
 \label{eq:higher-genus-Jb-jet}
\end{equation}
and we  set $s=z/(2\pi^2)$. 
 Equation~\eqref{eq:all-cusp-hprime-u0} again tells us:
\begin{equation}
 u_0'\!\left(\frac{z}{2\pi^2}\right)=-2h''(z)\ ,
\end{equation}
and then just as in the genus zero case, starting with {\it e.g.}~(\ref{eq:all-cusp-h1-second}), that signature Bessel factor  will be how the effect of the boundary enters all functions through:
\begin{equation}
\Jcal_b h^\prime(z)=h''(z)
 I_0\!\left(\frac{b}{\pi}\sqrt{h'(z)}\right)\ ,
 \label{eq:higher-genus-Jb-action-on-h1}
\end{equation} 
and since $\Jcal_b$ commutes with $z$-derivatives:
\begin{equation}
 \Jcal_b h^{(r)}(z)
 =
 \frac{d^{\,r-1}}{dz^{\,r-1}}
 \left[
 h''(z)
 I_0\!\left(\frac{b}{\pi}\sqrt{h'(z)}\right)
 \right],
 \qquad r\geq1\ .
 \label{eq:higher-genus-Jb-action-on-h2}
\end{equation}

So we can extend the genus zero notation
$h_1(z;b)$ of equation~\eqref{eq:all-cusp-h-h1} by defining
\begin{equation}
 H_{g,1}(z;b)
 =
 \sum_{\substack{n\geq1\\2g-2+n>0}}
 \frac{V_{g,n}(b,0,\ldots,0)}
 {(2\pi^2)^{3g-3+n}(n-1)!}\,z^{n-1}.
 \label{eq:higher-genus-h1g-def}
\end{equation}
For $g{=}0$ this is the function $h_1(z;b)$ from before.
For $g{\geq1}$,  $H_{g,1}(z)$ is a
differential polynomial in 
$h''(z),h'''(z),\ldots$.  Equation
\eqref{eq:higher-genus-Jb-action-on-h2} allows us to write a very 
compact exact formula using the chain rule:
\begin{widetext}
\begin{equation}
 H_{g,1}(z;b)
 =
 \sum_{r\geq2}
 \frac{\partial H_g}{\partial h^{(r)}}
 \frac{\dd^{r-1}}{\dd z^{r-1}}
 \left[
 h''(z)I_0\!\left(\frac{b}{\pi}\sqrt{h'(z)}\right)
 \right],
 \qquad g\geq1\ ,
 \label{eq:higher-genus-h1g-master}
\end{equation}
where $h^{(r)}{\equiv}\dd^rh/\dd z^r$. 
At any fixed genus the sum is finite. For genus zero the corresponding statement is the formula already derived
in equation~\eqref{eq:all-cusp-h1-second}.  At
$b{=}0$ the expression in square brackets reduces to $h''(z)$, so the result collapses to $H_{g,1}(z;0)=\frac{\dd H_g(z)}{\dd z}$,
as required when the last finite boundary is turned into one more cusp.
Let's unpack this a bit for genus one again gives a particularly simple illustration.  We have from  equation~\eqref{eq:higher-genus-h1} that 
 $H_1(z)=\frac1{24}\log h''(z)$,
so only the derivative with respect to $h''$ is needed, and therefore
 equation~\eqref{eq:higher-genus-h1g-master} becomes:
\begin{equation}
\begin{aligned}
 H_{1,1}(z;b)
 ={}&\frac1{24}\frac{h'''(z)}{h''(z)}
 I_0\!\left(\frac{b}{\pi}\sqrt{h'(z)}\right)
 +\frac{b}{48\pi}\frac{h''(z)}{\sqrt{h'(z)}}
 I_1\!\left(\frac{b}{\pi}\sqrt{h'(z)}\right), 
\end{aligned}
 \label{eq:higher-genus-h11-exact}
\end{equation}
the higher-genus
analogue of equation~\eqref{eq:all-cusp-h1-second}.
So that we're not getting too lost, it is useful to write this back in terms of volumes and $u_0(x)$ using~\eqref{eq:higher-genus-h1g-def}:
\begin{equation}
\begin{aligned}
 \sum_{n=1}^{\infty}
 V_{1,n}(b,0,\ldots,0)\frac{x^{n-1}}{(n-1)!}
 ={}&
 \frac1{24}\left[
 \frac{u_0''(x)}{u_0'(x)}
 J_0\!\left(b\sqrt{u_0(x)}\right)
 \right.
 \left.
 -\,\frac{b\,u_0'(x)}{2\sqrt{u_0(x)}}
 J_1\!\left(b\sqrt{u_0(x)}\right)
 \right].
 \label{eq:higher-genus-h11-u0}
\end{aligned}
\end{equation}
Expanding, orders $x^0$  and $x^1$ readily yield:\footnote{Lowenstein~\cite{Lowenstein:2024fji}, using the more involved sum-over-all-times KdV approach mentioned in footnote~\ref{fn:lowenstein}, derived a general  formula for the genus one case that almost matches this when specialized to this one-finite boundary case. After accounting for the overall normalization, the first two Bessel structures agree while the $J_2$ term appears smaller by a factor of two, so the volumes don't come out quite right in the final result. The approach is a useful contrast to note, nevertheless.}  
\begin{equation}
 V_{1,1}(b)=\frac{b^2+4\pi^2}{48}\ ,\quad \text{and}
 \quad
 V_{1,2}(b,0)
 =\frac{(b^2+4\pi^2)(b^2+12\pi^2)}{192}\ ,
 \label{eq:higher-genus-h11-checks}
\end{equation}
in agreement with the known volumes.
It is useful to expand the Bessel factor in
 equation~\eqref{eq:higher-genus-h1g-master} to get an exact 
formula for every coefficient in the  polynomial in $b^2$ within the generating function:
\begin{equation}
 \left[H_{g,1}(z;b)\right]_{b^{2\ell}}
 =
 \frac{1}{4^\ell\pi^{2\ell}(\ell!)^2}
 \sum_{r\geq2}
 \frac{\partial H_g}{\partial h^{(r)}}
 \frac{\dd^{r-1}}{\dd z^{r-1}}
 \left[(h'(z))^\ell h''(z)\right]\ .
 \label{eq:higher-genus-boundary-coefficients}
\end{equation}
\end{widetext}
At genus zero, recall that the same Bessel expansion led to the closed functions
$\Phi_\ell$ in equations~\eqref{eq:HT-h1-Bessel-expansion}--\eqref{eq:HT-Phi-closed}.  Equation~\eqref{eq:higher-genus-boundary-coefficients} supplies their natural
higher genus counterparts.  

We can return now to the large-$n$ theorem of Hide and
Thomas~\cite{HideThomas2025}.  We note from 
 equation~\eqref{eq:higher-genus-h1g-master} that $H_{g,1}(z;b)$ starts out as:
\begin{equation}
\begin{aligned}
 H_{g,1}(z;b)
 ={}&
 I_0\!\left(\frac{b}{\pi}\sqrt{h'(z)}\right)
 \frac{\dd H_g(z)}{\dd z}+\cdots
 \label{eq:higher-genus-HT-split}
\end{aligned}
\end{equation}
where the ellipsis represents terms where there are  $z$-derivative acts on the
 $I_0$ factor.
It is the first term that most readily communicates the effects  of the 
 singularity signalling the large order behaviour of the $n$ expansion in the (mostly cusp-dominated) volumes $V_{g,n}$.  As reviewed around equations
\eqref{eq:Zograf-Bessel-z-h}--\eqref{eq:HT-Zograf-branch}, that singularity
occurs at
 $h'(z)=\frac{j_0^2}{4}$,
where $j_0$ is the first positive zero of $J_0$.  Again we see that  the exact
boundary dependence 
become entirely $I_0\!\left(\frac{j_0b}{2\pi}\right)$ as we saw in the genus zero case around 
 equation~\eqref{eq:HT-asymptotic-factor}, but we can see that it dominates at all genus, consistent with the old all-cusp observations of Manin and Zograf~\cite{Manin:1999wj}.  This is because in equation~\eqref{eq:higher-genus-HT-split}, terms in
which a $z$-derivative hits the  Bessel factor are
correspondingly less singular than the first term.  The leading singular
part at every fixed genus is therefore:
\begin{equation}
 H_{g,1}(z;b)
 \sim
 I_0\!\left(\frac{j_0b}{2\pi}\right)
 \frac{\dd H_g(z)}{\dd z}
 \quad \text{as}\quad z\longrightarrow \frac{j_0}{2}J_1(j_0)\ ,
 \label{eq:higher-genus-HT-leading}
\end{equation}
 and so we get as $n\longrightarrow\infty$, for any $g$:
\begin{equation}
 \frac{V_{g,n}(b,0,\ldots,0)}{V_{g,n}(0,\ldots,0)}
 \longrightarrow
 I_0\!\left(\frac{j_0b}{2\pi}\right)+\cdots\ ,
 \label{eq:higher-genus-HT-ratio}
\end{equation}  This is the one-finite-boundary case of
 equation~\eqref{eq:HT-large-n}.  

 We can  then expect that adding another  fixed finite boundary
will bring in another such Bessel factor, with the same  leading effect.  (We will confirm this next subsection.) Iterating this for $k$ fixed boundaries of lengths~$b_i$ we get the factorized boundary dependence in
 equation~\eqref{eq:HT-large-n}.  Of course, ref.~\cite{HideThomas2025} proves not just  this leading piece but also the subleading  estimate of the correction, which we have not attempted to do here.  The point here was that through the simple use of the identity~(\ref{eq:WP-full-identity}) we can quickly  understand how the universal
$I_0(j_0b/2\pi)$ factor emerges at any genus,  for each added finite boundary.

\subsubsection{General result for $k$ finite boundaries, any genus}
\label{sec:WP-k-boundaries-arbitrary-genus}

It is not hard to see what to do next. Looking again at
equation~\eqref{eq:higher-genus-Jb-jet}, the point is that the same formula
that told us how $\Jcal_b$ acts on the entire $u_0(x)$ jet can simply be
used again after the first finite boundary has been inserted.  So for a
second boundary of length $b_2$ we act with $\Jcal_{b_2}$ on the $u_0(x)$ jet data
already present in the one-boundary answer.

It is easiest to see this explicitly at genus zero.  From the one-boundary
result we have:
\begin{equation}
 h_1''(z;b_1)
 =
 h''(z)
 I_0\!\left(
 \frac{b_1}{\pi}\sqrt{h'(z)}
 \right)
 \ ,
 \label{eq:higher-genus-h1-second-repeat}
\end{equation}
and, after integrating once,
\begin{equation}
 h_1'(z;b_1)
 =
 \frac{2\pi}{b_1}\sqrt{h'(z)}
 I_1\!\left(
 \frac{b_1}{\pi}\sqrt{h'(z)}
 \right)
 \ .
 \label{eq:higher-genus-h1-first-repeat}
\end{equation}
Notice that $h_1'$ depends on $z$ only through $h'(z)$.  Also,  we have from~\eqref{eq:higher-genus-Jb-action-on-h1} that:
\begin{equation}
 \Jcal_{b_2}h'
 =
 h''
 I_0\!\left(
 \frac{b_2}{\pi}\sqrt{h'}
 \right)
 \ .
 \label{eq:higher-genus-Jb-hprime}
\end{equation}
Therefore the second boundary insertion acts by the  chain rule:
\begin{widetext}
\begin{align}
 h_2''(z;b_1,b_2)
 =
 \Jcal_{b_2}h_1'(z;b_1)
 =
 \frac{\partial h_1'}{\partial h'}\,
 \Jcal_{b_2}h'
 =
 h''(z)
 I_0\!\left(
 \frac{b_1}{\pi}\sqrt{h'(z)}
 \right)
 I_0\!\left(
 \frac{b_2}{\pi}\sqrt{h'(z)}
 \right)
 \ ,
 \label{eq:higher-genus-h2-second}
\end{align} where in the last step we used:
\begin{equation}
 \frac{\partial}{\partial h'}
 \left[
 \frac{2\pi}{b_1}\sqrt{h'}\,
 I_1\!\left(
 \frac{b_1}{\pi}\sqrt{h'}
 \right)
 \right]
 =
 I_0\!\left(
 \frac{b_1}{\pi}\sqrt{h'}
 \right)
 \ .
\end{equation}
 This pleasing structure makes it natural to expect the pattern to continue for each
successive finite boundary.
Using $\dd h'=h''\dd z$, equation~\eqref{eq:higher-genus-h2-second} can also
be integrated once to give $h_2'$, but we won't need it here.

Let us pause to see explicitly how a volume emerges, focusing on the case of $V_{0,5}(b_1,b_2,0,0,0)$. The all-cusp function's first few terms  is in equation~\eqref{eq:h-expansion}, and we'll only need these derivatives:
\begin{equation}
 h'(z)=z+\frac12z^2+O(z^3)\ ,
 \qquad
 h''(z)=1+z+\frac54z^2+O(z^3)\ .
 \label{eq:higher-genus-h-low-expansion}
\end{equation}
For each boundary the Bessel factor is:
\begin{align}
 I_0\!\left(\frac{b_i}{\pi}\sqrt{h'}\right)
 ={}&
 1+\frac{b_i^2}{4\pi^2}z
+\left(
 \frac{b_i^2}{8\pi^2}
 +\frac{b_i^4}{64\pi^4}
 \right)z^2
 +O(z^3)\ .
 \label{eq:higher-genus-I0-low-expansion}
\end{align}
Substituting these into equation~\eqref{eq:higher-genus-h2-second} gives:
\begin{align}
 h_2''(z;b_1,b_2)
 ={}&
 1+\left[
 1+\frac{b_1^2+b_2^2}{4\pi^2}
 \right]z
 +\left[
 \frac54
 +\frac{3(b_1^2+b_2^2)}{8\pi^2}
 +\frac{b_1^4+b_2^4}{64\pi^4}
 +\frac{b_1^2b_2^2}{16\pi^4}
 \right]z^2
 +O(z^3)\ .
 \label{eq:higher-genus-h2-low-expansion}
\end{align}
Since the natural definition of $h_2$ has $V_{0,n}/[(n-1)!(2\pi^2)^{n-3}]$ emerging at order $z^4$, two derivatives of $h_2$ place $V_{0,5}$ at quadratic order and we must multiply by $8\pi^4$, giving the correct standard result (with three lengths set to zero):
\begin{align}
 V_{0,5}(b_1,b_2,0,0,0)
 ={}&
 10\pi^4
 +3\pi^2(b_1^2+b_2^2)
 +\frac18(b_1^4+b_2^4)
 +\frac12b_1^2b_2^2
 \ ,
 \label{eq:higher-genus-V05-k2}
\end{align}
and we see the correct mixed term directly arising from the product of the two Bessel
functions.  It is natural to see how successive insertions simply continue filling in the dependence on the finite lengths one by one!
\end{widetext}
So we can write down the general genus zero
pattern immediately.  Defining the $k$-finite-boundary, $(n-k)$-cusp volumes' generating function as: 
\begin{equation}
 h_k(z;b_1,\ldots,b_k)
 =\!\!
 \sum_{n\geq\max(3,k)}
 \frac{
 V_{0,n}(b_1,\ldots,b_k,0,\ldots,0)
 }{
 (2\pi^2)^{n-3}(n-1)!
 }
 z^{n-1}
 \ ,
 \label{eq:higher-genus-hk-def}
\end{equation}
repeated boundary insertion gives simply:
\begin{equation}
 h_k''(z;b_1,\ldots,b_k)
 =
 h''(z)
 \prod_{i=1}^k
 I_0\!\left(
 \frac{b_i}{\pi}\sqrt{h'(z)}
 \right)
 \ .
 \label{eq:higher-genus-hk-product}
\end{equation}
Equivalently,
\begin{equation}
 h_k'(z;b_1,\ldots,b_k)
 =
 \int_0^{h'(z)}
 \prod_{i=1}^k
 I_0\!\left(
 \frac{b_i}{\pi}\sqrt{t}
 \right)
 \dd t
 \ .
 \label{eq:higher-genus-hk-first-integral}
\end{equation}
So the full dependence on any fixed set of finite boundary lengths is built
by multiplying together the same Bessel factors that appeared in the
one-boundary problem, just as we anticipated in the discussion in the previous subsection.

Notice that this gives a extremely compact form  of the content expressed through non-linear recursion among  the $k$-finite-boundary $(n-k)$-cusp volumes in ref.~\cite{ancona2026recursionvolumemodulispace}.\footnote{There is likely much to learn by  converting between the two  approaches, as is already clear from preliminary explorations.}

The same simple approach to adding more finite boundaries extends to arbitrary genus, although now there is one important
difference.  The all-cusp higher genus function $H_g(z)$ is not itself just a function of
$h'(z)$ and $h''(z)$, but depends on more and more higher $h$-derivatives as $g$ increases.  Each new finite boundary therefore acts on all occurrences
of those derivatives, with precisely the Bessel-modified derivatives suggested
by equation~\eqref{eq:higher-genus-Jb-jet}. 

To make things explicit, we  define the genus~$g$ $k$-finite-boundary $(n{-}k)$-cusp analogue of our one-finite-boundary generating functions~\eqref{eq:higher-genus-h1g-def} from before:
\begin{equation}
 H_{g,k}(z;b_1,\ldots,b_k)
 =\!\!\!
 \sum_{\substack{n\geq k\\2g-2+n>0}}\!\!\!
 \frac{
 V_{g,n}(b_1,\ldots,b_k,0,\ldots,0)
 }{
 (2\pi^2)^{3g-3+n}(n-k)!
 }
 z^{n-k}
 \ .
 \label{eq:higher-genus-Hgk-def}
\end{equation}
The multiple finite-boundary result can  be written:
\begin{equation}
 H_{g,k}(z;b_1,\ldots,b_k)
 =
 \Jcal_{b_k}\cdots
 \Jcal_{b_2}\Jcal_{b_1}H_g(z)
 \ ,
 \qquad g\geq1\ .
 \label{eq:higher-genus-Hgk-master}
\end{equation}
Here each $\Jcal_b$ acts on  $h(z)$ and its $z$-derivatives according to
equation~\eqref{eq:higher-genus-Jb-action-on-h2}.  More generally, for any differential function ${\cal F}$ of 
$h$'s jet-data,
\begin{equation}
 \Jcal_b{\cal F}
 =
 \sum_{r\geq1}
 \frac{\partial{\cal F}}{\partial h^{(r)}}
 \frac{\dd^{r-1}}{\dd z^{r-1}}
 \left[
 h''(z)
 I_0\!\left(\frac{b}{\pi}\sqrt{h'(z)}\right)
 \right]\ .
 \label{eq:higher-genus-Jb-general-action}
\end{equation}
Setting any $b{=}0$ collapsing to just a $z$-derivative, which is appropriate to a cusp insertion, as noted below~(\ref{eq:higher-genus-h1g-master}).

For the original all-cusp function $H_g$ there is no explicit dependence
on $h'$, so the $r{=}1$ term is absent and this reduces to
equation~\eqref{eq:higher-genus-h1g-master}.  After the first finite
boundary has been inserted, explicit $h'$ dependence is present through
the Bessel functions, and so the $r{=}1$ term contributes to all the subsequent
insertions. Note also that since $[\Jcal_a,\Jcal_b]{=}0$ (which follows straightforwardly  from their realization as KdV flows, or by explicitly working them out as actions on $h$-jet data), the order in which the $\Jcal_{b_i}$ are applied in equation~\eqref{eq:higher-genus-Hgk-master} does not matter.

It is again worthwhile to pause to see how  this works explicitly. Let's do genus 1,  and study 
two successive finite-boundary insertions. So we start again with $H_1(z)=\frac1{24}\log h''(z)$ and try to extract $V_{1,2}(b_1,b_2)$, saturating the boundary insertions. According to~\eqref{eq:higher-genus-Hgk-def}, we must look at the constant term for it. Acting twice on $H_1(z)$ gives, before setting $z{=}0$:
\begin{widetext}
\begin{align}
 H_{1,2}(z;b_1,b_2)
 =\frac1{24}\Bigg\{&
 \left(
 \frac{h''''}{h''}
 -\frac{(h''')^2}{(h'')^2}
 \right)
 I_0\!\left(\frac{b_1}{\pi}\sqrt{h'}\right)
 I_0\!\left(\frac{b_2}{\pi}\sqrt{h'}\right)
 +2h'''
 \left[
 I_0\!\left(\frac{b_1}{\pi}\sqrt{h'}\right)
 \frac{\partial}{\partial h'}
 I_0\!\left(\frac{b_2}{\pi}\sqrt{h'}\right)
 +(b_1\leftrightarrow b_2)
 \right]
 \nonumber\\
 &+(h'')^2
 \left[
 I_0\!\left(\frac{b_1}{\pi}\sqrt{h'}\right)
 \frac{\partial^2}{\partial (h')^2}
 I_0\!\left(\frac{b_2}{\pi}\sqrt{h'}\right)
 +(b_1\leftrightarrow b_2)
 \right.
\left.
 +\frac{\partial}{\partial h'}
 I_0\!\left(\frac{b_1}{\pi}\sqrt{h'}\right)
 \frac{\partial}{\partial h'}
 I_0\!\left(\frac{b_2}{\pi}\sqrt{h'}\right)
 \right]
 \Bigg\}\ .
 \label{eq:higher-genus-H12-explicit}
\end{align}
    \end{widetext}
The derivatives with respect to $h'$ 
must be done  before evaluating
at the endpoint, but they are left on display here to show the structure of the computation. The  Bessel functions are expanded next, and   
at  $z{=}0$ we use:
\begin{equation}
 h'(0)=0\ ,\,
 h''(0)=1\ ,\,
 h'''(0)=1\ ,\,
 h''''(0)=\frac52\ ,
 \label{eq:higher-genus-h-endpoint-low}
\end{equation}
and after some algebra we recover:
\begin{equation}
 H_{1,2}(0;b_1,b_2)
 =
 \frac1{24}
 \left[
 \frac32
 +\frac{b_1^2+b_2^2}{2\pi^2}
 +\frac{b_1^4+b_2^4}{32\pi^4}
 +\frac{b_1^2b_2^2}{16\pi^4}
 \right]
 \ .
 \label{eq:higher-genus-H12-endpoint}
\end{equation}
Since the constant term in the definition~(\ref{eq:higher-genus-Hgk-def}) of $H_{1,2}(z;b_1,b_2)$ is ${V_{1,2}(b_1,b_2)}/{(2\pi^2)^2}$,
 we obtain the familiar result~\eqref{eq:intro-11example}, repeated here:
\begin{equation}
 V_{1,2}(b_1,b_2)
 =
 \frac{
 \left(4\pi^2+b_1^2+b_2^2\right)
 \left(12\pi^2+b_1^2+b_2^2\right)
 }{192}
 \ .
 \label{eq:higher-genus-V12-full}
\end{equation}
This is somewhat more of  a robust check than the genus zero case because  the second boundary insertion acts on all of the non-trivial $h(z)$ derivatives  created by the
first insertion.

A third insertion was explored to check  that it gives the complete three finite-boundary result for $V_{1,3}(b_1,b_2,b_3)$.
%
This was successful, but  the details, while straightforward, are lengthy and hence are not shown here.
In particular, the fully mixed
$b_1^2b_2^2b_3^2$ term comes with the correct coefficient, which is a highly non-trivial outcome.

Equations~\eqref{eq:higher-genus-hk-product} and
\eqref{eq:higher-genus-Hgk-master} therefore give a compact, but explicit, all-genus
construction of the ordinary Weil--Petersson volumes with any fixed number
of finite geodesic boundaries and any remaining boundaries taken to be
cusps.  Taking $k{=}n$ eventually fills in the complete finite-length volume
polynomial. It is fascinating that the Weil-Petersson volumes can be constructed  entirely out of this simple Bessel-type of action on the basic ``closed string'' off-shell free energy $F_g$.\footnote{Note that the construction applies with relatively little modification to the ${\cal N}{=}2$ and ${\cal N}{=}4$ cases we will discuss in later sections: $u_0(x)$ and hence $h(z)$ instead satisfy a different leading string equation.}  There is considerably more structure to be explored in this remarkable formula, but that shall be left for future work.

\section{${\cal N}{=}1$ Weil-Petersson Volumes}
\label{sec:boundary-addition-WP-N1}

\subsection{The Specialization}

Right at the outset,  we note that the ${\cal N}{=}1$ case differs sharply from the ordinary Weil--Petersson
case in a major feature: The
boundary-addition identity will not have the integral term.  The  origin of this is the fact that the function $u_0(x)$ will be zero in this case, so the   term responsible for the integral in
equation~\eqref{eq:master-offshell} does not arise since it is controlled by $u_0^\prime(\mu)$, which vanishes. 

Put differently, this  
is a hard
edge system, for which the analogue of the $z_i$ coordinates are:~\cite{Johnson:2026twg}
\begin{equation}
 \widehat z_i^2=-E_i\ .
 \label{eq:N1-hard-coordinate}
\end{equation}
They are independent of $u_0(x)$, and hence inert under the action of $\delta_E$ (see the derivation in Section~\ref{sec:boundary-addition-derivation}).

 So the 
boundary-addition identity reduces to just the term involving the action of $\Jcal_b$:
\begin{equation}
 \Vcal_{g,n+1}(\bm b,b;x)
 =\Jcal_b^{\rm h}\Vcal_{g,n}(\bm b;x)\ ,
 \label{eq:N1-full-offshell}
\end{equation}
where the length-space boundary operator is now:
\begin{equation}
 \Jcal_b^{\rm h}
 \equiv
 \mathcal L^{-1}_{\widehat z\to b}
 \left[2\widehat z\,\deltacal_E\right].
 \label{eq:N1-Jhard-def}
\end{equation}
The ${\cal N}{=}1$ JT background is set by~\cite{Johnson:2020heh}:
\begin{equation}
 x\equiv t_0=\mu=1\ ,
 \qquad
 t_k^{{\cal N}=1}=\frac{\pi^{2k}}{(k!)^2},
 \qquad k\geq1\ .
 \label{eq:N1-background}
\end{equation}
The  string equation, coming  from the appropriate family of {\it positive} multicritical random matrix models~\cite{Dalley:1991qg,Dalley:1991vr,Dalley:1992br}, is given by:
\begin{equation}
 u\Rcal^2-\frac{\hbar^2}{2}\Rcal\Rcal''
 +\frac{\hbar^2}{4}(\Rcal')^2
 =\hbar^2\Gamma^2\ ,
 \quad
 \Rcal\equiv\sum_{k=1}^{\infty}t_kR_k[u]+x\ .
 \label{eq:big-string-equation}
\end{equation}
Perturbation theory is developed around the $x{>}0$ regime, for which
$u_0(x){=}0$,
a quite different situation from the bosonic case. The function $u(x)$ then has expansion that begins:
\begin{equation}
u(x)=\hbar^2u_2(x)+\hbar^4u_4(x)+\hbar^6u_6(x)+\cdots ,
 \label{eq:N1-higher-u}
\end{equation}
and solving the string equation perturbatively gives:
\begin{align}
 u_2(x)
 &=\frac{\Gamma^2-\frac14}{x^2}\ ,
 \label{eq:N1-u2}\\
 u_4(x)
 &=-\frac{2t_1}{x^5}
 \left(\Gamma^2-\frac14\right)
 \left(\Gamma^2-\frac94\right)\ ,
 \label{eq:N1-u4}\\
 u_6(x)
 &=
 \left(\Gamma^2-\frac14\right)
 \left(\Gamma^2-\frac94\right)
 \Bigg[
 \frac{7t_1^2}{x^8}
 \left(\Gamma^2-\frac{21}{4}\right)
 \nonumber\\
 &\hspace{39mm}
 -\frac{2t_2}{x^7}
 \left(\Gamma^2-\frac{25}{4}\right)
 \Bigg]\ ,
 \label{eq:N1-u6}
\end{align}
which will be enough for studying up to genus $g{=}3$. Note that 
the familiar (non-zero) Neveu-Schwarz ${\cal N}{=}1$ Weil-Petersson volumes arise at $\Gamma=0$,
however leaving~$\Gamma$ arbitrary will be useful, since the entire construction will also compute results for the addition of Ramond punctures as well, as explained in refs.~\cite{Johnson:2026twg,Johnson:2026jls}.

In the absence of $u_0$, we must study how the boundary operator acts on the $u_{2g}$. Happily, ref.~\cite{Johnson:2026twg}  showed that it can be written elegantly in terms of the  Gel'fand--Dikii resolvent $\widehat R(x,E)$,
which obeys the equation~\cite{Gelfand:1975rn}:
\begin{equation}
 4(u-E)\widehat R^2
 -2\hbar^2\widehat R\,\widehat R''
 +\hbar^2(\widehat R')^2=1\ .
 \label{eq:N1-GD-resolvent}
\end{equation}
The general action of the loop operator on $u(x)$ is simply:
\begin{equation}
 \deltacal_Eu
 =2\frac{\partial}{\partial x}\widehat R(x,E)\ ,
 \label{eq:N1-loop-resolvent}
\end{equation}
so all we need is the genus expansion of ${\widehat R}(x,E)$, which can be obtained by recursively expanding~(\ref{eq:N1-GD-resolvent}). With $u_0$ and all of its derivatives vanishing on the hard-edge branch, the expansion is particularly simple.  Writing
$E=-\widehat z^2$, its first three non-trivial orders are:\footnote{These differential polynomials in the $u_{2g}$ and derivatives are in fact descendants of the Gel'fand--Dikii
polynomials $R_k[u]$, in terms of which ${\widehat R}(x,E)$ has an expansion. Further expanding 
$u=\sum_g\hbar^{2g}u_{2g}$ yields these mixed objects.}
\begin{align}
 \widehat R_1
 &=\frac{u_2}{4\widehat z^3}\ ,
 \label{eq:N1-Rhat1}\\
 \widehat R_2
 &=\frac{u_4}{4\widehat z^3}
 +\frac{u_2''-3u_2^2}{16\widehat z^5}\ ,
 \label{eq:N1-Rhat2}\\
 \widehat R_3
 &=\frac{u_6}{4\widehat z^3}
 +\frac{u_4''-6u_2u_4}{16\widehat z^5}
 \nonumber\\
 &\quad
 +\frac{
 u_2^{(4)}-10u_2u_2''
 -5(u_2')^2+10u_2^3
 }{64\widehat z^7}\ .
 \label{eq:N1-Rhat3}
\end{align}
With the length-space operator $\Jcal_b$ given in~(\ref{eq:N1-Jhard-def}),
equations~\eqref{eq:N1-loop-resolvent}--\eqref{eq:N1-Rhat3}
immediately give:
\begin{align}
 \Jcal_b^{\rm h}u_2
 &=u_2',
 \label{eq:N1-hard-u2}\\
 \Jcal_b^{\rm h}u_4
 &=\frac{\partial}{\partial x}
 \left[
 u_4+\frac{b^2}{24}
 \left(u_2''-3u_2^2\right)
 \right],
 \label{eq:N1-hard-u4}\\
 \Jcal_b^{\rm h}u_6
 &=\frac{\partial}{\partial x}
 \Bigg[
 u_6
 +\frac{b^2}{24}
 \left(u_4''-6u_2u_4\right)
 \nonumber\\
 &\hspace{9mm}
 +\frac{b^4}{1920}
 \left(
 u_2^{(4)}-10u_2u_2''
 -5(u_2')^2+10u_2^3
 \right)
 \Bigg]\ .
 \label{eq:N1-hard-u6}
\end{align}
and once again at fixed genus we have  finite polynomials in $b^2$. At this point, the structure for working out successive volumes by adding boundaries is extremely similar to the computations done for this class of models in ref.~\cite{Johnson:2026twg}. A number of surprising and consequential patterns arose upon looking at the how the computations were subsequently structured.

In retrospect, there is an even more efficient way to organize the computation.
Recall that when we can set $u_0{=}0$, the boundary operator in length space, can be written nicely, \eqref{eq:intro-Jb-t},  in terms of the KdV time derivatives. So we have:
\begin{equation}
 \Jcal_b^{\rm h}
 =
 \sum_{k=0}^{\infty}
 \frac{(-1)^k b^{2k}}{4^k(k!)^2}
 \frac{\partial}{\partial t_k}\ ,
 \qquad t_0\equiv x\ .
 \label{eq:N1-Jhard-times}
\end{equation}
While it is the same structure of operator as before, it is going to act on quite different data.

Now recall that the  $n$-boundary volume at genus $g$ is simply obtained by acting repeatedly on $F_g$, and so in length space, this results in the compact relation:
\begin{equation}
 V_{g,n}(b_1,\ldots,b_n)
 =
 -\left.
 \Jcal^{\rm h}_{b_1}\cdots\Jcal^{\rm h}_{b_n}F_g
 \right|_{x=1}\ .
 \label{eq:N1-Fg-master}
\end{equation}
But the $F_g$ are simply obtained by twice integrating the $u_{2g}$, and so applying this to
equations~\eqref{eq:N1-u2}--\eqref{eq:N1-u6} we get:
\begin{align}
 F_1
 &=-\frac12
 \left(\Gamma^2-\frac14\right)\log x\ ,
 \label{eq:N1-F1}\\
 F_2
 &=-\frac{t_1}{12x^3}
 \left(\Gamma^2-\frac14\right)
 \left(\Gamma^2-\frac94\right)\ ,
 \label{eq:N1-F2}\\
 F_3
 &=
 \left(\Gamma^2-\frac14\right)
 \left(\Gamma^2-\frac94\right)
 \Bigg[
 \frac{t_1^2}{12x^6}
 \left(\Gamma^2-\frac{21}{4}\right)
 \nonumber\\
 &\hspace{31mm}
 -\frac{t_2}{30x^5}
 \left(\Gamma^2-\frac{25}{4}\right)
 \Bigg]\ .
 \label{eq:N1-F3}
\end{align}
The $t_k$ and $x$ dependence is manifest, and so we can simply act with operator~(\ref{eq:N1-Jhard-times}) as many times as desired.  It is this that is at the heart of ref.~\cite{Johnson:2026twg}'s observation that this formalism allows for swift derivation of closed form formulae for $V_{g,n}$ for the ${\cal N}{=}1$ case. In fact, this direct action on the free energy makes the derivations even swifter! Let's see.

At genus one there is no dependence on any $t_k$ with $k{\geq1}$, so
every boundary insertion in equation~\eqref{eq:N1-Fg-master} simply
contributes an $x$-derivative.  Hence we immediately get:
\begin{equation}
 V_{1,n}
 =
 -\left.
 \frac{\partial^nF_1}{\partial x^n}
 \right|_{x=1}
 =
 (-1)^{n-1}\frac{(n-1)!}{2}
 \left(\Gamma^2-\frac14\right)\ .
 \label{eq:N1-g1-Gamma}
\end{equation}
Setting $\Gamma=0$ reproduces the well-known result:
\begin{equation}
 V_{1,n}=(-1)^n\frac{(n-1)!}{8}\ .
 \label{eq:N1-g1}
\end{equation}
In fact, there is more information since the $\Gamma^2$ term yields the genus zero result for $n$ NS-boundaries and 2 Ramond punctures (since $\widetilde\Gamma\equiv\hbar\Gamma$ counts such punctures~\cite{Johnson:2026twg,Johnson:2026jls}).

At genus two only the first two terms of
equation~\eqref{eq:N1-Jhard-times} can contribute:
\begin{equation}
 \Jcal_b^{\rm h}
 =
 \frac{\partial}{\partial x}
 -\frac{b^2}{4}\frac{\partial}{\partial t_1}.
 \label{eq:N1-Jhard-g2}
\end{equation}
Since $F_2$ is linear in $t_1$, in the product of $n$ boundary
operators either every boundary contributes an $x$-derivative, or
exactly one supplies a $t_1$-derivative.  The complete answer therefore
follows in one line:
\begin{equation}
 \begin{aligned}
 V_{2,n}(\bm b)
 ={}&(-1)^n\frac{(n+1)!}{96}
 \left(\Gamma^2-\frac14\right)
 \left(\Gamma^2-\frac94\right)
 \\
 &\times
 \left[
 4t_1(n+2)+\sum_{i=1}^n b_i^2
 \right].
 \end{aligned}
 \label{eq:N1-g2-Gamma}
\end{equation}
For $\Gamma{=}0$ and $t_1{=}\pi^2$ this becomes
the known result for~$n$ NS-boundaries:
\begin{equation}
 V_{2,n}
 =
 (-1)^n\frac{3(n+1)!}{2\cdot4^4}
 \left[
 (2\pi)^2(n+2)+\sum_{i=1}^n b_i^2
 \right],
 \label{eq:N1-g2}
\end{equation}
first derived by  Norbury~\cite{Norbury:2020vyi} (along with (\ref{eq:N1-g1})) using intersection theory and topological recursion. In the more general result we have here, the non-zero $\Gamma$ terms also encode closed form formulae volumes for two and four Ramond punctures at genus one and zero respectively, originally derived  in ref.~\cite{norbury2024superweilpeterssonmeasuresmoduli}. At higher genus,  many more such closed-form formulae were derived using these methods in ref.~\cite{Johnson:2026jls}.

Moving to genus three the appropriate operator is:
\begin{equation}
 \Jcal_b^{\rm h}
 =
 \frac{\partial}{\partial x}
 -\frac{b^2}{4}\frac{\partial}{\partial t_1}
 +\frac{b^4}{64}\frac{\partial}{\partial t_2}\ ,
 \label{eq:N1-Jhard-g3}
\end{equation} and from it we can readily anticipate the structure of the resulting volumes, by noticing how  the two couplings appear in $F_3$,  as $t_1^2$ and $t_2$:
The answer can
contain only
\begin{equation}
 1\ ,\quad
 \sum_i b_i^2\ ,\quad
 \sum_i b_i^4\ ,\quad\text{and}\quad
 \sum_{i<j}b_i^2b_j^2\ .
 \label{eq:N1-g3-structures}
\end{equation}
Applying equation~\eqref{eq:N1-Fg-master} gives:
\begin{widetext}
\begin{equation}
\begin{aligned}
 V_{3,n}(\bm b;\Gamma)
 ={}&(-1)^{n+1}\frac{(n+3)!}{5760}
 \left(\Gamma^2-\frac14\right)
 \left(\Gamma^2-\frac94\right)
 \Bigg[
 4(n+4)
 \left\{
 (n+5)\left(\Gamma^2-\frac{21}{4}\right)t_1^2
 -2\left(\Gamma^2-\frac{25}{4}\right)t_2
 \right\}
 \\
 &+2(n+4)\left(\Gamma^2-\frac{21}{4}\right)t_1
 \sum_{i=1}^n b_i^2
+\frac18\left(\Gamma^2-\frac{25}{4}\right)
 \sum_{i=1}^n b_i^4
 +\frac12\left(\Gamma^2-\frac{21}{4}\right)
 \sum_{i<j}b_i^2b_j^2
 \Bigg]\ ,
\end{aligned}
\label{eq:N1-g3-Gamma}
\end{equation}
and at $\Gamma=0$, with
$t_1=\pi^2$ and $t_2=\pi^4/4$, this reduces directly to:
\begin{equation}
\begin{aligned}
 V_{3,n}(\bm b)
 ={}&(-1)^n\frac{(n+3)!}{5\cdot4^8}
 \Bigg[
 (2\pi)^4(n+4)(42n+185)
 +84(2\pi)^2(n+4)\sum_i b_i^2
 +25\sum_i b_i^4
 +84\sum_{i<j}b_i^2b_j^2
 \Bigg],
\end{aligned}
\label{eq:N1-g3}
\end{equation}
the third of ref.~\cite{Norbury:2020vyi}'s closed-form formulae obtained by other means. Again, the $\Gamma^2$ dependence, once turned into a~${\widetilde\Gamma^2}$ dependence, readily gives  results involving even numbers of Ramond punctures at lower genera. In all cases, working with arbitrary $\Gamma$ computes results of having additional Ramond punctures, trading more punctures for genus according to:
\begin{equation}
 V^{(2m)}_{g-m,n}
 =
 \Gamma^{2m}\text{ coefficient in } V_{g,n}(\Gamma)\ ,
 \label{eq:N1-R-extraction}
\end{equation} where the superscript indicates $(2m)$ Ramond punctures. Ref.~\cite{Johnson:2026jls} explores many aspects of such volumes using these methods, complementing the  approaches used in  refs.~\cite{norbury2024superweilpeterssonmeasuresmoduli,Alexandrov:2024kuj}.
\end{widetext}

Overall, this simple way of organizing the calculation nicely explains their polynomial
structure.  At fixed genus the hard-edge $F_g$ contains only finitely
many terms in the KdV couplings, while a factor
$\partial/\partial t_k$ in a the  boundary insertion operator supplies a factor
$b^{2k}$.  The possible symmetric polynomials in the boundary lengths
are therefore  encoded in the coupling terms occurring in
$F_g$.  

The especially nice thing about this approach is that it immediately proves  why at {\it any genus} there will be a closed form formula for $V_{g,n}$. The genus 4 case was already worked out in ref.~\cite{Johnson:2026twg} and this approach readily achieves the same result. Increasing complexity comes from doing the combinatorics of partitioning the derivatives correctly, rather than matters of structure. It is straightforward, with the aid of {\tt Maple}, {\tt Mathematica}, or similar tools, to use this procedure to swiftly generate higher genus closed form formulae.

\subsection{Exact $T$-expansion at the removable cone point}

The form of the hard-edge boundary operator makes the removable cone point value
$b{=}2\pi\ii$ particularly interesting.  On the
${\cal N}{=}1$ background given by equation~\eqref{eq:N1-background},
equation~\eqref{eq:N1-Jhard-times} gives
\begin{equation}
 \Jcal_{2\pi\ii}^{\rm h}
 =
 \sum_{k=0}^{\infty}
 t_k^{{\cal N}=1}\frac{\partial}{\partial t_k},
 \qquad
 t_0^{{\cal N}=1}=1.
 \label{eq:N1-cone-point-Euler}
\end{equation}
So in terms of the classic  full Euler operator built from the
 KdV variables:
\begin{equation}
 {\cal E}_{\rm h}
 =
 x\frac{\partial}{\partial x}
 +\sum_{k=1}^{\infty}
 t_k\frac{\partial}{\partial t_k}.
 \label{eq:N1-Eh}
\end{equation}
we see that the ${\cal N}{=}1$ cone-point boundary operator is
precisely its restriction to the particular background ($x{=}t_0{=}1$, $t_k=t_k^{{\cal N}=1}$):
\begin{equation}
 \Jcal_{2\pi\ii}^{\rm h}
 =
 \left.{\cal E}_{\rm h}\right|_{{\cal N}=1}.
 \label{eq:N1-cone-point-Eh}
\end{equation}
Let's turn to what the operator does to $u$. This can be worked out from the string equation~\eqref{eq:big-string-equation} itself.
If we assign the scaling dimensions
\begin{equation}
 [x]=[t_k]=[\hbar]=1,\qquad
 [u]=[\Gamma]=0,\qquad
 [\partial_x]=-1,
\end{equation}
every Gel'fand--Dikii polynomial
$R_k[u]$ has weight zero, and every term in equation~\eqref{eq:big-string-equation} has
weight two:
\begin{equation}
 [u\Rcal^2]
 =
 [\hbar^2\Rcal\Rcal'']
 =
 [\hbar^2(\Rcal')^2]
 =
 [\hbar^2\Gamma^2]
 =2\ .
\end{equation}
The equation is invariant under
\begin{equation}
 x\longrightarrow\lambda x\ ,\qquad
 t_k\longrightarrow\lambda t_k\ ,\qquad
 \hbar\longrightarrow\lambda\hbar\ ,
\end{equation}
 So solutions
obey
\begin{equation}
 u(\lambda x,\{\lambda t_k\},\lambda\hbar)
 =
 u(x,\{t_k\},\hbar)\ .
\end{equation}
Differentiating with respect to $\lambda$ at $\lambda=1$ gives
\begin{equation}
 \left(
 x\frac{\partial}{\partial x}
 +\sum_{k=1}^{\infty}t_k\frac{\partial}{\partial t_k}
 \right)u
 =
 -\hbar\frac{\partial u}{\partial\hbar}\ .
 \label{eq:N1-scaling}
\end{equation}

Using
$u=\sum_{g=1}^{\infty}\hbar^{2g}u_{2g}$ therefore gives:
\begin{equation}
 {\cal E}_{\rm h}u_{2g}=-2g\,u_{2g}\ ,
 \qquad
 \left.
 \Jcal_{2\pi\ii}^{\rm h}u_{2g}
 \right|_{{\cal N}=1,x=1}
 =-2g\,u_{2g}(1)\ .
 \label{eq:N1-cone-point-u2g}
\end{equation}
From $ u_{2g}=2\frac{\partial^2F_g}{\partial x^2}$, we can deduce for the free energy:
\begin{equation}
 {\cal E}_{\rm h}F_g=-(2g-2)F_g\ . \label{eq:N1-cone-point-Fg}
\end{equation}
Now we combine this  with the representation
of the volumes derived in equation~\eqref{eq:N1-Fg-master}.  Noting that:
\begin{equation}
 \left[{\cal E}_{\rm h},\Jcal_b^{\rm h}\right]
 =-\Jcal_b^{\rm h}\ ,
 \label{eq:N1-Eh-J-comm}
\end{equation}
we can commute ${\cal E}_{\rm h}$ through the $n$ boundary insertions in
\eqref{eq:N1-Fg-master} to give:
\begin{equation}
 \left.
 \Jcal_{2\pi\ii}^{\rm h}\Vcal_{g,n}
 \right|_{{\cal N}=1,x=1}
 =
 -(2g-2+n)V_{g,n}\ ,
 \label{eq:N1-cone-point-volume}
\end{equation}
and finally the boundary-addition identity at the removable-cone point is:
\begin{equation}
 V_{g,n+1}(\bm b,2\pi\ii)
 =-(2g-2+n)V_{g,n}(\bm b)\ .
 \label{eq:N1-removable}
\end{equation}
We see that at this special point, the content of  the boundary operator  is simply a
scaling statement  on the space of couplings, amounting to a measure of Euler characteristic $\chi{=}2{-}2g{-}n$.

This removable-cone point relation
 was already obtained  by Norbury in ref.~\cite{Norbury:2020vyi}, in the conventions used there.
As noted there, it is actually much more powerful than its ordinary counterpart. A modification of the polynomial remainder argument given above equation~\eqref{eq:cone-ambiguity-condition}, just using this identity shows that, since the ${\cal N}=1$ volumes have degree $g-1$ in the squared boundary lengths, the additional requirement of  symmetry determines all higher $V_{g,n}$ once $V_{g,g-1}$ is known. This goes some way to showing why (as  observed  above and in ref.~\cite{Johnson:2026twg}) closed form formulae for all $n$ can emerge.

Just as for the ordinary Weil-Petersson volume identity, we can go  beyond this, using the removable-cone point
as a natural origin for an exact reorganization of the full
arbitrary-$b$ boundary insertion into a tower of derivative identities.  Introducing again 
$T{=}b^2+4\pi^2$, the universal expansion was given in
equation~(\ref{eq:J-T-expansion}), with (repeating for convenience):
\begin{equation}
 {\cal B}_m
 =
 \sum_{j=0}^{\infty}
 \frac{\pi^{2j}}{j!(j+m)!}
 \frac{\partial}{\partial t_{j+m}}\ ,
 \label{eq:N1-B-general}
\end{equation}
but now for the  ${\cal N}=1$ background~(\ref{eq:N1-background})
and so now
\begin{equation}
 \left.{\cal B}_m\right|_{{\cal N}=1}
 =
 \frac{1}{\pi^{2m}}
 \sum_{k=m}^{\infty}
 \frac{k!}{(k-m)!}\,
 t_k^{{\cal N}=1}
 \frac{\partial}{\partial t_k}\ .
 \label{eq:N1-Bm-special}
\end{equation}
The first three  are therefore
\begin{align}
 {\cal B}_0
 &=
 \sum_{k=0}^{\infty}
 t_k^{{\cal N}=1}
 \frac{\partial}{\partial t_k}
 =
 \left.{\cal E}_{\rm h}\right|_{{\cal N}=1},
 \label{eq:N1-B0}\\
 {\cal B}_1
 &=
 \frac1{\pi^2}
 \sum_{k=1}^{\infty}
 k\,t_k^{{\cal N}=1}
 \frac{\partial}{\partial t_k},
 \label{eq:N1-B1}\\
 {\cal B}_2
 &=
 \frac1{\pi^4}
 \sum_{k=2}^{\infty}
 k(k-1)t_k^{{\cal N}=1}
 \frac{\partial}{\partial t_k}.
 \label{eq:N1-B2}
\end{align}
The contrast with the ordinary WP case is interesting. For ordinary WP the ${\cal B}_0$ became the $L_{-1}$ operator, while  here it is  scaling.

Let's interpret the next operator.
In analogy with what happened with operator (\ref{eq:B2-Euler}), we see it is tracking changes under:
\begin{equation}
  t_k^{{\cal N}=1}\to t_k(s)=s^k t_k^{{\cal N}=1}\ ,
 \end{equation}
which again amounts to the replacement $\pi^2\to s\pi^2$ in the volumes. So it is  ${\cal B}_1$ that counts powers of $\pi^2$ in the
${\cal N}=1$ specialization.

The  expressions (\ref{eq:N1-F1})--(\ref{eq:N1-F3})  already show the weighted grading. Assigning $t_k$ degree $k$,  the genus $g$ free energy   $F_g$ has degree $g-1$.\footnote{This follows from the structure of the $x>0$ expansion, which gives us our solutions.  Recall that $u_0=0$ and the perturbative solution begins with $u_2=O(\hbar^2)$.  So the $k$th Gel'fand--Dikii term $t_kR_k[u]$ first enters the perturbative expansion at order $\hbar^{2k}$. This means that so the natural genus-counting assignment of degree to~$t_k$ is $k$.  A term with  $\prod_k t_k^{N_k}$ contributing to~$F_g$ therefore obeys $\sum_k kN_k{=}g-1$ (since we integrate twice and divide by~$\hbar^2$ to get $F$ from $u$).  (This is quite different from the ordinary bosonic soft-edge case, where $u_0\neq0$ and all $R_k[u_0]$ already participate in the classical string equation.) The number $g{-}1$ also has intersection theory meaning.
The complex dimension of $\Mbar_{g,n}$ is $3g-3+n$, while Norbury's $\Theta$-class~\cite{Norbury:2020vyi},
$\Theta_{g,n}$ has complex  degree
$2g-2+n$.  The remaining degree available to the full Weil--Petersson  class is therefore $g{-}1$.}
 Every action
$\partial/\partial t_k$ is accompanied in the boundary operator by a
factor $b^{2k}$.  Hence the physical volume has total degree $g-1$
when each power of $\pi^2$ and each power of $b_i^2$ is assigned
degree one.
Putting it all together as before we see that:
equation~\eqref{eq:N1-B1}  gives
\begin{equation}
 \left.
 {\cal B}_1\Vcal_{g,n}
 \right|_{{\cal N}=1,x=1}
 =
 \frac1{\pi^2}
 \left(
 g-1-\frac12{\cal E}_b
 \right)V_{g,n}\ .
 \label{eq:N1-B1-volume}
\end{equation}
So the first two non-vanishing orders of the  $T$-polynomial are:
\begin{equation}
\begin{aligned}
 V_{g,n+1}(\bm b,b)
 ={}&
 -(2g-2+n)V_{g,n}(\bm b)
 \\
 &-\frac{T}{4\pi^2}
 \left(
 g-1-\frac12{\cal E}_b
 \right)V_{g,n}(\bm b)
 +O(T^2)\ .
\end{aligned}
 \label{eq:N1-cone-first-correction}
\end{equation}
The higher ${\cal B}_m$, given by equation~\eqref{eq:N1-Bm-special},  involve
the successive falling factorial weights
$k(k-1)\cdots(k-m+1)$ on the KdV couplings.

\subsection{Intersection theory interpretation}
\label{sec:intersection-theory-N=1-WP}
There is a particularly clean intersection theory interpretation of this entire
tower, which can be compared directly with that of the ordinary
Weil--Petersson case discussed above in Section~\ref{sec:intersection-theory-WP}.  Norbury's $\Theta$-class~\cite{Norbury:2017eih}, which has degree $2g{-}2{+}n$, 
obeys the forgetful relation
\begin{equation}
 \Theta_{g,n+1}
 =
 \psi_{n+1}\,\pi^*\Theta_{g,n}\ .
 \label{eq:Theta-forget}
\end{equation}
It also has the special property that its restriction to a boundary
component containing a genus zero piece vanishes.  So it vanishes on the
stabilization divisors $D_{i,n+1}$ encountered in the ordinary
Weil--Petersson calculation in Section~\ref{sec:intersection-theory-WP}.  The divisor contributions that
produced the integral term there are therefore immediately absent in the
$\Theta$-weighted theory.

The (positive) ${\cal N}{=}1$ Weil-Petersson volume is~\cite{Norbury:2017eih}:
\begin{equation}
 \widehat V_{g,n}(\bm b)
 =
 \int_{\Mbar_{g,n}}
 \Theta_{g,n}
 \exp\!\left(
 2\pi^2\kappa_1+\frac12\sum_i b_i^2\psi_i
 \right).
 \label{eq:N1-Vhat}
\end{equation}
As in the ordinary Weil--Petersson calculation, we have the forgetful class relations~(\ref{eq:forgetful-class-relations}).
Here, every contribution containing $D_{i,n+1}$ vanishes
after multiplication by $\Theta_{g,n+1}$.  Using
equation~\eqref{eq:Theta-forget}, the $(n+1)$-boundary volume can
therefore be written as:
\begin{widetext}
\begin{align}
 \widehat V_{g,n+1}(\bm b,b)
 &=
 \int_{\Mbar_{g,n+1}}
 \psi_{n+1}\,
 \pi^*\!\left[
 \Theta_{g,n}
 \exp\!\left(
 2\pi^2\kappa_1+\frac12\sum_i b_i^2\psi_i
 \right)
 \right]
 \exp\!\left[
 \frac{b^2+4\pi^2}{2}\psi_{n+1}
 \right].
 \label{eq:N1-forgetful-step}
\end{align}
\end{widetext}
Expanding the final exponential and again using the relation between $\kappa$-classes and $\psi$-classes~(\ref{eq:psi-push-kappa}) gives the hierarchy:
\begin{equation}
 \widehat V_{g,n+1}(\bm b,b)
 =
 \sum_{m=0}^{\infty}
 \frac{T^m}{2^m m!}
 \widehat\Kcal_{g,n}^{(m)}\ .
 \label{eq:N1-T}
\end{equation}
where $T=b^2+4\pi^2$, and the $\kappa$-decorated super volumes are:
\begin{equation}
 \widehat\Kcal_{g,n}^{(m)}
 =
 \int_{\Mbar_{g,n}}
 \Theta_{g,n}\kappa_m
 \exp\!\left(
 2\pi^2\kappa_1+\frac12\sum_i b_i^2\psi_i
 \right).
 \label{eq:N1-Khat}
\end{equation}
The hierarchy truncates at the degree allowed by the moduli space dimension.   Changing conventions to the ones we use here:
\begin{equation}
 \widehat V_{g,n}=(-1)^n V_{g,n},
\end{equation}
equation~\eqref{eq:N1-full-offshell} becomes:
\begin{equation}
 \widehat V_{g,n+1}
 =
 -\Jcal_b^{\rm h}\widehat V_{g,n}.
 \label{eq:N1-signless-addition}
\end{equation}
Comparison of equations~\eqref{eq:J-T-expansion} and
\eqref{eq:N1-T} then gives the dictionary:
\begin{equation}
 \widehat\Kcal_{g,n}^{(m)}
 =
 \frac{(-1)^{m+1}}{2^m}
 \left.
 {\cal B}_m\widehat\Vcal_{g,n}
 \right|_{{\cal N}=1,x=1}\ ,
 \label{eq:N1-B-kappa-dictionary}
\end{equation}
where this time the ${\cal B}_m$ are  working on  the ${\cal N}{=}1$ background, as explored in the previous subsection.
In particular:
\begin{align}
 \widehat\Kcal_{g,n}^{(0)}
 &=
 -{\cal B}_0\widehat V_{g,n}
 =
 (2g-2+n)\widehat V_{g,n}\ ,
 \label{eq:N1-K0-B0}\\
 \widehat\Kcal_{g,n}^{(1)}
 &=
 \frac12{\cal B}_1\widehat V_{g,n}
 =
 \frac1{2\pi^2}
 \left(
 g-1-\frac12{\cal E}_b
 \right)\widehat V_{g,n}\ ,
 \label{eq:N1-K1-B1}\\
 \widehat\Kcal_{g,n}^{(2)}
 &=
 -\frac14{\cal B}_2\widehat V_{g,n}\ ,
 \label{eq:N1-K2-B2}
\end{align}
and similarly at every higher order.

At $T=0$, only the $m=0$ term in
equation~\eqref{eq:N1-T} remains.  Since
$\kappa_0=2g-2+n$, the first of these relations gives
\begin{equation}
 \widehat V_{g,n+1}(\bm b,2\pi\ii)
 =
 (2g-2+n)\widehat V_{g,n}(\bm b),
 \label{eq:N1-removable-signless}
\end{equation}
which is exactly equation~\eqref{eq:N1-removable} after converting
back to the $(-1)^n$ convention.  More generally,
equation~\eqref{eq:N1-B-kappa-dictionary} shows that the exact
$T$-expansion of the hard-edge boundary operator about the removable
cone point is the KdV realization of the sequence
$\kappa_0,\kappa_1,\kappa_2,\ldots$ generated by the
$\Theta$-class forgetful relation.

The contrast with ordinary Weil--Petersson theory is very clear in this form.  In both cases, adding and subsequently
forgetting the extra marked point generates the~$\kappa_m$ tower.
For ordinary Weil--Petersson volumes, the stabilization divisors
$D_{i,n+1}$ survive and generate the additional integral term.  For
the super-volume, the presence of the $\Theta$-class
removes  those contributions, leaving the  $\kappa_m$-decorated
tower.

\section{${\cal N}{=}2$ Weil-Petersson Volumes}
\label{sec:boundary-addition-WP-N2}

\subsection{The Specialization}

For ${\cal N}{=}2$ JT supergravity, which has a $U(1)$ R-symmetry, Turiaci and Witten~\cite{Turiaci:2023jfa}  gave a random matrix model description of the sectors with fixed R-charge~$q$. The leading spectral density has  a threshold energy~$E_0$.  The generalization of the Weil-Petersson volumes naturally emerges  in this fixed charge sector. The multicritical description of the model developed in ref.~\cite{Johnson:2023ofr} provides a natural language for their exploration~\cite{Johnson:2025oty,Ahmed:2025lxe}.

For non-zero $E_0$, this is again a soft-edge problem, in contrast to the
${\cal N}{=}1$ case. The  endpoint now  lies at $u_0(\mu)=E_0$ instead of $u_0=0$. 

The $t_k$ for use in string equation~(\ref{eq:big-string-equation}) are~\cite{Johnson:2023ofr}:
\begin{equation}
 t_k(E_0)=
 \frac{\pi^{k-1}J_k(2\pi\sqrt{E_0})}
 {2(2k+1)k!E_0^{k/2}}\ ,\quad \mu=t_0=
 \frac{J_0(2\pi\sqrt{E_0})}{2\pi}\ ,
 \label{eq:N2-tk}
\end{equation} and moreover a large number $\Gamma$ of BPS states are turned on such that $\widetilde\Gamma=\hbar\Gamma$ is finite in the $\hbar\to0$ classical limit. Its value is~\cite{Turiaci:2023jfa,Mertens:2017mtv,Stanford:2017thb}:
\begin{align}\quad 
 \widetilde\Gamma=
 \frac{\sin(2\pi\sqrt{E_0})}{4\pi^2}\ .
 \label{eq:N2-background}
\end{align}
The classical limit has $R_k[u]\to u_0^k$ in the string equation, as usual. It is useful to define:
\begin{equation}
G_0(u_0)=\sum_{k=1}^{\infty}t_k(E_0)u_0^k\ ,
 \label{eq:N2-G0}
\end{equation}
and then the leading string equation obtained from~\eqref{eq:big-string-equation}
may then be written:
\begin{equation}
 x=-{\widetilde G}(u_0)\quad \text{where}\quad {\widetilde G}(u_0)\equiv G_0(u_0)- \frac{\widetilde\Gamma}{\sqrt u_0}\ .
 \label{eq:N2-F}
\end{equation}
The endpoint condition is
$ {\widetilde G}(E_0){+}\mu{=}0$, and the $u_0$-derivative data are readily obtained by successive differentiations, $ u_0'{=}-1/\dot {\widetilde G}$, 
$ u_0''{=}-{\ddot {\widetilde G}}/(\dot {\widetilde G})^3$, {\it etc.,} 
where a dot is a $u_0$ derivative.  Evaluating the  sums at $u{=}E_0$ gives:
\begin{align}
 u_0'(\mu)&=-4\pi E_0\ ,
 \nonumber\\
 u_0''(\mu)&=
 24\pi^2E_0-16\pi^4E_0^2\ ,
 \nonumber\\
 u_0'''(\mu)&=
 -192\pi^3E_0
 +416\pi^5E_0^2
 -160\pi^7E_0^3\ .
 \label{eq:N2-jets}
\end{align}
The model dependent normalization discussed earlier above~\eqref{eq:master-physical} is~\cite{Ahmed:2025lxe}:
\begin{align}
 \left.\Vcal_{g,n}\right|_{{\cal N}=2}
 =K_{g,n}V^{\mathcal N=2}_{g,n}\ ,
\quad
 K_{g,n}=(2\pi)^{-\chi} \ ,
  \label{eq:N2-K}
\end{align}
 where  $\chi{=}2{-}2g{-}n$. Using
 ${K_{g,n}}/{K_{g,n+1}}{=}1/{2\pi}$,
and   $u_0'(\mu){=}-4\pi E_0$,  
equation~\eqref{eq:master-physical} gives the full  ${\cal N}{=}2$ boundary-addition identity for general $b$:
\begin{widetext}
\begin{equation}
\begin{aligned}
 V^{\mathcal N=2}_{g,n+1}(\bm b,b)
 ={}
 E_0\sum_{i=1}^{n}\int_0^{b_i}y\,
 V^{\mathcal N=2}_{g,n}(\ldots,y,\ldots)\dd y
 +
 \frac1{K_{g,n+1}}
 \left.
 \Jcal_b\Vcal_{g,n}(\bm b;x)
 \right|_{{\cal N}=2}.
\end{aligned}
 \label{eq:N2-full-identity}
\end{equation}      
\end{widetext}
The two universal off-shell formulae in
equations~\eqref{eq:intro-V03-V11-off},  evaluated with~\eqref{eq:N2-jets} and~\eqref{eq:N2-K} give:
\begin{equation}
 V^{\mathcal N=2}_{0,3}=E_0,
 \qquad
 V^{\mathcal N=2}_{1,1}(b)
 =
 -\frac18+\frac{b^2+4\pi^2}{48}E_0\ .
 \label{eq:N2-first-physical-volumes}
\end{equation}
after using that $K_{0,3}=K_{1,1}=2\pi$. We can get $V_{0,4}$ and $V_{1,2}$ using~\eqref{eq:N2-full-identity} as follows. Starting from
$\Vcal_{0,3}=-u_0'/2$ and
equation~\eqref{eq:softedge-first-jet-actions},
\begin{equation}
 \frac1{K_{0,4}}
 \left.
 \Jcal_b\Vcal_{0,3}
 \right|_{{\cal N}=2}
 =
 -3E_0
 +\frac{E_0^2}{2}(b^2+4\pi^2)\ .
 \label{eq:N2-V03-residual}
\end{equation}
The integral term supplies
$\frac{E_0^2}{2}
 \sum_{i=1}^{3}b_i^2$, and
adding the two pieces gives the known result
\cite{Turiaci:2023jfa,Ahmed:2025lxe}:
\begin{equation}
 V^{\mathcal N=2}_{0,4}
 =
 -3E_0
 +\frac{E_0^2}{2}
 \left(
 4\pi^2+\sum_{i=1}^{4}b_i^2
 \right)\ .
 \label{eq:N2-V04}
\end{equation}
The boundary action on $\Vcal_{1,1}$ is
\begin{equation}
\begin{aligned}
 \Jcal_{b_2}\Vcal_{1,1}(b_1)
 ={}&
 \frac1{24}
 \left[
 \frac{\Jcal_{b_2}u_0''}{u_0'}
 -
 \frac{u_0''}{(u_0')^2}
 \Jcal_{b_2}u_0'
 \right]
 -
 \frac{b_1^2}{96}
 \Jcal_{b_2}u_0'\ .
\end{aligned}
 \label{eq:soft-JV11}
\end{equation}
Substitution of equations~\eqref{eq:N2-jets} and
\eqref{eq:softedge-first-jet-actions}, together with the integral term in
equation~\eqref{eq:N2-full-identity}, gives the known result~\cite{Ahmed:2025lxe}:
\begin{equation}
\begin{aligned}
 V^{\mathcal N=2}_{1,2}(b_1,b_2)
 ={}&
 \frac18
 -
 \frac{
 3(b_1^2+b_2^2)+14\pi^2
 }{24}E_0\nonumber\\
 {}&+
 \frac{
 (4\pi^2+b_1^2+b_2^2)
 (12\pi^2+b_1^2+b_2^2)
 }{192}E_0^2\ .
\end{aligned}
 \label{eq:N2-V12}
\end{equation}

\subsection{Exact $T$-expansion at the removable cone point}

Just as in the previous cases, it is useful to reorganize the complete
arbitrary-$b$ boundary insertion around the removable cone point. Here
the endpoint value $u_0(\mu)$ is nonzero, so the simple $t_k$-space form of the ${\cal B}_m$
operators is not going to work here. Instead the exact Bessel $T$-expansion~\eqref{eq:softedge-exact-T}, introduced during  the ordinary WP discussion,
applies directly. We have:
\begin{equation}
 T=b^2+4\pi^2\ ,
 \qquad
 \Delta(s)=U(s)-E_0\ .
 \label{eq:N2-T-Delta}
\end{equation}
Equation~\eqref{eq:softedge-exact-T} therefore becomes:
\begin{widetext}
\begin{equation}
 \begin{aligned}
 \sum_{m=0}^{\infty}\frac{s^m}{m!}
 \Jcal_bu_0^{(m)}(\mu)
 ={}&
 \sum_{r=0}^{\infty}
 \frac{(-T/4)^r}{r!}\,
 U'(s)
 \frac{\Delta(s)^{r/2}}{\pi^r}
 I_r\!\left(2\pi\sqrt{\Delta(s)}\right)
 \ ,
 \end{aligned}
 \label{eq:N2-exact-T}
\end{equation}
\end{widetext}
an exact polynomial in $T$, rewriting  the general $b$ result.  It truncates at finite order
in $T$ on every fixed endpoint derivative and hence on a fixed genus
volume.
The zeroth level of this exact expansion is of course the cone point itself, $b{=}2\pi \ii$.
Here, the action simplifies much further.

The $r{=}0$ term in equation~\eqref{eq:N2-exact-T} at $b{=}2\pi \ii$ is:
\begin{equation}
 \sum_{m=0}^{\infty}\frac{s^m}{m!}
  \Jcal_{2\pi\ii}u_0^{(m)}(\mu)
 =
 U'(s)
 J_0\!\left(
 2\pi\sqrt{E_0-U(s)}
 \right)\ .
 \label{eq:N2-generating-star}
\end{equation} This can be simplified because the string equation allows the resummation (for the form of the $t_k$ in~\eqref{eq:N2-tk}):
\begin{align}
 &2\pi
 \left[
 \mu+G_0+2u_0{\dot G_0}
 \right]
 =
 2\pi[\mu+{\widetilde G}(u_0)+2u_0{\dot{\widetilde G}}(u_0)]\nonumber\\
 &\hskip2cm=J_0(2\pi\sqrt{E_0})
 +
 \sum_{k=1}^{\infty}
 \frac{
 \pi^kJ_k(2\pi\sqrt{E_0})
 }{
 k!E_0^{k/2}
 }u_0^k
 \nonumber\\&\hskip2cm=
 J_0\!\left(
 2\pi\sqrt{E_0-u_0}
 \right).
 \label{eq:N2-Bessel}
\end{align}
Notice that this can also be written as:
\begin{equation}
 2\pi\left(2D_{u_0}+1\right)
 \bigl(\widetilde G(u_0)+\mu\bigr)
 =
 J_0\!\left(
 2\pi\sqrt{E_0-u_0}
 \right)\ ,
 \label{eq:euler-form}
\end{equation}
where we've defined:
\begin{equation}
 D_{u_0}
 \equiv
 u_0\frac{\partial}{\partial u_0}\ ,
 \label{eq:Euler-u0}
\end{equation} and so $2D_{u_0}+1$
naturally removes the denominator $2k{+}1$ in every $t_k$ in the sum, making the resumming into Bessel possible. (Note that $(2D_{u_0}+1)$ removes the $u_0^{-1/2}$ part of ${\widetilde G}(u_0)$.)
As an aside, the combination on the right hand side is interesting. At fixed $x$,  a variation of the inverse string equation $x{+}{\widetilde G}(u_0(x)){=}0$, together with ${\dot{\widetilde G}}(u_0)u_0'{=}{-}1$, gives:
\begin{equation}
 \delta_E{\widetilde G}(u_0)
 =
 \frac1{2(u_0-E)^{3/2}}\ ,
 \label{eq:N2-delta-FJ}
\end{equation}
which inverse Laplace
transforms into:
\begin{equation}
 \Jcal_b{\widetilde G}(u_0)
 =
 J_0\!\left(
 b\sqrt{u_0-E}
 \right)\ ,
 \label{eq:N2-JF-general}
\end{equation}
and so at $b{=}2\pi\ii$ we have the interesting form:
\begin{equation}
 2\pi\left(2D_{u_0}+1\right)
 \bigl(\widetilde G(u_0)+\mu\bigr)
 =
 \Jcal_{2\pi\ii}{\widetilde G}(u_0)\ ,
 \label{eq:N2-string-function-relation}
\end{equation}
to which we will return later. Carrying on with the calculation, evaluating at $u_0(\mu{+}s){=}U(s)$ and using $ {\widetilde G}(U(s)){=}{-}\mu{-}s$ and
$U'(s){\dot {\widetilde G}}(U(s)){=}-1$
gives:
\begin{align}
 &U'(s)
 J_0\!\left(
 2\pi\sqrt{E_0-U(s)}
 \right)
 =
 -2\pi
 \left[
 sU'(s)+2U(s)
 \right]\nonumber\\
 &\hskip2cm =
 -2\pi\sum_{m=0}^{\infty}
 (m+2)u_0^{(m)}(\mu)
 \frac{s^m}{m!}\ ,
 \label{eq:N2-generating-final}
\end{align}
and so we see from~\eqref{eq:N2-generating-star} that:
\begin{equation}
 \Jcal_{2\pi\ii}u_0^{(m)}(\mu)
 =
 -2\pi(m+2)u_0^{(m)}(\mu)\ .
 \label{eq:N2-all-jet}
\end{equation} 
Just as we saw with ordinary and ${\cal N}{=}1$ Weil-Petersson,  this  extracts a particular property of  a general
off-shell volume, since they are built from the $u_0^{(m)}$.  
The natural object emerging here is the  derivative-weight operator we'll denote as:
\begin{equation}
 {\cal H}_2
 =
 \sum_{m=0}^{\infty}
 (m+2)u_0^{(m)}
 \frac{\partial}{\partial u_0^{(m)}}\ ,
 \label{eq:N2-H2}
\end{equation}
in terms of which 
\begin{equation}
 \Jcal_{2\pi\ii}=-2\pi{\cal H}_2\ .
 \label{eq:N2-Jstar-H2}
\end{equation}
On the other hand at fixed genus the volumes' length grading allow us to write:
\begin{equation}
 \left(
 {\cal H}_2-\Ecal_b
 \right)\Vcal_{g,n}
 =
 n\,\Vcal_{g,n}\ ,
 \qquad
 \Ecal_b=
 \sum_{i=1}^{n}
 b_i\frac{\partial}{\partial b_i}\ .
 \label{eq:N2-jet-homogeneity}
\end{equation}
This grading is already visible in the lowest examples.  For
$\Vcal_{0,3}=-u_0'/2$, the derivative weight is three, equal to
$n=3$.  In $\Vcal_{1,1}$, the term $u_0''/u_0'$ has derivative
weight one and no boundary length weight, while the term $b^2u_0'$ has derivative weight three and
boundary length degree two; both therefore have
${\cal H}_2-\Ecal_b$ weight one.  The same balance persists in the
universal higher-genus expressions: additional powers of boundary
length accompany the corresponding increase in derivative weight.

Combining equations~\eqref{eq:N2-K},
\eqref{eq:N2-Jstar-H2}, and
\eqref{eq:N2-jet-homogeneity} gives
\begin{equation}
 \frac1{K_{g,n+1}}
 \left.
 \Jcal_{2\pi\ii}\Vcal_{g,n}
 \right|_{{\cal N}=2}
 =
 -\left(n+\Ecal_b\right)
 V^{\mathcal N=2}_{g,n}.
 \label{eq:N2-star-residual-general}
\end{equation}
and the full boundary-addition identity at the removable-cone point is therefore:
\begin{widetext}
\begin{equation}
\begin{aligned}
 V^{\mathcal N=2}_{g,n+1}
 (\bm b,2\pi\ii)
 ={}&
 E_0\sum_{i=1}^{n}
 \int_0^{b_i}y\,
 V^{\mathcal N=2}_{g,n}
 (\ldots,y,\ldots)\dd y
 -
 \left(n+\Ecal_b\right)
 V^{\mathcal N=2}_{g,n}(\bm b)\ .
\end{aligned}
 \label{eq:N2-special}
\end{equation}
The removable-cone point boundary action is this time  neither annihilating, as in
ordinary WP volumes (see equation~\eqref{eq:Jstar-jets-JT}), nor acting as the hard-edge Euler dilatation in coupling space (measuring Euler characteristic) as it was for
the ${\cal N}=1$ case (see equation~\eqref{eq:N1-cone-point-volume}). Instead it becomes more akin to a dilatation operator acting on boundaries, measuring their number combined with boundary-length scaling.

As an interesting example, for $V_{0,3}=E_0$,
equation~\eqref{eq:N2-special} gives:
\begin{equation}
 V^{\mathcal N=2}_{0,4}
 (b_1,b_2,b_3,2\pi\ii)
 =
 \frac{E_0^2}{2}
 \sum_{i=1}^{3}b_i^2-3E_0
 \ ,
 \label{eq:N2-V04-star}
\end{equation}
which is  equation~\eqref{eq:N2-V04} at the removable cone.
 For the genus one case, equation~\eqref{eq:N2-special}
gives:
\begin{equation}
 V^{\mathcal N=2}_{1,2}(b,2\pi\ii)
 =
 E_0\int_0^b
 yV^{\mathcal N=2}_{1,1}(y)\dd y
 -
 \left(
 1+b\frac{\partial}{\partial b}
 \right)
 V^{\mathcal N=2}_{1,1}(b)
 =
\frac{
24
-\left(24b^2+16\pi^2\right)E_0
+\left(b^4+8\pi^2b^2\right)E_0^2
}{192}.
\end{equation}
Direct substitution of $b_2=2\pi\ii$ into
equation~\eqref{eq:N2-V12} gives  the same result.
    
\end{widetext}

The higher powers of $T$ in equation~\eqref{eq:N2-exact-T} give the
successive contributions  away from the removable cone point, and just we with the ordinary Weil-Petersson case, they supplement the $T=0$ result and  reconstruct the complete  boundary insertion for arbitrary $b$. We won't explore them explicitly much further here, although it is clearly interesting to do so further. Following what happened for the ordinary and ${\cal N}{=}1$ Weil-Petersson cases, it is natural to expect that they have an interpretation as $\kappa$-deformed volumes in some appropriate intersection theory description of the ${\cal N}{=}2$ volumes. Such a description does not seem to have been provided in the literature so far,  so  this leaves us with the interesting task of deducing what we can about it using what we've learned from the boundary-addition identity.

\subsection{Intersection theory and ${\cal N}{=}2$ WP Volumes}
\label{sec:N2-intersection-theory}

There are some rather suggestive clues to help  deduce  an intersection theory definition of the ${\cal N}{=}2$ Weil-Petersson volumes that would sit alongside the definition~\eqref{eq:intersection-volume} for ordinary WP and definition~\eqref{eq:N1-Vhat} for the~${\cal N}{=}1$ case.  

The boundary-addition identity itself~\eqref{eq:N2-full-identity} provides the first big lead.  The integral term,
by direct comparison with what we learned of its forgetful-map origin in the ordinary Weil-Petersson case (and its absence in the ${\cal N}{=}1$ case), ought to have  an interpretation again in terms of the 
stabilization of divisors. That interpretation should also explain the $E_0$ factor.

Meanwhile the $\Jcal_b$ contribution will then come from (again by analogy with the previous cases) a tower of $\kappa$-class decorated volumes associated with the smooth (non-divisor) part of the
forgetful map.   However, the precise nature of such objects can't be guessed without information about what goes into the construction of an {\it undecorated} volume. So our main mission for this Subsection (and the analogous one for small ${\cal N}{=}4$ later) will be to find this information.

In pursuit of this goal, there is an independent clue from the simple ``universal''  model
identified in ref.~\cite{Johnson:2026plw} as playing the role of the core topological model at the heart of the extended supersymmetric volumes. To put this idea in context, note that in a sense the Airy model, involving  insertions of the $\psi$-class,  is the (Kontsevich-Witten~\cite{Witten:1989ig,Witten:1990hr,Kontsevich:1992ti}) topological core for ordinary Weil-Petersson, while the Bessel model is the (Do-Norbury~\cite{Do:2016odu}) topological core for ${\cal N}{=}1$ case. These models are then, roughly speaking, ``gravitationally dressed'' by Wolpert's $\kappa_1$-class factor $\exp(2\pi^2\kappa_1)$ to make the full volume. 

The  model discussed in ref.~\cite{Johnson:2026plw} is a type of gapped interpolation between Airy and Bessel,  where the  gap parameter plays the
role of an $E_0$-controlled deformation away from the Bessel theory toward Airy. 
A core point emphasized in the work is that it is a perfect toy model of the structure seen in ${\cal N}{=}2$ and small ${\cal N}{=}4$ models where the volumes have expansions in the deformation parameter ($E_0$ or $J$) whose lowest order (undeformed) term is Bessel, and the highest order is the bosonic WP volume. In the toy, bosonic WP is played by the analogous Airy quantity. So a route to finding the description of ${\cal N}{=}2$ and ${\cal N}{=}4$ volumes is to understand the toy model, and then seek the appropriate gravitational dressing.

The topological model has a description in terms of the  $r{=}2$ deformed $\Theta$-class of Chidambaram, Garcia-Failde, and Giacchetto~\cite{Chidambaram:2022cqc}.  This deformed $\Theta$-class has an especially useful representation in terms of $\kappa$-classes:
\begin{equation}
 \Theta^{2,\epsilon}_{g,n}
 =
 (-\epsilon^2)^{2g-2+n}
 \exp\!\left[
 \sum_{m\geq1}
 s_m(-\epsilon^2)^{-m}\kappa_m
 \right]\ ,
 \label{eq:N2-deformed-Theta}
\end{equation}
where the numbers $s_m$ are defined by the relation:
\begin{equation}
 \exp\!\left(
 -\sum_{m\geq1}s_m v^m
 \right)
 =
 \sum_{k\geq0}
 (-1)^k(2k+1)!!\,v^k\ .
 \label{eq:N2-deformed-Theta-times}
\end{equation}
The first few coefficients are:
\begin{equation}
 s_1=3\ ,\qquad
 s_2=-\frac{21}{2}\ ,\qquad
 s_3=69\ ,\qquad
 s_4=-\frac{2529}{4}\ .
 \label{eq:N2-deformed-Theta-first}
\end{equation}
Since $\kappa_m$ has degree $m$ this class has a highest degree of $d=2g-2+n$, along with  terms with lower degrees coming from expanding the exponential, where the terms of degree $d-q$ come with coefficient $(-\epsilon^2)^{d-q}$.

The undeformed limit $\epsilon\to0$ keeps the   highest-degree
component of $\exp\left(\sum_{i\geq1} s_i\kappa_i\right)$, which is in fact Norbury's $\Theta$ class~\cite{Kazarian:2021zqh,Chidambaram:2022cqc,Norbury:2017eih}, and indeed describes the Bessel model (before gravitational dressing). The  natural identification of~$E_0$ with the deformation parameter is
$\epsilon^2=E_0$, together with $\Omega^{\rm AB}_{g,n}=(-1)^{2g-2+n}\Theta^{2,\epsilon}_{g,n}$ in the volume convention used here.  
This gives the basic description for the pure topological theory as an insertion of the  class (``AB" for Airy-Bessel):
\begin{equation}
 \Omega^{\rm AB}_{g,n}(E_0)
 =
 E_0^{\,2g-2+n}
 \exp\!\left[
 \sum_{i\geq1}
 \sigma_i^{\rm AB}(E_0)\kappa_i
 \right]\ ,
 \label{eq:N2-bare-class}
\end{equation}
with, for example:
\begin{equation}
 \sigma_1^{\rm AB}
 =
 -\frac3{E_0}\ ,\qquad
 \sigma_2^{\rm AB}
 =
 -\frac{21}{2E_0^2}\ ,\qquad
 \sigma_3^{\rm AB}
 =
 -\frac{69}{E_0^3}\ .
 \label{eq:N2-bare-sigma}
\end{equation}
It is useful to see explicitly how this works alongside the computations for  the topological model done in ref.~\cite{Johnson:2026plw},  focusing initially on  the correlators $W_{g,1}(z)$ for one boundary.  They'd be expected to be:
\begin{equation}
 W^{\rm AB}_{g,1}(z)
 =
 \sum_{d\geq0}
 \frac{(2d+1)!!}{z^{2d+2}}
 \int_{\overline{\mathcal M}_{g,1}}
 \Omega^{\rm AB}_{g,1}(E_0)\,
 \psi_1^d
 \ ,
 \label{eq:N2-bare-Wg1}
\end{equation}
and we must remember that the highest degree $d$ of $\Omega^{\rm AB}$ is $2g-2+n$, while the dimension of ${\mathcal M}_{g,n}$ is $3g-3+n$.
For $(g,n)=(1,1)$, we have that (since $d=1$):
\begin{equation}
 \Omega^{\rm AB}_{1,1}
 =
 E_0-3\kappa_1\ .
\end{equation}
Using the results (see Appendix~\ref{app:intersection-integrals}):
\begin{equation}
 \int_{\overline{\mathcal M}_{1,1}}\psi_1
 =
 \int_{\overline{\mathcal M}_{1,1}}\kappa_1
 =
 \frac1{24}\ ,
\end{equation}
equation~\eqref{eq:N2-bare-Wg1} gives:
\begin{align}
 W^{\rm AB}_{1,1}(z)
 =
 \frac{3!!\,E_0}{z^4}\frac1{24}
 -
 \frac{3}{z^2}\frac1{24}
 =
 -\frac1{8z^2}
 +\frac{E_0}{8z^4}\ ,
 \label{eq:N2-bare-W11-check}
\end{align}
confirming what was computed in ref~\cite{Johnson:2026plw} using random matrix model methods.
At genus two we  have terms through
$d=3$:
\begin{align}
 \Omega^{\rm AB}_{2,1}
 ={}&
 E_0^3
 -3E_0^2\kappa_1
 +E_0\left(
 \frac92\kappa_1^2-\frac{21}{2}\kappa_2
 \right)
 \nonumber\\
 &\quad
 -\frac92\kappa_1^3
 +\frac{63}{2}\kappa_1\kappa_2
 -69\kappa_3\ .
 \label{eq:N2-bare-Omega21}
\end{align}
Consulting handy references on  intersection numbers (such as refs.~\cite{KockPsiClasses,Kaufmann:1996aa}, or Appendix~\ref{app:intersection-integrals}) gives:
\begin{align}
 &\int_{\overline{\mathcal M}_{2,1}}\kappa_1^2\psi_1^2
 =
 \frac{139}{5760}\ ,
 \quad
 \int_{\overline{\mathcal M}_{2,1}}\kappa_1\kappa_2\psi_1
 =
 \frac{101}{5760}\ ,
 \nonumber
 \\
 &\int_{\overline{\mathcal M}_{2,1}}\kappa_1^3\psi_1
 =
 \frac{169}{1920}\ ,\quad 
 \int_{\overline{\mathcal M}_{2,1}}\kappa_2\psi_1^2 =
 \frac{29}{5760}\ ,\nonumber\\
 &\int_{\overline{\mathcal M}_{2,1}}\kappa_3\psi_1 =
 \frac{1}{384}\ ,
\end{align}
and so:
\begin{align}
 &\int
 \left(
 \frac92\kappa_1^2-\frac{21}{2}\kappa_2
 \right)\psi_1^2
 =
 \frac{107}{1920}\ ,
 \nonumber\\
 &\int
 \left(
 -\frac92\kappa_1^3
 +\frac{63}{2}\kappa_1\kappa_2
 -69\kappa_3
 \right)\psi_1
 =
 -\frac3{128}\ ,
\end{align}
and then substitution into equation~\eqref{eq:N2-bare-Wg1}  gives:
\begin{align}
 W^{\rm AB}_{2,1}(z)
 ={}&
 -\frac{9}{128z^4}
 +\frac{107E_0}{128z^6}
 -\frac{203E_0^2}{128z^8}
 +\frac{105E_0^3}{128z^{10}}\ ,
 \label{eq:N2-bare-W21-check}
\end{align}
the random matrix model result of
ref.~\cite{Johnson:2026plw}.

A first natural guess for how to gravitationally dress this topological theory is to mimic what happens for ordinary 
Weil--Petersson (see equation~\eqref{eq:intersection-volume}), simply  multiplying 
\eqref{eq:N2-bare-class} by $\exp(2\pi^2\kappa_1)$.  This would give:
\begin{equation}
 \Omega^{\rm naive}_{g,n}(E_0)
 =
 \Omega^{\rm AB}_{g,n}(E_0)
 \exp(2\pi^2\kappa_1)\ .
 \label{eq:N2-naive-class}
\end{equation}
The dressing shifts  the  numbers in~(\ref{eq:N2-bare-sigma}), affecting how all the $\kappa_m$ involve couple into the model. For example:
\begin{equation}
 \sigma_1^{\rm naive}
 =
 2\pi^2-\frac3{E_0}\ ,
 \label{eq:N2-naive-sigma1}
\end{equation} and since this affects   $\kappa_1$, it allows us to see if we can  construct 
$V^{\mathcal N=2}_{0,3}$, $V^{\mathcal N=2}_{0,4}$ and
$V^{\mathcal N=2}_{1,1}$ (for which the dimension of moduli space is at most 1). The guess turns out to be successful for those  checks.   Moving forward, a   volume capable of seeing the~$\kappa_2$ direction is $V^{\mathcal N=2}_{1,2}$, which we successfully computed from boundary-addition identity in
equation~\eqref{eq:N2-V12}, and a computation shows that the naive guess fails:
\begin{equation}
 V^{\mathcal N=2}_{1,2}
 -
 V^{\rm naive}_{1,2}
 =
 \frac{\pi^2E_0}{6}\ .
 \label{eq:N2-naive-failure}
\end{equation}
The discrepancy can be repaired by adjusting the $\kappa_2$ contribution by adding:
 \begin{equation}
\delta\sigma_2(E_0)
 =
 \frac{4\pi^2}{E_0}\ ,
 \label{eq:N2-sigma2-correction}
\end{equation}
which then fixes $V_{1,2}$, but of course there will be adjustments to be made on the infinite set of $\kappa_m$. The main lesson here is already  that the gravitational dressing cannot consist only of the
familiar $\exp(2\pi^2\kappa_1)$ factor.  An entire additional family of $\kappa$-class modifications needs to be determined. Successive computation of more examples of volumes is in principle the way to the answer,  but clearly a systematic approach is needed. 

Happily, we have built just the right set of tools for the job in earlier sections.
In Subsection~\ref{sec:WP-all-cusps-genus-zero} we noted that at  genus zero, the volumes for the case where all boundaries are zero ({\it i.e.,} all-cusp volumes)
 are generated simply by successive  $x$-derivatives of the 
string equation function~$u_0(x)$. This is a universal result, and so if we use the appropriate $u_0(x)$ for this ${\cal N}{=}2$ case, we'll have all the cuspy volumes $V_{0,n}^{{\cal N}{=}2}({\bf 0})$ contained in one portable device. The dimension of the moduli space  grows with $n$, meaning that successively higher $\kappa_i$ will be able to contribute on the intersection theory side of this computation, allowing us to constrain it with our generating function coefficients. 

First we fix the normalization. In the  ${\cal N}{=}2$ normalization we discussed earlier we have:
\begin{equation}
 V^{\mathcal N=2}_{0,n}(0,\ldots,0)
 =
 -\frac{1}{2(2\pi)^{n-2}}
 \left.
 \frac{\partial^{n-2}u_0}{\partial x^{n-2}}
 \right|_{x=\mu}\ ,
 \quad n\geq3.
 \label{eq:N2-all-cusp-genus-zero}
\end{equation}
To parallel the all-genus cusp generating functions introduced earlier for
ordinary Weil--Petersson volumes, define
\begin{equation}
 \mathscr H_0(z;E_0)
 =
 \sum_{n=3}^{\infty}
 \frac{V^{\mathcal N=2}_{0,n}(0,\ldots,0)}
 {E_0^{\,n-2}n!}\,z^n\ ,
 \label{eq:N2-scriptH0-def}
\end{equation}
(using a script font for this supersymmetric context). $\mathscr H_0$'s  first derivative plays the role of the genus-zero function $h(z)$ used
earlier (as was the case for the ordinary~$H_{0}$ of the $H_{g}$ family~(\ref{eq:higher-genus-hg-def})).  It is useful to introduce the natural endpoint variables:\footnote{These variables are normalized for $E_0{\neq}0$, so taking the limit
$E_0{\to}0$ to try to recover the ordinary WP case should not be done at fixed $z$.  To take such a limit, first undo the rescaling
$z{=}E_0\xi$, with $\xi{=}2\pi(x{-}\mu)$, and define
$\widehat{\mathscr H}_0(\xi){=}E_0^{-2}\mathscr H_0(E_0\xi)$.
Then we have
$\partial_\xi^2\widehat{\mathscr H}_0
{=}[E_0-u_0(\mu+\xi/(2\pi))]/2$, which behaves well as
$E_0{\to}0$.}
\begin{equation}
 z=2\pi E_0(x-\mu)\ ,
 \qquad
 y=\frac{E_0-u_0(x)}{2}\ ,
 \label{eq:N2-zy-def}
\end{equation}
Equation~\eqref{eq:N2-all-cusp-genus-zero} then gives simply:
\begin{equation}
 y=\frac{\partial^2\mathscr H_0}{\partial z^2}\ .
 \label{eq:N2-scriptH0-y}
\end{equation}
with $y=z+O(z^2)$ near the endpoint.

Now we are ready  to determine the complete gravitational/$\kappa$-dressing.  Write:
\begin{equation}
 \Omega^{({\cal N}=2)}_{g,n}(E_0)
 =
 E_0^{\,2g-2+n}
 \exp\!\left[
 \sum_{i\geq1}\sigma_i(E_0)\kappa_i
 \right]\ .
 \label{eq:N2-full-intersection-class}
\end{equation}
We wish to study $V_{0,m+3}({\bm 0})$, for $m{\geq}1$, and  the moduli space $\Mbar_{0,m+3}$ has complex dimension $m$.  This means that
$V_{0,m+3}({\bm 0})$ contains $\sigma_m$ linearly, with coefficient
\begin{equation}
 \int_{\Mbar_{0,m+3}}\kappa_m=1\ ,
\end{equation}
while every other contribution involves the
$\sigma_1,\ldots,\sigma_{m-1}$, allowing us to determine the 
 $\sigma_m$ recursively.

Let's look at how it works for $m{=}2$. Using~\eqref{eq:N2-all-cusp-genus-zero}  and~\eqref{eq:N2-jets} gives:
\begin{equation}
 V^{\mathcal N=2}_{0,5}({\bm 0})
 =
 12E_0-26\pi^2E_0^2+10\pi^4E_0^3\ .
 \label{eq:N2-V05-cusp}
\end{equation}
Meanwhile, using results listed in Appendix~\ref{app:intersection-integrals}:
\begin{equation}
 \int_{\Mbar_{0,5}}\kappa_2=1\ ,
 \qquad
 \int_{\Mbar_{0,5}}\kappa_1^2
 =5\ ,
 \label{eq:M05-intersections}
\end{equation}
we have from~\eqref{eq:N2-full-intersection-class}:
\begin{equation}
{V^{\mathcal N=2}_{0,5}(0^5)}
 =
 {E_0^3}\left(\sigma_2+\frac52\sigma_1^2\right)\ .
\end{equation}
Using equation~\eqref{eq:N2-naive-sigma1} for $\sigma_1$ therefore gives
\begin{equation}
 \sigma_2(E_0)
 =
 \frac{4\pi^2}{E_0}
 -\frac{21}{2E_0^2}\ ,
 \label{eq:N2-sigma2-cusp}
\end{equation}
and so we recover  the additional $4\pi^2/E_0$ of ~\eqref{eq:N2-sigma2-correction}, required for getting $V_{1,2}$ correct is reproduced
independently from the genus zero all-cusp~$V_{0,5}$!

Thus emboldened, let's write the entire tower of needed couplings in closed form.  Let us consider $z$ as a function of $y$
through equation~\eqref{eq:N2-zy-def}, and expand the inverse relation as:
\begin{equation}
 z(y;E_0)
 =
 \sum_{k\geq0}
 a_k(E_0)\frac{y^{k+1}}{(k+1)!}\ ,
 \qquad
 a_0=1\ .
 \label{eq:N2-ak-def}
\end{equation}
These $a_k$ are a local description of how the ${\cal N}{=}2$ background is built from basic KdV ingredients. They are analogues of the multicritical $t_k$ of equation~\eqref{eq:N2-tk}, but adapted to $u(x){=}E_0$ instead of $u(x){=}0$.

We can compute them using the Bessel re-summed form we used earlier for the leading string equation~\eqref{eq:N2-Bessel}. Using  with
$x{=}\mu{+}z/(2\pi E_0)$ and $u_0{=}E_0{-}2y$ it gives:
\begin{equation}
 (E_0-2y)\frac{\partial z}{\partial y}-z
 =
 E_0J_0\!\left(2\pi\sqrt{2y}\right)\ .
 \label{eq:N2-z-y-ODE}
\end{equation}
Substituting $z$ into this and comparing coefficients gives:
\begin{equation}
 a_k(E_0)
 =
 (2k+1)!!
 \sum_{j=0}^{k}
 \frac{(-1)^j(2\pi)^{2j}}
 {(2j+1)!}\,
 E_0^{-(k-j)}\ .
 \label{eq:N2-ak-closed}
\end{equation}
For example:
\begin{equation}
 a_1
 =
 \frac{3}{E_0}-2\pi^2\ ,
 \quad
 a_2
 =
 \frac{15}{E_0^2}
 -\frac{10\pi^2}{E_0}
 +2\pi^4\ .
 \label{eq:a1-a2}
\end{equation}
Following the usual dictionary, the   $a_k(E_0)$ are the right basis to be interpreted as  $\psi$-class  insertions. On the other hand~\eqref{eq:N2-full-intersection-class} is describing  the same background as $\kappa$-insertions, with  coefficients $\sigma_m(E_0)$. 
The computations behind the various intersection theory results we've used, involving $\kappa$ insertions, employ repeated use of the forgetful construction
relating appearance of $\kappa$  to more $\psi$ insertions. So the $\psi$-insertions are really descendants of that process, with multiple integrals resulting from single ones.  The standard way to track the combinatorics is to write the relation:
\begin{equation}
 1+\sum_{k\geq1}a_k(E_0)v^k
 =
 \exp\!\left[
 -\sum_{i\geq1}\sigma_i(E_0)v^i
 \right]\ ,
 \label{eq:N2-ak-sigma}
\end{equation} connecting the individual $\kappa$ insertions (connected) to the  
logarithm of descendant  KdV/$\psi$-class insertions (disconnected). So for example, 
$ a_1{=}-\sigma_1$ and 
$ a_2{=}\frac12\sigma_1^2-\sigma_2$, giving
$ \sigma_1{=}-a_1$  and $
 \sigma_2{=}\frac12a_1^2-a_2$.
Using the coefficients~\eqref{eq:a1-a2}  gives the $\sigma_1$ and $\sigma_2$ we computed earlier in~\eqref{eq:N2-naive-sigma1} and~\eqref{eq:N2-sigma2-cusp}.\footnote{The descent part of the computation there is hidden in the fact that in~(\ref{eq:M05-intersections}), the $5=6-1$, the six coming from two $\psi$ insertions and the~1 from a single $\psi$. See Appendix~\ref{app:intersection-integrals} for more.} 

The relation~\eqref{eq:N2-ak-sigma} fully
determines the complete tower of $\sigma_i(E_0)$ from the classical
genus zero cusp data.
We've therefore determined the full gravitational
$\kappa$ dressing for a description of  ${\cal N}{=}2$ WP volumes.  The first five coefficients are:
\begin{align}
 \sigma_1(E_0)
 &=
 2\pi^2-\frac3{E_0}\ ,
 \quad
 \sigma_2(E_0)
 =
 \frac{4\pi^2}{E_0}-\frac{21}{2E_0^2}\ ,
 \nonumber\\
 \sigma_3(E_0)
 &=
 \frac{28\pi^2}{E_0^2}-\frac{69}{E_0^3}\ ,
 \nonumber\\
 \sigma_4(E_0)
 &=
 -\frac{8\pi^4}{E_0^2}
 +\frac{276\pi^2}{E_0^3}
 -\frac{2529}{4E_0^4}\ ,\nonumber\\
  \sigma_5(E_0)
 &=
 -\frac{176\pi^4}{E_0^3}
 +\frac{3372\pi^2}{E_0^4}
 -\frac{36243}{5E_0^5}\ .
 \label{eq:N2-full-sigma-first}
\end{align}

We did this using  the genus-zero class, constraining it using genus zero volumes. It is natural to worry that maybe our answer is genus-dependent, which would be unsatisfactory. The all-cusp higher genus construction of
subsection~\ref{sec:WP-all-cusps-arbitrary-genus} dispels these worries. The same universal  off-shell construction gives that, in the
physical ${\cal N}{=}2$ normalization, the  all-cusp volumes at genus $g$ are:
\begin{equation}
 V^{\mathcal N=2}_{g,n}(0,\ldots,0)
 =
 -\frac{1}{(2\pi)^{2g+n-2}}
 \left.
 \frac{\partial^nF_g}{\partial x^n}
 \right|_{x=\mu}\ ,
 \label{eq:N2-all-cusp-Fg}
\end{equation}
where $F_g(x)$ is the genus $g$ off-shell free energy expression discussed earlier, built out of derivatives of~$u_0(x)$
(at genus zero this is the same statement as
$u_0{=}2F_0''$).  We extend the notation above by defining:
\begin{equation}
 \mathscr H_g(z;E_0)
 =
 \sum_{\substack{n\geq0\\2g-2+n>0}}
 \frac{V^{\mathcal N=2}_{g,n}(0,\ldots,0)}
 {E_0^{\,2g-2+n}n!}\,z^n\ .
 \label{eq:N2-scriptHg-def}
\end{equation}
For $g\geq2$, equation~\eqref{eq:N2-all-cusp-Fg} resums to:
\begin{equation}
 \mathscr H_g(z;E_0)
 =
 -(2\pi E_0)^{2-2g}
 F_g\!\left(
 \mu+\frac{z}{2\pi E_0}
 \right)\ ,
 \qquad g\geq2\ ,
 \label{eq:N2-scriptHg-Fg}
\end{equation}
while as before, at genus one the unstable $n=0$ term is removed:
\begin{equation}
 \mathscr H_1(z;E_0)
 =
 -F_1\!\left(
 \mu+\frac{z}{2\pi E_0}
 \right)
 +F_1(\mu)\ .
 \label{eq:N2-scriptH1-F1}
\end{equation}
Now come the crucial points. The relation~\eqref{eq:N2-ak-def} between~$z$ and $y$, determining the coefficients $a_k(E_0)$, is an expression of the content of $u_0(x)$, though the use of the string equation. Those $a_k(E_0)$ given in equation~\eqref{eq:N2-ak-closed} {\it do not change} therefore. Similarly, the intersection theory descent technology relating $\kappa$-insertions to the $\psi$-insertions is entirely genus independent. Hence the $\sigma_i(E_0)$ resulting from~\eqref{eq:N2-ak-sigma} also do not change. So the point is that \eqref{eq:N2-full-intersection-class} is the answer  for all stable $(g,n)$. But the volumes at higher $g$ are different, so what changes? 

What changes, on both sides, is the machinery for making volumes at a given genus $g$. On the one hand,~\eqref{eq:N2-all-cusp-Fg} instructs us to take derivatives on $F_g(x)$ to make the volume of interest. The result is again made of derivatives of $u_0(x)$, and so will be built from  new combinations of the~$a_k$ data and hence $\sigma_i$ data. On the other hand,~\eqref{eq:N2-full-intersection-class} will provide different combinations of the $\kappa_i$ because the dimension of $\Mbar_{g,n}$ will have changed. Spectacularly, the two machines again agree. 

At risk of belabouring the point, we can look at the case of $V_{1,2}$. 
Taking $(g,n)=(1,2)$ in equation~\eqref{eq:N2-all-cusp-Fg}, and using
$F_1=-\frac1{24}\log u_0'$, gives:
\begin{equation}
 V^{\mathcal N=2}_{1,2}(0,0)
 =
 \frac{1}{96\pi^2}
 \left[
 \frac{u_0'''}{u_0'}
 -
 \left(\frac{u_0''}{u_0'}\right)^2
 \right]_{x=\mu}\ .
 \label{eq:N2-V12-cusp-F1}
\end{equation}
Using the on-shell endpoint derivatives in
equation~\eqref{eq:N2-jets} then gives
\begin{equation}
 V^{\mathcal N=2}_{1,2}(0,0)
 =
 \frac18
 -\frac{7\pi^2}{12}E_0
 +\frac{\pi^4}{4}E_0^2\ ,
 \label{eq:N2-V12-cusp}
\end{equation}
in agreement with the zero-boundary specialization of
equation~\eqref{eq:N2-V12}.

Turning to our class~\eqref{eq:N2-full-intersection-class}, since the dimension of $\Mbar_{1,2}$ is two, only
the terms of degree two  contribute, giving:
\begin{equation}
 V^{\mathcal N=2}_{1,2}(0,0)
 =
 E_0^2
 \int_{\Mbar_{1,2}}
 \left(
 \sigma_2\kappa_2
 +\frac12\sigma_1^2\kappa_1^2
 \right)\ .
 \label{eq:N2-V12-cusp-kappa}
\end{equation}
Using the results:
\begin{equation}
 \int_{\Mbar_{1,2}}\kappa_2=\frac1{24}\ ,
 \qquad
 \int_{\Mbar_{1,2}}\kappa_1^2=\frac18\ ,
\end{equation}
together with equations~\eqref{eq:N2-full-sigma-first}, yields:
\begin{equation}
 V^{\mathcal N=2}_{1,2}(0,0)
 =
 E_0^2
 \left(
 \frac{\sigma_2}{24}
 +\frac{\sigma_1^2}{16}
 \right)
 =
 \frac18
 -\frac{7\pi^2}{12}E_0
 +\frac{\pi^4}{4}E_0^2\ ,
\end{equation}
in pleasing agreement with equation~\eqref{eq:N2-V12-cusp}.

Finally then, including the usual boundary descendants, the intersection theory description of the  ${\cal N}{=}2$ Weil-Petersson  volumes is written as:
\begin{equation}
 V^{\mathcal N=2}_{g,n}(\bm b)
 =
 \int_{\Mbar_{g,n}}
 \Omega^{({\cal N}=2)}_{g,n}(E_0)
 \exp\!\left(
 \frac12\sum_{i=1}^n b_i^2\psi_i
 \right)\ ,
 \label{eq:N2-intersection-volume}
\end{equation}
with our deformed class given in equation~\eqref{eq:N2-full-intersection-class}, and the~$\sigma_i$ all given by~\eqref{eq:N2-ak-closed} and~\eqref{eq:N2-ak-sigma}.

Notice that  the overall prefactor  $E_0^{2g-2+n}$ of the
class~\eqref{eq:N2-full-intersection-class} satisfies a pair of related consistency checks. First, its dependence on $2g{-}2{+}n$ is a requirement for consistent
factorization of the class under degeneration and gluing of the underlying surfaces~\cite{Kaufmann:1996aa,Manin:1999wj}: For a genus~$g$ surface
with~$n$ boundaries it is possible to be  decomposed into $2g{-}2{+}n$ pairs of
pants, so  the normalization $A^{2g-2+n}$ is natural.  The value of $A$ is fixed  by~$V_{0,3}$, the normalization of the elementary pair of pants.
This fits with the fact that $\Mbar_{0,3}$ has dimension zero, so there can be no 
$\kappa$- or $\psi$-class insertions, and therefore the intersection formula gives
$V^{\mathcal N=2}_{0,3}{=}E_0$,  matching the computation in~\eqref{eq:N2-first-physical-volumes}. (In the present
case it happens that this normalization coincides with the
deformation scale $\epsilon^2{=}E_0$ appearing in ref.~\cite{Chidambaram:2022cqc}'s deformed
$\Theta$-class description, but this is not really a general requirement, as will become clear  in the next Section.)

 Two final remarks will suffice for this Subsection. We see that, as promised, we can now explain from this way of building volumes the factor of $E_0$ in the integral term of the boundary-addition
identity~\eqref{eq:N2-full-identity}.  Adding one marked
point increases the overall factor $E_0^{\,2g-2+n}$ in
\eqref{eq:N2-full-intersection-class} by one power, while  as before (in the ordinary WP case) on the
stabilization divisor the additional $\psi_{n+1}$ part of the
forgetful-map relation vanishes.  As a result, we get:
\begin{equation}
 \left.
 \Omega^{({\cal N}=2)}_{g,n+1}
 \right|_{D_{i,n+1}}
 =
 E_0\,
 \Omega^{({\cal N}=2)}_{g,n}\ ,
\end{equation}
and the subsequent steps taken  to compute the divisor contribution will yield the  integral term in~\eqref{eq:N2-full-identity}, with the factor 
of~$E_0$. It is also interesting to explore the rewriting around the removable-cone point in this language, and see that the action of $\Jcal_b$ again builds a tower of $\kappa$-decorated volumes, but it is high time we turned to the case of (small) ${\cal N}{=}4$ Weil-Petersson volumes.

\section{Small ${\cal N}{=}4$ Weil-Petersson Volumes}
\label{sec:boundary-addition-WP-N4}

\subsection{The Specialization}

The small ${\cal N}{=}4$ theory~\cite{Heydeman:2020hhw} has  $SU(2)$ R-symmetry (distinguishing it from the full ${\cal N}{=}4$ which has $SU(2){\times}SU(2)$). Turiaci and Witten argued~\cite{Turiaci:2023jfa} for a random matrix model description within each sector labelled by spin $J$.   A multicritical random matrix model description was constructed in ref.~\cite{Johnson:2024tgg}, with the
string equation analysis refined in
refs.~\cite{Johnson:2025oty,Ahmed:2025lxe}.\footnote{Ref.~\cite{Heydeman:2025vcc} has since developed a large ${\cal N}{=}4$ and an ${\cal N}{=}3$ generalization,  which were also studied in the matrix model approach in ref.~\cite{Johnson:2025oty,Ahmed:2025lxe}. We will not explore them here, except to note that it was observed in ref.~\cite{Johnson:2025oty} that  the large ${\cal N}{=}4$ theory's random matrix structure has much in common with  ${\cal N}{=}2$. So many of our remarks of the previous Section should be relevant  there.} 

For the choice of normalization we use here, the $t_k$ are:
\begin{align}
 t_k(J)=
 &\frac{
 8\pi^{k+1}
 }{
 J^k k!(4k^2+8k+3)
 }
 J_{k+1}(2\pi J)\ ,\quad
 k=0,1,2,\ldots\ ,
 \label{eq:N4-tk}
\end{align}
with $J\in\tfrac12\mathbb Z_{\ne0}$. 
We have $\widetilde\Gamma$=2, and: 
\begin{equation}
 \mu=t_0=
 \frac{8\pi}{3}J_1(2\pi J)\ .
 \label{eq:N4-mu}
\end{equation}
The leading string equation is:
\begin{equation}
 u_0\,[G_0(u_0)+x]^2
 =
 \widetilde\Gamma^2
 =
 4\ ,\,\,\text{where}\,\,\,  G_0(u)\equiv
 \sum_{k=1}^{\infty}t_k(J)u^k\ ,
 \label{eq:N4-positive-string}
\end{equation}
and the  endpoint/threshold is:
 $u_0(\mu)=J^2$.
On the physical branch it is useful to write:
\begin{equation}
 x=-{\widetilde G}_J(u)
 \equiv
 -G_0(u)-\frac{2\sigma_J}{\sqrt u},
 \qquad
 \sigma_J=
 \operatorname{sgn}(J)(-1)^{2J}\ ,
 \label{eq:N4-FJ}
\end{equation}
so that
$ {\widetilde G}_J(J^2){+}\mu{=}0$. Differentiating   relation~\eqref{eq:N4-FJ} gives the endpoint data:
\begin{align}
 u_0'(\mu)
 &=-\frac{J^3}{2},
 \nonumber\\
 u_0''(\mu)
 &=
 \frac{3J^4-\pi^2J^6}{4},
 \nonumber\\
 u_0'''(\mu)
 &=
 -\frac{63}{32}J^5
 +\frac{13\pi^2}{8}J^7
 -\frac{5\pi^4}{16}J^9\ .
 \label{eq:N4-jets}
\end{align}
The physical normalization when converting from off-shell volumes is~\cite{Ahmed:2025lxe}:
\begin{equation}
 \left.\Vcal_{g,n}\right|_J
 =
 K_{g,n}V^{(J)}_{g,n}\ ,
 \qquad
 K_{g,n}=4^\chi\ ,
 \label{eq:N4-K}
\end{equation}
and hence $\frac{K_{g,n}}{K_{g,n+1}}=4$.

Substituting this together with
$u_0'(\mu){=}-J^3/2$ into
equation~\eqref{eq:master-physical} gives the boundary-addition identity for ${\cal N}{=}4$ volumes:
\begin{equation}
\begin{aligned}
 V^{(J)}_{g,n+1}(\bm b,b)
 ={}&
 J^3\sum_{i=1}^{n}
 \int_0^{b_i}y\,
 V^{(J)}_{g,n}(\ldots,y,\ldots)\dd y
 \\
 &+
 \frac1{K_{g,n+1}}
 \left.
 \Jcal_b\Vcal_{g,n}(\bm b;x)
 \right|_J\ .
\end{aligned}
 \label{eq:N4-full-identity}
\end{equation}
Very similarly to the ${\cal N}{=}2$ case, since $u_0(\mu)$ is again non-zero, we'll use that the exact Bessel representation of the
boundary operator.  Defining:
\begin{equation}
 U_J(s)=u_0(\mu+s),
 \qquad
 U_J(0)=J^2\ ,
 \label{eq:N4-U}
\end{equation}
equation~\eqref{eq:softedge-generating} gives the action of $\Jcal_b$ as:
\begin{equation}
 \sum_{m=0}^{\infty}
 \frac{s^m}{m!}
 \Jcal_bu_0^{(m)}(\mu)
 =
 U_J'(s)
 J_0\!\left(
 b\sqrt{U_J(s)-J^2}
 \right)\ .
 \label{eq:N4-generating}
\end{equation}
The first universal off-shell volumes provide a few first examples.
From
$\Vcal_{0,3}=-u_0'/2$ one obtains:
\begin{equation}
 \left.\Vcal_{0,3}\right|_J
 =
 \frac{J^3}{4}
 =
 K_{0,3}V^{(J)}_{0,3}\ ,
\end{equation}
and hence:
\begin{equation}
 V^{(J)}_{0,3}=J^3\ .
 \label{eq:N4-V03}
\end{equation}
Similarly,
\begin{equation}
 \left.\Vcal_{1,1}(b)\right|_J
 =
 K_{1,1}
 \left[
 -\frac J4
 +\frac{b^2+4\pi^2}{48}J^3
 \right]\ ,
\end{equation}
and so:
\begin{equation}
 V^{(J)}_{1,1}(b)
 =
 -\frac J4
 +\frac{b^2+4\pi^2}{48}J^3\ .
 \label{eq:N4-V11}
\end{equation}
Exploring the identity to get $V_{0,4}$ we find that
equations~\eqref{eq:N4-jets} and
\eqref{eq:softedge-first-jet-actions} give:
\begin{align}
 \frac1{K_{0,4}}
 \left.
 \Jcal_b\Vcal_{0,3}
 \right|_J
 &=
 -\frac1{2K_{0,4}}
 \left[
 u_0''
 -\frac{b^2}{4}(u_0')^2
 \right]_{\!J}
 \nonumber\\
 &=
 -6J^4
 +\frac{J^6}{2}(b^2+4\pi^2)\ .
 \label{eq:N4-V03-residual-general}
\end{align}
Meanwhile the integral term is:
\begin{equation}
 J^3
 \sum_{i=1}^{3}
 \int_0^{b_i}yJ^3\,\dd y
 =
 \frac{J^6}{2}
 \sum_{i=1}^{3}b_i^2\ .
\end{equation}
The full identity therefore gives:
\begin{equation}
 V^{(J)}_{0,4}
 =
 -6J^4
 +\frac{J^6}{2}
 \left(
 4\pi^2+\sum_{i=1}^{4}b_i^2
 \right)\ ,
 \label{eq:N4-V04}
\end{equation}
reproducing a result of ref.~\cite{Ahmed:2025lxe}.

The next genus one example is not much more involved.
Using equation~\eqref{eq:soft-JV11} with the endpoint data
\eqref{eq:N4-jets} and adding the integral term in
equation~\eqref{eq:N4-full-identity} gives the symmetric polynomial:
\begin{widetext}
\begin{equation}
 \begin{aligned}
 V^{(J)}_{1,2}(b_1,b_2)
 ={}&
 \frac{J^2}{192}
 \Big[
 216
 -224\pi^2J^2
 -48J^2(b_1^2+b_2^2)
 +J^4(b_1^2+b_2^2+4\pi^2)(b_1^2+b_2^2+12\pi^2)
 \Big]\  ,
 \end{aligned}
 \label{eq:N4-V12}
\end{equation}
providing a useful example of how  the different powers of $J$ are organized. 

\subsection{Exact $T$-expansion at the removable cone point}

The exact $\Jcal_b$ action in equation~\eqref{eq:N4-generating} can be
organized around the removable cone point in precisely the same way as
for the ${\cal N}=2$ case. Writing the analogous objects:
\begin{equation}
 T=b^2+4\pi^2\ ,
 \qquad
 \Delta_J(s)=U_J(s)-J^2\ ,
 \label{eq:N4-T-Delta}
\end{equation}
 the exact Bessel $T$-expansion
\eqref{eq:softedge-exact-T} gives:
\begin{equation}
 \begin{aligned}
 \sum_{m=0}^{\infty}
 \frac{s^m}{m!}
 \Jcal_bu_0^{(m)}(\mu)
 ={}&
 \sum_{r=0}^{\infty}
 \frac{(-T/4)^r}{r!}\,
 U_J'(s)
 \frac{\Delta_J(s)^{r/2}}{\pi^r}
 I_r\!\left(2\pi\sqrt{\Delta_J(s)}\right)
 \ .
 \end{aligned}
 \label{eq:N4-exact-T}
\end{equation}
\end{widetext}
As before, this is simply a finite polynomial in $T$, with  only finitely many powers contributing once applied to a given off-shell volume.

The zeroth level of the expansion is the removable-cone value
$T{=}0$, {\it i.e.,} 
 $b=2\pi\ii$.
The $r{=}0$ term in equation~\eqref{eq:N4-exact-T}, equivalently
equation~\eqref{eq:N4-generating} at this value is:
\begin{equation}
 \sum_{m=0}^{\infty}
 \frac{s^m}{m!}
 \Jcal_{2\pi \ii}u_0^{(m)}(\mu)
 =
 U_J'(s)
 J_0\!\left(
 2\pi\sqrt{J^2-U_J(s)}
 \right)\ .
 \label{eq:N4-generating-star}
\end{equation}
Following what we learned in the ${\cal N}{=}2$ case is to again use the Euler operator~\eqref{eq:Euler-u0}, and  note that the denominator in the $t_k$ in~\eqref{eq:N4-tk} factorizes as:
\begin{equation}
 4k^2+8k+3=(2k+1)(2k+3)\ ,
 \label{eq:N4-den-factor}
\end{equation} so this time it is the second order Euler operator
$(2D_{u_0}{+}1)(2D_{u_0}{+}3)$ that removes the denominator factors in the sum. The resulting
Bessel sum  over $k$ gives:
\begin{widetext}
\begin{equation}
 (2D_{u_0}+1)(2D_{u_0}+3)
 \bigl({\widetilde G}_J(u)+\mu\bigr)
 =
 8J\frac{\partial}{\partial u}
 J_0\!\left(
 2\pi\sqrt{J^2-u}
 \right)\ ,
 \label{eq:N4-FJ-Bessel}
\end{equation}
\end{widetext}
and similar steps to what went before yield: 
\begin{equation}
 (2D_{u_0}+1)(2D_{u_0}+3)
 \bigl({\widetilde G}_J(u)+\mu\bigr)
 =
 8J\frac{\partial}{\partial u}
 \left[
 \Jcal_{2\pi \ii}{\widetilde G}_J(u)
 \right]\ .
 \label{eq:N4-string-function-relation}
\end{equation}

This is the small ${\cal N}{=}4$ counterpart of the first order
${\cal N}{=}2$ relation
\eqref{eq:N2-string-function-relation}.  That first order structure produced the simple eigenvalue relation~\eqref{eq:N2-all-jet}.  Here the string equation
instead supplies a second-order Euler relation and fixes the
$u$-derivative of the $\Jcal_{2\pi\ii}$ action. The resulting non-linearity means that there is no simple analogous relation this time.  
Expanding the Euler operator gives, after replacing $u_0$ derivatives of ${\widetilde G(u_0)}$ with $x$ derivatives of $u_0$, and then substituting for $U_J{=}u_0(\mu+s)$:
\begin{equation}
 4U_J^2\frac{U_J''}{(U_J')^3}
 -\frac{12U_J}{U_J'}
 -3s
 =
 \frac{8\pi J}{
 \sqrt{J^2-U_J}}
 J_1\!\left(
 2\pi\sqrt{J^2-U_J}
 \right)\ ,
 \label{eq:N4-UJ-ODE}
\end{equation}
which is a second order ODE with initial conditions:
$ U_J(0){=}J^2$ and
 $U_J'(0){=}-\frac{J^3}{2}$.
With a solution $U(s)$, the diagonal law is replaced by using the
generating identity~\eqref{eq:N4-all-jet}  two write:
\begin{equation}
 \Jcal_{2\pi \ii}u_0^{(m)}(\mu)
 =
 \left.
 \frac{\partial^m}{\partial s^m}
 \left[
 U_J'(s)
 J_0\!\left(
 2\pi\sqrt{J^2-U_J(s)}
 \right)
 \right]
 \right|_{s=0}\ .
 \label{eq:N4-all-jet}
\end{equation}
together with~\eqref{eq:N4-UJ-ODE}.
The ODE determines $U_J(s)$ recursively about
$s=0$,  and its first few terms are
\begin{eqnarray}
 &&U_J(s)
 =
 J^2
 -\frac{J^3}{2}s
 +\frac{3J^4-\pi^2J^6}{8}s^2
 \\
 &&\hskip1.2cm+\left(
 -\frac{21}{64}J^5
 +\frac{13\pi^2}{48}J^7
 -\frac{5\pi^4}{96}J^9
 \right)s^3
 +\cdots\ .
 \nonumber
\end{eqnarray}
Equation~\eqref{eq:N4-all-jet} then gives the first few derivatives:
\begin{align}
 \Jcal_{2\pi \ii}u_0
 &=-\frac{J^3}{2},
 \nonumber\\
 \Jcal_{2\pi \ii}u_0'
 &=
 u_0''+\pi^2(u_0')^2
 =
 \frac34J^4,
 \nonumber\\
 \Jcal_{2\pi \ii}u_0''
 &=
 u_0'''
 +3\pi^2u_0'u_0''
 +\frac{\pi^4}{2}(u_0')^3
 \nonumber\\
 &=
 -\frac{63}{32}J^5
 +\frac{\pi^2}{2}J^7\ .
 \label{eq:N4-star-jets}
\end{align}
The first two examples start out looking like a pattern of eigenvalue relations again,  but things fail at subsequent levels when the richer second-order structure takes hold.

A lesson here is that although the overall structure of the small ${\cal N}{=}4$ boundary-addition identity looks superficially similar to that of ${\cal N}{=}2$, it is already clear that the boundary operators $\Jcal_b$ contain much more complexity. Even at the cone point the non-linearity that sets in means that the  geometrical interpretation of its operation on the volumes is considerably less simple. 

While it would be interesting to delve more deeply into this, we'll change tack and explore the intersection theory construction of the ${\cal N}{=}4$ volumes.

\subsection{Intersection theory and ${\cal N}{=}4$ WP Volumes}
\label{sec:N4-intersection-theory}
Given what was done for the ${\cal N}{=}2$ case in Subsection~\ref{sec:N2-intersection-theory}, along with the accompanying discussion, it is clear what needs to be done to define the intersection theory description of small ${\cal N}=4$ Weil-Petersson volumes. In the language used there, it is simply a matter of defining the deformation parameter and specifying the gravitational dressing appropriate for the case in hand.

In the physical normalization of the small ${\cal N}=4$ volumes,
$K_{g,n}=4^{2-2g-n}$, repeated cusp insertion gives
\begin{equation}
 V^{(J)}_{0,n}(0,\ldots,0)
 =
 -\frac12\,4^{\,n-2}
 \left.
 \frac{\partial^{n-2}u_0}{\partial x^{n-2}}
 \right|_{x=\mu}\ ,
 \qquad n\geq3\ .
 \label{eq:N4-all-cusp-genus-zero}
\end{equation}
The analogous  normalized endpoint variables are:
\begin{equation}
 z=\frac{J^3}{4}(x-\mu)\ ,
 \qquad
 y=\frac{J^2-u_0(x)}{2}\ .
 \label{eq:N4-zy}
\end{equation}
Since $u_0'(\mu)=-J^3/2$, these obey $y=z+O(z^2)$ near the endpoint.
Regard $z$ as a function of $y$ and we again write:
\begin{equation}
 z(y;J)
 =
 \sum_{k\geq0}
 a_k(J)\frac{y^{k+1}}{(k+1)!}\ ,
 \qquad
 a_0=1\ .
 \label{eq:N4-ak-def}
\end{equation}

The inverse string equation derived above,
\eqref{eq:N4-FJ-Bessel}
becomes, on using
$u=J^2-2y$ and ${\widetilde G}_J(u)+\mu=-4z/J^3$,
\begin{equation}
 (J^2-2y)^2\frac{\partial^2z}{\partial y^2}
 -6(J^2-2y)\frac{\partial z}{\partial y}
 +3z
 =
 J^4\frac{\partial}{\partial y}
 J_0\!\left(2\pi\sqrt{2y}\right)\ .
 \label{eq:N4-z-y-ODE}
\end{equation}
Substituting equation~\eqref{eq:N4-ak-def} gives the recurrence
\begin{align}
 J^4a_{k+1}
 &-(4k+6)J^2a_k
 +(2k+1)(2k+3)a_{k-1}
 \nonumber\\
 &\hskip1cm=
 J^4
 \frac{(-1)^{k+1}(2\pi^2)^{k+1}}{(k+1)!}\ ,
 \label{eq:N4-ak-recurrence}
\end{align}
with $a_{-1}=0$.  The solution is:
\begin{equation}
 a_k(J)
 =
 (2k+1)!!
 \sum_{j=0}^{k}
 (k-j+1)
 \frac{(-1)^j(2\pi)^{2j}}{(2j+1)!}\,
 J^{-2(k-j)}\ .
 \label{eq:N4-ak-closed}
\end{equation}
For example:
\begin{equation}
 a_1
 =
 \frac6{J^2}-2\pi^2\ ,
 \qquad
 a_2
 =
 \frac{45}{J^4}
 -\frac{20\pi^2}{J^2}
 +2\pi^4\ .
 \label{eq:N4-a1-a2}
\end{equation}
This result for $a_k$ here should be contrasted with the result~\eqref{eq:N2-ak-closed} for ${\cal N}{=}2$. The extra $(k-j+1)$ in the sum is perhaps an avatar of the onset of non-linearity we saw in the previous section.

Just as in the ${\cal N}{=}2$, case we  determine the corresponding $\sigma_i(J)$ through:
\begin{equation}
 1+\sum_{k\geq1}a_k(J)v^k
 =
 \exp\!\left[
 -\sum_{i\geq1}\sigma_i(J)v^i
 \right]\ ,
 \label{eq:N4-ak-sigma}
\end{equation}
and the first several are:
\begin{align}
 \sigma_1(J)
 &=
 2\pi^2-\frac6{J^2}\ ,
 \quad
 \sigma_2(J)
 =
 \frac{8\pi^2}{J^2}-\frac{27}{J^4}\ ,
 \nonumber\\
 \sigma_3(J)
 &=
 \frac{72\pi^2}{J^4}-\frac{222}{J^6}\ ,
 \nonumber\\
 \sigma_4(J)
 &=
 -\frac{16\pi^4}{J^4}
 +\frac{888\pi^2}{J^6}
 -\frac{4977}{2J^8}\nonumber\\
  \sigma_5(J)
 &=
 -\frac{480\pi^4}{J^6}
 +\frac{13272\pi^2}{J^8}
 -\frac{171126}{5J^{10}}\ .
 \label{eq:N4-sigma-first}
\end{align}
The small ${\cal N}{=}4$ class takes these as input and is therefore given by:
\begin{equation}
 \Omega^{(J)}_{g,n}
 =
 J^{3(2g-2+n)}
 \exp\!\left[
 \sum_{m\geq1}\sigma_m(J)\kappa_m
 \right]\ ,
 \label{eq:N4-intersection-class}
\end{equation}
and the full finite-boundary volumes are:
\begin{equation}
 V^{(J)}_{g,n}(\bm b)
 =
 \int_{\Mbar_{g,n}}
 \Omega^{(J)}_{g,n}
 \exp\!\left(
 \frac12\sum_{i=1}^n b_i^2\psi_i
 \right)\ .
 \label{eq:N4-intersection-volume}
\end{equation}
The overall  normalization of the class~\eqref{eq:N4-intersection-class} is fixed the same way
as in the ${\cal N}{=}2$ case, {\it i.e.,} consistent with the result from~(\ref{eq:N4-V03}) that $V^{(J)}_{0,3}{=}J^3$. It is interesting  that here  the gravitational dressing appearing in~\eqref{eq:N4-sigma-first} is
organized in inverse powers of~$J^2$, rather than $J^3$. This is in contrast
to the ${\cal N}=2$ case, where the normalization and the  dressing deformation (inverse) scale were the same,  $E_0$.

A few low-order checks are in order.  At $(g,n){=}(0,4)$ and $(1,1)$,
the class only probes $\sigma_1$ and gives:
\begin{align}
 V^{(J)}_{0,4}(\bm b)
 &=
 -6J^4
 +\frac{J^6}{2}
 \left(
 4\pi^2+\sum_{i=1}^4b_i^2
 \right)\ ,
 \nonumber\\
 V^{(J)}_{1,1}(b)
 &=
 -\frac J4
 +\frac{J^3}{48}
 \left(b^2+4\pi^2\right)\ ,
 \label{eq:N4-intersection-low-checks}
\end{align}
reproducing equations~\eqref{eq:N4-V04} and \eqref{eq:N4-V11}. A similar check works for  $V_{1,2}$, involving $\sigma_2$.  Independently, at genus zero the same $\sigma_2$ is tested by the $V_{0,5}({\bm 0})$.  Using the
$\Mbar_{0,5}$ intersections already encountered in equation~(\ref{eq:M05-intersections}),
\begin{align}
 V^{(J)}_{0,5}({\bm 0})
 &=
 J^9
 \left(
 \sigma_2+\frac52\sigma_1^2
 \right)
 \nonumber\\
 &=
 63J^5-52\pi^2J^7+10\pi^4J^9\ ,
 \label{eq:N4-intersection-V05-check}
\end{align}
in agreement with equation~\eqref{eq:N4-jets} and the all-cusp relation
\eqref{eq:N4-all-cusp-genus-zero}.

Just as before, the use of the cusp genus-zero volumes was just a swift means to an end. The same reasons this prescription all readily persists to all genus that were given in the ${\cal N}{=}2$ section apply here.  Finally, again note that the definition of the class itself means that changing from $(n+1)$ boundaries to $n$ brings in a factor $J^3$, precisely explaining (from this intersection theory perspective of handling divisor contributions from the forgetful map) the $J^3$ in the  integral term in the boundary-addition identity~(\ref{eq:N4-full-identity}). Detailed exploration of the  forgetful map origin of the tower of $\kappa$-decorated volumes that encapsulated by the $\Jcal_b \Vcal_{g,n}(x)$ term is left for another time.

\section{Discussion}
\label{sec:discussion}

Over the course of the paper, many specific  interesting further research avenues presented themselves, and were commented on in the body of the paper, so they will not be repeated here. Instead we will close with some broader remarks.

The boundary-addition identity~(\ref{eq:master-offshell})  presented in this paper has a generality to it that is worth emphasizing. At the level of ``off-shell'' quantities, it is the same formula that underlies the ordinary, ${\cal N}{=}1$, ${\cal N}{=}2$,  and small ${\cal N}{=}4$ Weil-Petersson cases. 
Putting in the data that specialized the identity to a specific one of those models is a simple matter (by just changing the function $u_0(x)$, essentially), and led to a range of phenomena, which this paper explored. 

In this sense we've provided a unifying framework that puts these various systems, quite different in some respects, all on the same footing. This was achieved, building on the work in  ref.~\cite{Johnson:2026twg}, by finding a simple way of harnessing their common underlying KdV integrable organizational structure.

A lot of the focus, including various applications, was directed to the ordinary Weil-Petersson case, including the many formulae for general genus generating functions for volumes, their consequences for asymptotic behaviour, and so on. It should be noted, however, that the extended supersymmetric versions of these formulae can be developed with equal ease. In fact, again in most instances, since things were written essentially in terms of~$u_0(x)$, little needs to be changed to adapt results to those cases. 

It is to be expected therefore that there is still a lot that can be swiftly achieved with this formalism, for both physical and mathematical applications. Indeed, just the simplest  cases of writing the supersymmetric generalization of the all-cusp volume generating functions were enough to allow us to fully solve the problem of constructing the intersection theory description of ${\cal N}{=}2$  and small ${\cal N}{=}4$ Weil-Petersson volumes, results that had been missing from the literature. Hopefully, these new intersection theory constructions  will be useful in applications further afield.

As a final remark, it is worth noting again that the formalism on which this is all based, the framework fully outlined in ref.~\cite{Johnson:2026twg}, applies to a much more general range of systems than just the Weil-Petersson volumes discussed here. There are likely to be new tools to be found within in at that are least as powerful as the boundary-addition identity that was the focus of this paper.

\section*{Acknowledgments}
This work was partly supported by  the  US Department of Energy (under award \#\protect{DE-SC} 0011702).  CVJ also thanks the University of California Santa Barbara for support, and Amelia for her support and patience. 

\appendix

\begin{widetext}

\section{Intersection-theory integrals used in the text}
\label{app:intersection-integrals}

For convenience, we collect here some of the intersection theory results used at
various points in the text, along  with (it is hoped) enough of the standard forgetful map
machinery to get a sense of  how the mixed $\kappa$-$\psi$ insertions reduce to ordinary
$\psi$-class intersections.  What follows is no substitute for good pedagogical sources such as refs.~\cite{KockPsiClasses,Zvonkine2012}. (Ref.~\cite{Do:2008Tourist} is also excellent as broader guide that  continues to be useful over the years.) We use throughout that the complex dimension of
$\Mbar_{g,n}$ is $3g-3+n$ and the convention:
\begin{equation}
 \kappa_a
 =
 \pi_*\!\left(\psi_{n+1}^{a+1}\right),
 \qquad
 \pi:\Mbar_{g,n+1}\longrightarrow\Mbar_{g,n}\ .
 \label{eq:app-IT-kappa-definition}
\end{equation}
The origin of the descent/simplification mechanism is roughly that the pullback of an existing  class obeys the following relation:
$\pi^*\psi_i=\psi_i-D_{i,n+1}$, where $D$ is the divisor discussed in the body of the text.  Since
$\psi_{n+1}$ restricts to zero on $D_{i,n+1}$, the divisor correction is
annihilated whenever the new marked point carries the positive power of
$\psi_{n+1}$ appearing in~\eqref{eq:app-IT-kappa-definition}.  The upshot is that a
single $\kappa$ insertion may be removed  by simply adding one marked point:
\begin{equation}
 \int_{\Mbar_{g,n}}
 \kappa_a\prod_{i=1}^n\psi_i^{d_i}
 =
 \int_{\Mbar_{g,n+1}}
 \psi_{n+1}^{a+1}\prod_{i=1}^n\psi_i^{d_i}\ .
 \label{eq:app-IT-single-kappa}
\end{equation}
For products of several $\kappa$-classes, repeated use of the same
relation generates subtraction terms when the auxiliary marked
points are merged.  Since only a relative few  cases are needed here, we'll 
write those reductions explicitly below.  It is worth noting that these kinds of compuations can be readily looked up using for example the SageMath package \texttt{admcycles} of Delecroix, Schmitt and van
Zelm~\cite{Delecroix_2021}. It is also available through browser-based
SageMath interfaces.  We also use
 $\kappa_0=(2g-2+n)$,
and at genus zero:
\begin{equation}
 \int_{\Mbar_{0,n}}
 \prod_{i=1}^n\psi_i^{d_i}
 =
 \frac{(n-3)!}{\prod_i d_i!}\ ,
 \qquad
 \sum_i d_i=n-3\ .
 \label{eq:app-IT-genus-zero-psi}
\end{equation}
A useful consequence, employed in the ${\cal N}{=}2$ discussion, follows
immediately from equations~\eqref{eq:app-IT-single-kappa} and
\eqref{eq:app-IT-genus-zero-psi}:
\begin{equation}
 \int_{\Mbar_{0,m+3}}\kappa_m
 =
 \int_{\Mbar_{0,m+4}}\psi_{m+4}^{m+1}
 =
 \frac{(m+1)!}{(m+1)!}
 =1\ .
 \label{eq:app-IT-kappam-genus-zero}
\end{equation}
In particular, the two genus-zero integrals used in the five-cusp computation are:
\begin{equation}
 \begin{aligned}
 \int_{\Mbar_{0,5}}\kappa_2
 &=1\ ,\\
 \int_{\Mbar_{0,5}}\kappa_1^2
 &=
 \int_{\Mbar_{0,7}}\psi_6^2\psi_7^2
 -
 \int_{\Mbar_{0,6}}\psi_6^3
 =6-1=5\ .
 \end{aligned}
 \label{eq:app-IT-M05}
\end{equation}
The second line is the simplest example of the subtraction generated when
more than one $\kappa$ insertion is reduced by successive forgetful maps.
At genus one, the basic one-point integrals used in the ordinary and
supersymmetric computations are:
\begin{equation}
 \int_{\Mbar_{1,1}}\psi_1
 =
 \frac1{24}\ ,
 \qquad
 \int_{\Mbar_{1,1}}\kappa_1
 =
 \int_{\Mbar_{1,2}}\psi_2^2
 =
 \frac1{24}\ .
 \label{eq:app-IT-M11}
\end{equation}
For the genus one two-point computation in the ${\cal N}{=}2$ section we need:
\begin{equation}
 \begin{aligned}
 \int_{\Mbar_{1,2}}\kappa_2
 &=
 \int_{\Mbar_{1,3}}\psi_3^3
 =\frac1{24}\ ,\\
 \int_{\Mbar_{1,2}}\kappa_1^2
 &=
 \int_{\Mbar_{1,4}}\psi_3^2\psi_4^2
 -
 \int_{\Mbar_{1,3}}\psi_3^3
 =
 \frac16-\frac1{24}
 =\frac18\ .
 \end{aligned}
 \label{eq:app-IT-M12}
\end{equation}
Finally, the genus two one-point computation in the topological
${\cal N}{=}2$ model uses a small collection of mixed intersections on
$\Mbar_{2,1}$.  The various pure descendant seeds needed are:
\begin{equation}
 \int_{\Mbar_{2,1}}\psi_1^4
 =\frac1{1152}\ ,
 \qquad
 \int_{\Mbar_{2,2}}\psi_1^2\psi_2^3
 =\frac{29}{5760}\ ,
 \qquad
 \int_{\Mbar_{2,3}}\psi_1^2\psi_2^2\psi_3^2
 =\frac7{240}\ ,
 \label{eq:app-IT-g2-psi-seeds}
\end{equation}
as well as:
\begin{equation}
 \int_{\Mbar_{2,2}}\psi_1\psi_2^4
 =\frac1{384}\ ,
 \qquad
 \int_{\Mbar_{2,3}}\psi_1\psi_2^2\psi_3^3
 =\frac{29}{1440}\ ,
 \qquad
 \int_{\Mbar_{2,4}}\psi_1\psi_2^2\psi_3^2\psi_4^2
 =\frac7{48}\ .
 \label{eq:app-IT-g2-additional-descendants}
\end{equation}
Equation~\eqref{eq:app-IT-single-kappa} gives the single-$\kappa$ cases:
\begin{equation}
 \begin{aligned}
 \int_{\Mbar_{2,1}}\kappa_1\psi_1^3
 &=
 \int_{\Mbar_{2,2}}\psi_1^3\psi_2^2
 =\frac{29}{5760}\ ,\\
 \int_{\Mbar_{2,1}}\kappa_2\psi_1^2
 &=
 \int_{\Mbar_{2,2}}\psi_1^2\psi_2^3
 =\frac{29}{5760}\ ,\\
 \int_{\Mbar_{2,1}}\kappa_3\psi_1
 &=
 \int_{\Mbar_{2,2}}\psi_1\psi_2^4
 =\frac1{384}\ .
 \end{aligned}
 \label{eq:app-IT-M21-single-kappa}
\end{equation}
Successive use of the forgetful-map relation gives the multiple-$\kappa$
cases:
\begin{equation}
 \begin{aligned}
 \int_{\Mbar_{2,1}}\kappa_1^2\psi_1^2
 &=
 \int_{\Mbar_{2,3}}\psi_1^2\psi_2^2\psi_3^2
 -
 \int_{\Mbar_{2,2}}\psi_1^2\psi_2^3
 =
 \frac{139}{5760}\ ,\\
 \int_{\Mbar_{2,1}}\kappa_1\kappa_2\psi_1
 &=
 \int_{\Mbar_{2,3}}\psi_1\psi_2^2\psi_3^3
 -
 \int_{\Mbar_{2,2}}\psi_1\psi_2^4
 =
 \frac{101}{5760}\ .
 \end{aligned}
 \label{eq:app-IT-M21-two-kappa}
\end{equation}
and:
\begin{equation}
 \begin{aligned}
 \int_{\Mbar_{2,1}}\kappa_1^3\psi_1
 ={}&
 \int_{\Mbar_{2,4}}\psi_1\psi_2^2\psi_3^2\psi_4^2
 -3\int_{\Mbar_{2,3}}\psi_1\psi_2^2\psi_3^3
 +\int_{\Mbar_{2,2}}\psi_1\psi_2^4
 \\
 ={}&
 \frac7{48}
 -3\left(\frac{29}{1440}\right)
 +\frac1{384}
 =\frac{169}{1920}\ .
 \end{aligned}
 \label{eq:app-IT-M21-three-kappa}
\end{equation}

\vfill\eject
\end{widetext}

\newpage

\bibliographystyle{apsrev4-1}
\bibliography{references}

\begin{thebibliography}{68}%
\makeatletter
\providecommand \@ifxundefined [1]{%
 \@ifx{#1\undefined}
}%
\providecommand \@ifnum [1]{%
 \ifnum #1\expandafter \@firstoftwo
 \else \expandafter \@secondoftwo
 \fi
}%
\providecommand \@ifx [1]{%
 \ifx #1\expandafter \@firstoftwo
 \else \expandafter \@secondoftwo
 \fi
}%
\providecommand \natexlab [1]{#1}%
\providecommand \enquote  [1]{``#1''}%
\providecommand \bibnamefont  [1]{#1}%
\providecommand \bibfnamefont [1]{#1}%
\providecommand \citenamefont [1]{#1}%
\providecommand \href@noop [0]{\@secondoftwo}%
\providecommand \href [0]{\begingroup \@sanitize@url \@href}%
\providecommand \@href[1]{\@@startlink{#1}\@@href}%
\providecommand \@@href[1]{\endgroup#1\@@endlink}%
\providecommand \@sanitize@url [0]{\catcode `\\12\catcode `\$12\catcode `\&12\catcode `\#12\catcode `\^12\catcode `\_12\catcode `\%12\relax}%
\providecommand \@@startlink[1]{}%
\providecommand \@@endlink[0]{}%
\providecommand \url  [0]{\begingroup\@sanitize@url \@url }%
\providecommand \@url [1]{\endgroup\@href {#1}{\urlprefix }}%
\providecommand \urlprefix  [0]{URL }%
\providecommand \Eprint [0]{\href }%
\providecommand \doibase [0]{http://dx.doi.org/}%
\providecommand \selectlanguage [0]{\@gobble}%
\providecommand \bibinfo  [0]{\@secondoftwo}%
\providecommand \bibfield  [0]{\@secondoftwo}%
\providecommand \translation [1]{[#1]}%
\providecommand \BibitemOpen [0]{}%
\providecommand \bibitemStop [0]{}%
\providecommand \bibitemNoStop [0]{.\EOS\space}%
\providecommand \EOS [0]{\spacefactor3000\relax}%
\providecommand \BibitemShut  [1]{\csname bibitem#1\endcsname}%
\let\auto@bib@innerbib\@empty
\bibitem [{\citenamefont {Wolpert}(2010)}]{Wolpert2010WeilPetersson}%
  \BibitemOpen
  \bibfield  {author} {\bibinfo {author} {\bibfnamefont {S.~A.}\ \bibnamefont {Wolpert}},\ }\href {\doibase 10.1090/cbms/113} {\emph {\bibinfo {title} {Families of Riemann Surfaces and Weil--Petersson Geometry}}},\ \bibinfo {series} {CBMS Regional Conference Series in Mathematics}, Vol.\ \bibinfo {volume} {113}\ (\bibinfo  {publisher} {American Mathematical Society},\ \bibinfo {address} {Providence, RI},\ \bibinfo {year} {2010})\BibitemShut {NoStop}%
\bibitem [{\citenamefont {Do}(2013)}]{Do:2011WPsurvey}%
  \BibitemOpen
  \bibfield  {author} {\bibinfo {author} {\bibfnamefont {N.}~\bibnamefont {Do}},\ }in\ \href@noop {} {\emph {\bibinfo {booktitle} {Handbook of Moduli, Volume I}}},\ \bibinfo {series} {Advanced Lectures in Mathematics}, Vol.~\bibinfo {volume} {24},\ \bibinfo {editor} {edited by\ \bibinfo {editor} {\bibfnamefont {G.}~\bibnamefont {Farkas}}\ and\ \bibinfo {editor} {\bibfnamefont {I.}~\bibnamefont {Morrison}}}\ (\bibinfo  {publisher} {International Press},\ \bibinfo {address} {Somerville, MA},\ \bibinfo {year} {2013})\ pp.\ \bibinfo {pages} {217--258},\ \Eprint {http://arxiv.org/abs/1103.4674} {arXiv:1103.4674 [math.GT]} \BibitemShut {NoStop}%
\bibitem [{\citenamefont {Do}(2008)}]{Do:2008Tourist}%
  \BibitemOpen
  \bibfield  {author} {\bibinfo {author} {\bibfnamefont {N.}~\bibnamefont {Do}},\ }\href@noop {} {\bibfield  {journal} {\bibinfo  {journal} {Gazette of the Australian Mathematical Society}\ }\textbf {\bibinfo {volume} {35}},\ \bibinfo {pages} {103} (\bibinfo {year} {2008})}\BibitemShut {NoStop}%
\bibitem [{\citenamefont {Jackiw}(1985)}]{Jackiw:1984je}%
  \BibitemOpen
  \bibfield  {author} {\bibinfo {author} {\bibfnamefont {R.}~\bibnamefont {Jackiw}},\ }\href {\doibase 10.1016/0550-3213(85)90448-1} {\bibfield  {journal} {\bibinfo  {journal} {Nucl. Phys.}\ }\textbf {\bibinfo {volume} {B252}},\ \bibinfo {pages} {343} (\bibinfo {year} {1985})}\BibitemShut {NoStop}%
\bibitem [{\citenamefont {Teitelboim}(1983)}]{Teitelboim:1983ux}%
  \BibitemOpen
  \bibfield  {author} {\bibinfo {author} {\bibfnamefont {C.}~\bibnamefont {Teitelboim}},\ }\href {\doibase 10.1016/0370-2693(83)90012-6} {\bibfield  {journal} {\bibinfo  {journal} {Phys. Lett.}\ }\textbf {\bibinfo {volume} {126B}},\ \bibinfo {pages} {41} (\bibinfo {year} {1983})}\BibitemShut {NoStop}%
\bibitem [{\citenamefont {Saad}\ \emph {et~al.}(2019)\citenamefont {Saad}, \citenamefont {Shenker},\ and\ \citenamefont {Stanford}}]{Saad:2019lba}%
  \BibitemOpen
  \bibfield  {author} {\bibinfo {author} {\bibfnamefont {P.}~\bibnamefont {Saad}}, \bibinfo {author} {\bibfnamefont {S.~H.}\ \bibnamefont {Shenker}}, \ and\ \bibinfo {author} {\bibfnamefont {D.}~\bibnamefont {Stanford}},\ }\href@noop {} {\  (\bibinfo {year} {2019})},\ \Eprint {http://arxiv.org/abs/1903.11115} {arXiv:1903.11115 [hep-th]} \BibitemShut {NoStop}%
\bibitem [{\citenamefont {Engels{\"o}y}\ \emph {et~al.}(2016)\citenamefont {Engels{\"o}y}, \citenamefont {Mertens},\ and\ \citenamefont {Verlinde}}]{Engelsoy:2016xyb}%
  \BibitemOpen
  \bibfield  {author} {\bibinfo {author} {\bibfnamefont {J.}~\bibnamefont {Engels{\"o}y}}, \bibinfo {author} {\bibfnamefont {T.~G.}\ \bibnamefont {Mertens}}, \ and\ \bibinfo {author} {\bibfnamefont {H.}~\bibnamefont {Verlinde}},\ }\href {\doibase 10.1007/JHEP07(2016)139} {\bibfield  {journal} {\bibinfo  {journal} {JHEP}\ }\textbf {\bibinfo {volume} {07}},\ \bibinfo {pages} {139} (\bibinfo {year} {2016})},\ \Eprint {http://arxiv.org/abs/1606.03438} {arXiv:1606.03438 [hep-th]} \BibitemShut {NoStop}%
\bibitem [{\citenamefont {Stanford}\ and\ \citenamefont {Witten}(2017)}]{Stanford:2017thb}%
  \BibitemOpen
  \bibfield  {author} {\bibinfo {author} {\bibfnamefont {D.}~\bibnamefont {Stanford}}\ and\ \bibinfo {author} {\bibfnamefont {E.}~\bibnamefont {Witten}},\ }\href {\doibase 10.1007/JHEP10(2017)008} {\bibfield  {journal} {\bibinfo  {journal} {JHEP}\ }\textbf {\bibinfo {volume} {10}},\ \bibinfo {pages} {008} (\bibinfo {year} {2017})},\ \Eprint {http://arxiv.org/abs/1703.04612} {arXiv:1703.04612 [hep-th]} \BibitemShut {NoStop}%
\bibitem [{\citenamefont {Stanford}\ and\ \citenamefont {Witten}(2020)}]{Stanford:2019vob}%
  \BibitemOpen
  \bibfield  {author} {\bibinfo {author} {\bibfnamefont {D.}~\bibnamefont {Stanford}}\ and\ \bibinfo {author} {\bibfnamefont {E.}~\bibnamefont {Witten}},\ }\href {\doibase 10.4310/ATMP.2020.v24.n6.a4} {\bibfield  {journal} {\bibinfo  {journal} {Adv. Theor. Math. Phys.}\ }\textbf {\bibinfo {volume} {24}},\ \bibinfo {pages} {1475} (\bibinfo {year} {2020})},\ \Eprint {http://arxiv.org/abs/1907.03363} {arXiv:1907.03363 [hep-th]} \BibitemShut {NoStop}%
\bibitem [{\citenamefont {Norbury}(2026)}]{Norbury:2020vyi}%
  \BibitemOpen
  \bibfield  {author} {\bibinfo {author} {\bibfnamefont {P.}~\bibnamefont {Norbury}},\ }\href {\doibase 10.1016/j.geomphys.2025.105750} {\bibfield  {journal} {\bibinfo  {journal} {J. Geom. Phys.}\ }\textbf {\bibinfo {volume} {222}},\ \bibinfo {pages} {105750} (\bibinfo {year} {2026})},\ \Eprint {http://arxiv.org/abs/2005.04378} {arXiv:2005.04378 [math.AG]} \BibitemShut {NoStop}%
\bibitem [{\citenamefont {Heydeman}\ \emph {et~al.}(2022)\citenamefont {Heydeman}, \citenamefont {Iliesiu}, \citenamefont {Turiaci},\ and\ \citenamefont {Zhao}}]{Heydeman:2020hhw}%
  \BibitemOpen
  \bibfield  {author} {\bibinfo {author} {\bibfnamefont {M.}~\bibnamefont {Heydeman}}, \bibinfo {author} {\bibfnamefont {L.~V.}\ \bibnamefont {Iliesiu}}, \bibinfo {author} {\bibfnamefont {G.~J.}\ \bibnamefont {Turiaci}}, \ and\ \bibinfo {author} {\bibfnamefont {W.}~\bibnamefont {Zhao}},\ }\href {\doibase 10.1088/1751-8121/ac3be9} {\bibfield  {journal} {\bibinfo  {journal} {J. Phys. A}\ }\textbf {\bibinfo {volume} {55}},\ \bibinfo {pages} {014004} (\bibinfo {year} {2022})},\ \Eprint {http://arxiv.org/abs/2011.01953} {arXiv:2011.01953 [hep-th]} \BibitemShut {NoStop}%
\bibitem [{\citenamefont {Turiaci}\ and\ \citenamefont {Witten}(2023)}]{Turiaci:2023jfa}%
  \BibitemOpen
  \bibfield  {author} {\bibinfo {author} {\bibfnamefont {G.~J.}\ \bibnamefont {Turiaci}}\ and\ \bibinfo {author} {\bibfnamefont {E.}~\bibnamefont {Witten}},\ }\href {\doibase 10.1007/JHEP12(2023)003} {\bibfield  {journal} {\bibinfo  {journal} {JHEP}\ }\textbf {\bibinfo {volume} {12}},\ \bibinfo {pages} {003} (\bibinfo {year} {2023})},\ \Eprint {http://arxiv.org/abs/2305.19438} {arXiv:2305.19438 [hep-th]} \BibitemShut {NoStop}%
\bibitem [{\citenamefont {Johnson}(2024)}]{Johnson:2023ofr}%
  \BibitemOpen
  \bibfield  {author} {\bibinfo {author} {\bibfnamefont {C.~V.}\ \bibnamefont {Johnson}},\ }\href {\doibase 10.1103/PhysRevD.110.106019} {\bibfield  {journal} {\bibinfo  {journal} {Phys. Rev. D}\ }\textbf {\bibinfo {volume} {110}},\ \bibinfo {pages} {106019} (\bibinfo {year} {2024})},\ \Eprint {http://arxiv.org/abs/2306.10139} {arXiv:2306.10139 [hep-th]} \BibitemShut {NoStop}%
\bibitem [{\citenamefont {Heydeman}\ \emph {et~al.}(2026)\citenamefont {Heydeman}, \citenamefont {Shi},\ and\ \citenamefont {Turiaci}}]{Heydeman:2025vcc}%
  \BibitemOpen
  \bibfield  {author} {\bibinfo {author} {\bibfnamefont {M.}~\bibnamefont {Heydeman}}, \bibinfo {author} {\bibfnamefont {X.}~\bibnamefont {Shi}}, \ and\ \bibinfo {author} {\bibfnamefont {G.~J.}\ \bibnamefont {Turiaci}},\ }\href {\doibase 10.1007/JHEP01(2026)054} {\bibfield  {journal} {\bibinfo  {journal} {JHEP}\ }\textbf {\bibinfo {volume} {01}},\ \bibinfo {pages} {054} (\bibinfo {year} {2026})},\ \Eprint {http://arxiv.org/abs/2504.20146} {arXiv:2504.20146 [hep-th]} \BibitemShut {NoStop}%
\bibitem [{\citenamefont {Ahmed}\ \emph {et~al.}(2026)\citenamefont {Ahmed}, \citenamefont {Johnson},\ and\ \citenamefont {Saraswat}}]{Ahmed:2025lxe}%
  \BibitemOpen
  \bibfield  {author} {\bibinfo {author} {\bibfnamefont {W.}~\bibnamefont {Ahmed}}, \bibinfo {author} {\bibfnamefont {C.~V.}\ \bibnamefont {Johnson}}, \ and\ \bibinfo {author} {\bibfnamefont {K.}~\bibnamefont {Saraswat}},\ }\href {\doibase 10.1007/JHEP02(2026)197} {\bibfield  {journal} {\bibinfo  {journal} {JHEP}\ }\textbf {\bibinfo {volume} {02}},\ \bibinfo {pages} {197} (\bibinfo {year} {2026})},\ \Eprint {http://arxiv.org/abs/2507.18715} {arXiv:2507.18715 [hep-th]} \BibitemShut {NoStop}%
\bibitem [{\citenamefont {Johnson}\ and\ \citenamefont {Kolanowski}(2026)}]{Johnson:2025oty}%
  \BibitemOpen
  \bibfield  {author} {\bibinfo {author} {\bibfnamefont {C.~V.}\ \bibnamefont {Johnson}}\ and\ \bibinfo {author} {\bibfnamefont {M.}~\bibnamefont {Kolanowski}},\ }\href {\doibase 10.1007/JHEP03(2026)126} {\bibfield  {journal} {\bibinfo  {journal} {JHEP}\ }\textbf {\bibinfo {volume} {03}},\ \bibinfo {pages} {126} (\bibinfo {year} {2026})},\ \Eprint {http://arxiv.org/abs/2507.07185} {arXiv:2507.07185 [hep-th]} \BibitemShut {NoStop}%
\bibitem [{\citenamefont {Mirzakhani}(2006)}]{Mirzakhani:2006fta}%
  \BibitemOpen
  \bibfield  {author} {\bibinfo {author} {\bibfnamefont {M.}~\bibnamefont {Mirzakhani}},\ }\href {\doibase 10.1007/s00222-006-0013-2} {\bibfield  {journal} {\bibinfo  {journal} {Invent. Math.}\ }\textbf {\bibinfo {volume} {167}},\ \bibinfo {pages} {179} (\bibinfo {year} {2006})}\BibitemShut {NoStop}%
\bibitem [{\citenamefont {Wolpert}(1983)}]{Wolpert:1983}%
  \BibitemOpen
  \bibfield  {author} {\bibinfo {author} {\bibfnamefont {S.}~\bibnamefont {Wolpert}},\ }\href {http://www.jstor.org/stable/2006980} {\bibfield  {journal} {\bibinfo  {journal} {Annals of Mathematics}\ }\textbf {\bibinfo {volume} {118}},\ \bibinfo {pages} {491} (\bibinfo {year} {1983})}\BibitemShut {NoStop}%
\bibitem [{\citenamefont {Mirzakhani}(2007)}]{Mirzakhani2007WPVolumesIntersection}%
  \BibitemOpen
  \bibfield  {author} {\bibinfo {author} {\bibfnamefont {M.}~\bibnamefont {Mirzakhani}},\ }\href {\doibase 10.1090/S0894-0347-06-00526-1} {\bibfield  {journal} {\bibinfo  {journal} {Journal of the American Mathematical Society}\ }\textbf {\bibinfo {volume} {20}},\ \bibinfo {pages} {1} (\bibinfo {year} {2007})}\BibitemShut {NoStop}%
\bibitem [{\citenamefont {Mulase}\ and\ \citenamefont {Safnuk}(2008)}]{MulaseSafnuk2008}%
  \BibitemOpen
  \bibfield  {author} {\bibinfo {author} {\bibfnamefont {M.}~\bibnamefont {Mulase}}\ and\ \bibinfo {author} {\bibfnamefont {B.}~\bibnamefont {Safnuk}},\ }\href@noop {} {\bibfield  {journal} {\bibinfo  {journal} {Indian Journal of Mathematics}\ }\textbf {\bibinfo {volume} {50}},\ \bibinfo {pages} {189} (\bibinfo {year} {2008})},\ \Eprint {http://arxiv.org/abs/math/0601194} {arXiv:math/0601194} \BibitemShut {NoStop}%
\bibitem [{\citenamefont {Eynard}\ and\ \citenamefont {Orantin}(2007{\natexlab{a}})}]{Eynard:2007fixed}%
  \BibitemOpen
  \bibfield  {author} {\bibinfo {author} {\bibfnamefont {B.}~\bibnamefont {Eynard}}\ and\ \bibinfo {author} {\bibfnamefont {N.}~\bibnamefont {Orantin}},\ }\href@noop {} {\bibfield  {journal} {\bibinfo  {journal} {{ arXiv:0705.3600}}\ } (\bibinfo {year} {2007}{\natexlab{a}})},\ \Eprint {http://arxiv.org/abs/0705.3600} {arXiv:0705.3600 [math-ph]} \BibitemShut {NoStop}%
\bibitem [{\citenamefont {Do}\ and\ \citenamefont {Norbury}(2009)}]{do:2009weil}%
  \BibitemOpen
  \bibfield  {author} {\bibinfo {author} {\bibfnamefont {N.}~\bibnamefont {Do}}\ and\ \bibinfo {author} {\bibfnamefont {P.}~\bibnamefont {Norbury}},\ }\href@noop {} {\bibfield  {journal} {\bibinfo  {journal} {Geometriae Dedicata}\ }\textbf {\bibinfo {volume} {141}},\ \bibinfo {pages} {93} (\bibinfo {year} {2009})},\ \Eprint {http://arxiv.org/abs/math/0603406} {arXiv:math/0603406 [math.AG]} \BibitemShut {NoStop}%
\bibitem [{\citenamefont {Ancona}\ and\ \citenamefont {Gayet}(2026)}]{ancona2026recursionvolumemodulispace}%
  \BibitemOpen
  \bibfield  {author} {\bibinfo {author} {\bibfnamefont {M.}~\bibnamefont {Ancona}}\ and\ \bibinfo {author} {\bibfnamefont {D.}~\bibnamefont {Gayet}},\ }\href {https://arxiv.org/abs/2605.19788} {\enquote {\bibinfo {title} {A recursion for the volume of the moduli space of hyperbolic spheres},}\ } (\bibinfo {year} {2026}),\ \Eprint {http://arxiv.org/abs/2605.19788} {arXiv:2605.19788 [math.AG]} \BibitemShut {NoStop}%
\bibitem [{\citenamefont {Zograf}(1993)}]{Zograf1993WeilPetersson}%
  \BibitemOpen
  \bibfield  {author} {\bibinfo {author} {\bibfnamefont {P.~G.}\ \bibnamefont {Zograf}},\ }in\ \href@noop {} {\emph {\bibinfo {booktitle} {Mapping class groups and moduli spaces of Riemann surfaces: proceedings of workshops}}},\ \bibinfo {series} {Contemporary Mathematics}, Vol.\ \bibinfo {volume} {150}\ (\bibinfo  {publisher} {American Mathematical Society},\ \bibinfo {address} {Providence, RI},\ \bibinfo {year} {1993})\ pp.\ \bibinfo {pages} {367--372}\BibitemShut {NoStop}%
\bibitem [{\citenamefont {Zograf}(1998{\natexlab{a}})}]{Zograf:1998ur}%
  \BibitemOpen
  \bibfield  {author} {\bibinfo {author} {\bibfnamefont {P.}~\bibnamefont {Zograf}},\ }\href@noop {} {\enquote {\bibinfo {title} {{Weil--Petersson} volumes of moduli spaces of curves and the genus expansion in two dimensional gravity},}\ } (\bibinfo {year} {1998}{\natexlab{a}}),\ \Eprint {http://arxiv.org/abs/math/9811026} {arXiv:math/9811026 [math.AG]} \BibitemShut {NoStop}%
\bibitem [{\citenamefont {Zograf}(1998{\natexlab{b}})}]{ZografP.G.1998Wvol}%
  \BibitemOpen
  \bibfield  {author} {\bibinfo {author} {\bibfnamefont {P.~G.}\ \bibnamefont {Zograf}},\ }\href@noop {} {\bibfield  {journal} {\bibinfo  {journal} {Functional analysis and its applications}\ }\textbf {\bibinfo {volume} {32}},\ \bibinfo {pages} {281} (\bibinfo {year} {1998}{\natexlab{b}})}\BibitemShut {NoStop}%
\bibitem [{\citenamefont {Manin}\ and\ \citenamefont {Zograf}(2000)}]{Manin:1999wj}%
  \BibitemOpen
  \bibfield  {author} {\bibinfo {author} {\bibfnamefont {Y.~I.}\ \bibnamefont {Manin}}\ and\ \bibinfo {author} {\bibfnamefont {P.}~\bibnamefont {Zograf}},\ }\href {\doibase 10.5802/aif.1764} {\bibfield  {journal} {\bibinfo  {journal} {Annales de l'Institut Fourier}\ }\textbf {\bibinfo {volume} {50}},\ \bibinfo {pages} {519} (\bibinfo {year} {2000})},\ \Eprint {http://arxiv.org/abs/math/9902051} {arXiv:math/9902051} \BibitemShut {NoStop}%
\bibitem [{\citenamefont {Hide}\ and\ \citenamefont {Thomas}(2025)}]{HideThomas2025}%
  \BibitemOpen
  \bibfield  {author} {\bibinfo {author} {\bibfnamefont {W.}~\bibnamefont {Hide}}\ and\ \bibinfo {author} {\bibfnamefont {J.}~\bibnamefont {Thomas}},\ }\href {\doibase 10.1007/s00220-025-05369-4} {\bibfield  {journal} {\bibinfo  {journal} {Commun. Math. Phys.}\ }\textbf {\bibinfo {volume} {406}},\ \bibinfo {pages} {203} (\bibinfo {year} {2025})},\ \Eprint {http://arxiv.org/abs/2312.11412} {arXiv:2312.11412 [math.DG]} \BibitemShut {NoStop}%
\bibitem [{\citenamefont {Kaufmann}\ \emph {et~al.}(1996)\citenamefont {Kaufmann}, \citenamefont {Manin},\ and\ \citenamefont {Zagier}}]{Kaufmann:1996aa}%
  \BibitemOpen
  \bibfield  {author} {\bibinfo {author} {\bibfnamefont {R.~M.}\ \bibnamefont {Kaufmann}}, \bibinfo {author} {\bibfnamefont {Y.~I.}\ \bibnamefont {Manin}}, \ and\ \bibinfo {author} {\bibfnamefont {D.~B.}\ \bibnamefont {Zagier}},\ }\href {\doibase 10.1007/BF02101297} {\bibfield  {journal} {\bibinfo  {journal} {Commun. Math. Phys.}\ }\textbf {\bibinfo {volume} {181}},\ \bibinfo {pages} {763} (\bibinfo {year} {1996})},\ \Eprint {http://arxiv.org/abs/alg-geom/9604001} {arXiv:alg-geom/9604001} \BibitemShut {NoStop}%
\bibitem [{\citenamefont {Johnson}(2026{\natexlab{a}})}]{Johnson:2026plw}%
  \BibitemOpen
  \bibfield  {author} {\bibinfo {author} {\bibfnamefont {C.~V.}\ \bibnamefont {Johnson}},\ }\href@noop {} {\  (\bibinfo {year} {2026}{\natexlab{a}})},\ \Eprint {http://arxiv.org/abs/2601.17122} {arXiv:2601.17122 [hep-th]} \BibitemShut {NoStop}%
\bibitem [{\citenamefont {Johnson}(2026{\natexlab{b}})}]{Johnson:2026twg}%
  \BibitemOpen
  \bibfield  {author} {\bibinfo {author} {\bibfnamefont {C.~V.}\ \bibnamefont {Johnson}},\ }\href@noop {} {\  (\bibinfo {year} {2026}{\natexlab{b}})},\ \Eprint {http://arxiv.org/abs/2604.11902} {arXiv:2604.11902 [hep-th]} \BibitemShut {NoStop}%
\bibitem [{\citenamefont {Tan}\ \emph {et~al.}(2006)\citenamefont {Tan}, \citenamefont {Wong},\ and\ \citenamefont {Zhang}}]{TanWongZhang2006}%
  \BibitemOpen
  \bibfield  {author} {\bibinfo {author} {\bibfnamefont {S.~P.}\ \bibnamefont {Tan}}, \bibinfo {author} {\bibfnamefont {Y.~L.}\ \bibnamefont {Wong}}, \ and\ \bibinfo {author} {\bibfnamefont {Y.}~\bibnamefont {Zhang}},\ }\href {\doibase 10.4310/jdg/1143593126} {\bibfield  {journal} {\bibinfo  {journal} {J. Differential Geom.}\ }\textbf {\bibinfo {volume} {72}},\ \bibinfo {pages} {73} (\bibinfo {year} {2006})},\ \Eprint {http://arxiv.org/abs/math/0404226} {arXiv:math/0404226 [math.GT]} \BibitemShut {NoStop}%
\bibitem [{\citenamefont {Johnson}(2026{\natexlab{c}})}]{Johnson:2026jls}%
  \BibitemOpen
  \bibfield  {author} {\bibinfo {author} {\bibfnamefont {C.~V.}\ \bibnamefont {Johnson}},\ }\href@noop {} {\  (\bibinfo {year} {2026}{\natexlab{c}})},\ \Eprint {http://arxiv.org/abs/2606.09990} {arXiv:2606.09990 [hep-th]} \BibitemShut {NoStop}%
\bibitem [{\citenamefont {Gel'fand}\ and\ \citenamefont {Dikii}(1975)}]{Gelfand:1975rn}%
  \BibitemOpen
  \bibfield  {author} {\bibinfo {author} {\bibfnamefont {I.~M.}\ \bibnamefont {Gel'fand}}\ and\ \bibinfo {author} {\bibfnamefont {L.~A.}\ \bibnamefont {Dikii}},\ }\href@noop {} {\bibfield  {journal} {\bibinfo  {journal} {Russ. Math. Surveys}\ }\textbf {\bibinfo {volume} {30}},\ \bibinfo {pages} {77} (\bibinfo {year} {1975})}\BibitemShut {NoStop}%
\bibitem [{\citenamefont {Brezin}\ and\ \citenamefont {Kazakov}(1990)}]{Brezin:1990rb}%
  \BibitemOpen
  \bibfield  {author} {\bibinfo {author} {\bibfnamefont {E.}~\bibnamefont {Brezin}}\ and\ \bibinfo {author} {\bibfnamefont {V.~A.}\ \bibnamefont {Kazakov}},\ }\href@noop {} {\bibfield  {journal} {\bibinfo  {journal} {Phys. Lett.}\ }\textbf {\bibinfo {volume} {B236}},\ \bibinfo {pages} {144} (\bibinfo {year} {1990})}\BibitemShut {NoStop}%
\bibitem [{\citenamefont {Gross}\ and\ \citenamefont {Migdal}(1990{\natexlab{a}})}]{Gross:1990vs}%
  \BibitemOpen
  \bibfield  {author} {\bibinfo {author} {\bibfnamefont {D.~J.}\ \bibnamefont {Gross}}\ and\ \bibinfo {author} {\bibfnamefont {A.~A.}\ \bibnamefont {Migdal}},\ }\href@noop {} {\bibfield  {journal} {\bibinfo  {journal} {Phys. Rev. Lett.}\ }\textbf {\bibinfo {volume} {64}},\ \bibinfo {pages} {127} (\bibinfo {year} {1990}{\natexlab{a}})}\BibitemShut {NoStop}%
\bibitem [{\citenamefont {Gross}\ and\ \citenamefont {Migdal}(1990{\natexlab{b}})}]{Gross:1990aw}%
  \BibitemOpen
  \bibfield  {author} {\bibinfo {author} {\bibfnamefont {D.~J.}\ \bibnamefont {Gross}}\ and\ \bibinfo {author} {\bibfnamefont {A.~A.}\ \bibnamefont {Migdal}},\ }\href@noop {} {\bibfield  {journal} {\bibinfo  {journal} {Nucl. Phys.}\ }\textbf {\bibinfo {volume} {B340}},\ \bibinfo {pages} {333} (\bibinfo {year} {1990}{\natexlab{b}})}\BibitemShut {NoStop}%
\bibitem [{\citenamefont {Douglas}\ and\ \citenamefont {Shenker}(1990)}]{Douglas:1990ve}%
  \BibitemOpen
  \bibfield  {author} {\bibinfo {author} {\bibfnamefont {M.~R.}\ \bibnamefont {Douglas}}\ and\ \bibinfo {author} {\bibfnamefont {S.~H.}\ \bibnamefont {Shenker}},\ }\href@noop {} {\bibfield  {journal} {\bibinfo  {journal} {Nucl. Phys.}\ }\textbf {\bibinfo {volume} {B335}},\ \bibinfo {pages} {635} (\bibinfo {year} {1990})}\BibitemShut {NoStop}%
\bibitem [{\citenamefont {Douglas}(1990)}]{Douglas:1990dd}%
  \BibitemOpen
  \bibfield  {author} {\bibinfo {author} {\bibfnamefont {M.~R.}\ \bibnamefont {Douglas}},\ }\href@noop {} {\bibfield  {journal} {\bibinfo  {journal} {Phys. Lett.}\ }\textbf {\bibinfo {volume} {B238}},\ \bibinfo {pages} {176} (\bibinfo {year} {1990})}\BibitemShut {NoStop}%
\bibitem [{\citenamefont {Banks}\ \emph {et~al.}(1990)\citenamefont {Banks}, \citenamefont {Douglas}, \citenamefont {Seiberg},\ and\ \citenamefont {Shenker}}]{Banks:1990df}%
  \BibitemOpen
  \bibfield  {author} {\bibinfo {author} {\bibfnamefont {T.}~\bibnamefont {Banks}}, \bibinfo {author} {\bibfnamefont {M.~R.}\ \bibnamefont {Douglas}}, \bibinfo {author} {\bibfnamefont {N.}~\bibnamefont {Seiberg}}, \ and\ \bibinfo {author} {\bibfnamefont {S.~H.}\ \bibnamefont {Shenker}},\ }\href@noop {} {\bibfield  {journal} {\bibinfo  {journal} {Phys. Lett.}\ }\textbf {\bibinfo {volume} {B238}},\ \bibinfo {pages} {279} (\bibinfo {year} {1990})}\BibitemShut {NoStop}%
\bibitem [{\citenamefont {Dijkgraaf}\ \emph {et~al.}(1991)\citenamefont {Dijkgraaf}, \citenamefont {Verlinde},\ and\ \citenamefont {Verlinde}}]{Dijkgraaf:1991rs}%
  \BibitemOpen
  \bibfield  {author} {\bibinfo {author} {\bibfnamefont {R.}~\bibnamefont {Dijkgraaf}}, \bibinfo {author} {\bibfnamefont {H.}~\bibnamefont {Verlinde}}, \ and\ \bibinfo {author} {\bibfnamefont {E.}~\bibnamefont {Verlinde}},\ }\href@noop {} {\bibfield  {journal} {\bibinfo  {journal} {Nucl. Phys.}\ }\textbf {\bibinfo {volume} {B348}},\ \bibinfo {pages} {435} (\bibinfo {year} {1991})}\BibitemShut {NoStop}%
\bibitem [{\citenamefont {Chekhov}\ \emph {et~al.}(2006)\citenamefont {Chekhov}, \citenamefont {Eynard},\ and\ \citenamefont {Orantin}}]{Chekhov:2006vd}%
  \BibitemOpen
  \bibfield  {author} {\bibinfo {author} {\bibfnamefont {L.}~\bibnamefont {Chekhov}}, \bibinfo {author} {\bibfnamefont {B.}~\bibnamefont {Eynard}}, \ and\ \bibinfo {author} {\bibfnamefont {N.}~\bibnamefont {Orantin}},\ }\href {\doibase 10.1088/1126-6708/2006/12/053} {\bibfield  {journal} {\bibinfo  {journal} {JHEP}\ }\textbf {\bibinfo {volume} {12}},\ \bibinfo {pages} {053} (\bibinfo {year} {2006})},\ \Eprint {http://arxiv.org/abs/math-ph/0603003} {arXiv:math-ph/0603003} \BibitemShut {NoStop}%
\bibitem [{\citenamefont {Eynard}\ and\ \citenamefont {Orantin}(2007{\natexlab{b}})}]{Eynard:2007kz}%
  \BibitemOpen
  \bibfield  {author} {\bibinfo {author} {\bibfnamefont {B.}~\bibnamefont {Eynard}}\ and\ \bibinfo {author} {\bibfnamefont {N.}~\bibnamefont {Orantin}},\ }\href {\doibase 10.4310/CNTP.2007.v1.n2.a4} {\bibfield  {journal} {\bibinfo  {journal} {Commun. Num. Theor. Phys.}\ }\textbf {\bibinfo {volume} {1}},\ \bibinfo {pages} {347} (\bibinfo {year} {2007}{\natexlab{b}})},\ \Eprint {http://arxiv.org/abs/math-ph/0702045} {arXiv:math-ph/0702045 [math-ph]} \BibitemShut {NoStop}%
\bibitem [{\citenamefont {Mertens}\ and\ \citenamefont {Turiaci}(2021)}]{Mertens:2020hbs}%
  \BibitemOpen
  \bibfield  {author} {\bibinfo {author} {\bibfnamefont {T.~G.}\ \bibnamefont {Mertens}}\ and\ \bibinfo {author} {\bibfnamefont {G.~J.}\ \bibnamefont {Turiaci}},\ }\href {\doibase 10.1007/JHEP01(2021)073} {\bibfield  {journal} {\bibinfo  {journal} {JHEP}\ }\textbf {\bibinfo {volume} {01}},\ \bibinfo {pages} {073} (\bibinfo {year} {2021})},\ \Eprint {http://arxiv.org/abs/2006.07072} {arXiv:2006.07072 [hep-th]} \BibitemShut {NoStop}%
\bibitem [{\citenamefont {Matone}(1997)}]{Matone:1994pz}%
  \BibitemOpen
  \bibfield  {author} {\bibinfo {author} {\bibfnamefont {M.}~\bibnamefont {Matone}},\ }\href {\doibase 10.1016/S0393-0440(96)00028-9} {\bibfield  {journal} {\bibinfo  {journal} {J. Geom. Phys.}\ }\textbf {\bibinfo {volume} {21}},\ \bibinfo {pages} {381} (\bibinfo {year} {1997})},\ \Eprint {http://arxiv.org/abs/hep-th/9402081} {arXiv:hep-th/9402081} \BibitemShut {NoStop}%
\bibitem [{\citenamefont {Kock}(2001)}]{KockPsiClasses}%
  \BibitemOpen
  \bibfield  {author} {\bibinfo {author} {\bibfnamefont {J.}~\bibnamefont {Kock}},\ }\href {https://www.mat.uab.es/~kock/GW/notes/psi-notes.pdf} {\enquote {\bibinfo {title} {Notes on psi classes},}\ } (\bibinfo {year} {2001}),\ \bibinfo {note} {preliminary version, 6 June 2001}\BibitemShut {NoStop}%
\bibitem [{\citenamefont {Zvonkine}(2012)}]{Zvonkine2012}%
  \BibitemOpen
  \bibfield  {author} {\bibinfo {author} {\bibfnamefont {D.}~\bibnamefont {Zvonkine}},\ }in\ \href {\doibase 10.4171/103-1/12} {\emph {\bibinfo {booktitle} {Handbook of Teichm{\"u}ller Theory, Volume III}}},\ \bibinfo {series} {IRMA Lectures in Mathematics and Theoretical Physics}, Vol.~\bibinfo {volume} {17},\ \bibinfo {editor} {edited by\ \bibinfo {editor} {\bibfnamefont {A.}~\bibnamefont {Papadopoulos}}}\ (\bibinfo  {publisher} {European Mathematical Society},\ \bibinfo {address} {Z{\"u}rich},\ \bibinfo {year} {2012})\ pp.\ \bibinfo {pages} {667--716}\BibitemShut {NoStop}%
\bibitem [{\citenamefont {Liu}\ and\ \citenamefont {Xu}(2009)}]{Liu:2007au}%
  \BibitemOpen
  \bibfield  {author} {\bibinfo {author} {\bibfnamefont {K.}~\bibnamefont {Liu}}\ and\ \bibinfo {author} {\bibfnamefont {H.}~\bibnamefont {Xu}},\ }\href {\doibase 10.1093/imrn/rnn148} {\bibfield  {journal} {\bibinfo  {journal} {Int. Math. Res. Not.}\ }\textbf {\bibinfo {volume} {2009}},\ \bibinfo {pages} {835} (\bibinfo {year} {2009})},\ \Eprint {http://arxiv.org/abs/0708.0565} {arXiv:0708.0565 [math.AG]} \BibitemShut {NoStop}%
\bibitem [{\citenamefont {Lowenstein}(2024)}]{Lowenstein:2024fji}%
  \BibitemOpen
  \bibfield  {author} {\bibinfo {author} {\bibfnamefont {A.}~\bibnamefont {Lowenstein}},\ }\href@noop {} {\  (\bibinfo {year} {2024})},\ \Eprint {http://arxiv.org/abs/2407.16039} {arXiv:2407.16039 [hep-th]} \BibitemShut {NoStop}%
\bibitem [{\citenamefont {Johnson}\ \emph {et~al.}(1992)\citenamefont {Johnson}, \citenamefont {Morris},\ and\ \citenamefont {White}}]{Johnson:1992wr}%
  \BibitemOpen
  \bibfield  {author} {\bibinfo {author} {\bibfnamefont {C.~V.}\ \bibnamefont {Johnson}}, \bibinfo {author} {\bibfnamefont {T.~R.}\ \bibnamefont {Morris}}, \ and\ \bibinfo {author} {\bibfnamefont {P.~L.}\ \bibnamefont {White}},\ }\href@noop {} {\bibfield  {journal} {\bibinfo  {journal} {Phys. Lett.}\ }\textbf {\bibinfo {volume} {B292}},\ \bibinfo {pages} {283} (\bibinfo {year} {1992})},\ \Eprint {http://arxiv.org/abs/hep-th/9206066} {hep-th/9206066} \BibitemShut {NoStop}%
\bibitem [{\citenamefont {Johnson}(1994)}]{Johnson:1993vk}%
  \BibitemOpen
  \bibfield  {author} {\bibinfo {author} {\bibfnamefont {C.~V.}\ \bibnamefont {Johnson}},\ }\href {\doibase 10.1016/0550-3213(94)90430-8} {\bibfield  {journal} {\bibinfo  {journal} {Nucl. Phys. B}\ }\textbf {\bibinfo {volume} {414}},\ \bibinfo {pages} {239} (\bibinfo {year} {1994})},\ \Eprint {http://arxiv.org/abs/hep-th/9301112} {arXiv:hep-th/9301112} \BibitemShut {NoStop}%
\bibitem [{\citenamefont {Itzykson}\ and\ \citenamefont {Zuber}(1992)}]{Itzykson:1992ya}%
  \BibitemOpen
  \bibfield  {author} {\bibinfo {author} {\bibfnamefont {C.}~\bibnamefont {Itzykson}}\ and\ \bibinfo {author} {\bibfnamefont {J.~B.}\ \bibnamefont {Zuber}},\ }\href {\doibase 10.1142/S0217751X92002581} {\bibfield  {journal} {\bibinfo  {journal} {Int. J. Mod. Phys. A}\ }\textbf {\bibinfo {volume} {7}},\ \bibinfo {pages} {5661} (\bibinfo {year} {1992})},\ \Eprint {http://arxiv.org/abs/hep-th/9201001} {arXiv:hep-th/9201001} \BibitemShut {NoStop}%
\bibitem [{\citenamefont {Johnson}(2021)}]{Johnson:2020heh}%
  \BibitemOpen
  \bibfield  {author} {\bibinfo {author} {\bibfnamefont {C.~V.}\ \bibnamefont {Johnson}},\ }\href {\doibase 10.1103/PhysRevD.103.046012} {\bibfield  {journal} {\bibinfo  {journal} {Phys. Rev. D}\ }\textbf {\bibinfo {volume} {103}},\ \bibinfo {pages} {046012} (\bibinfo {year} {2021})},\ \Eprint {http://arxiv.org/abs/2005.01893} {arXiv:2005.01893 [hep-th]} \BibitemShut {NoStop}%
\bibitem [{\citenamefont {Dalley}\ \emph {et~al.}(1992{\natexlab{a}})\citenamefont {Dalley}, \citenamefont {Johnson},\ and\ \citenamefont {Morris}}]{Dalley:1991qg}%
  \BibitemOpen
  \bibfield  {author} {\bibinfo {author} {\bibfnamefont {S.}~\bibnamefont {Dalley}}, \bibinfo {author} {\bibfnamefont {C.~V.}\ \bibnamefont {Johnson}}, \ and\ \bibinfo {author} {\bibfnamefont {T.}~\bibnamefont {Morris}},\ }\href@noop {} {\bibfield  {journal} {\bibinfo  {journal} {Nucl. Phys.}\ }\textbf {\bibinfo {volume} {B368}},\ \bibinfo {pages} {625} (\bibinfo {year} {1992}{\natexlab{a}})}\BibitemShut {NoStop}%
\bibitem [{\citenamefont {Dalley}\ \emph {et~al.}(1992{\natexlab{b}})\citenamefont {Dalley}, \citenamefont {Johnson},\ and\ \citenamefont {Morris}}]{Dalley:1991vr}%
  \BibitemOpen
  \bibfield  {author} {\bibinfo {author} {\bibfnamefont {S.}~\bibnamefont {Dalley}}, \bibinfo {author} {\bibfnamefont {C.~V.}\ \bibnamefont {Johnson}}, \ and\ \bibinfo {author} {\bibfnamefont {T.}~\bibnamefont {Morris}},\ }\href@noop {} {\bibfield  {journal} {\bibinfo  {journal} {Nucl. Phys.}\ }\textbf {\bibinfo {volume} {B368}},\ \bibinfo {pages} {655} (\bibinfo {year} {1992}{\natexlab{b}})}\BibitemShut {NoStop}%
\bibitem [{\citenamefont {Dalley}\ \emph {et~al.}(1992{\natexlab{c}})\citenamefont {Dalley}, \citenamefont {Johnson}, \citenamefont {Morris},\ and\ \citenamefont {Watterstam}}]{Dalley:1992br}%
  \BibitemOpen
  \bibfield  {author} {\bibinfo {author} {\bibfnamefont {S.}~\bibnamefont {Dalley}}, \bibinfo {author} {\bibfnamefont {C.~V.}\ \bibnamefont {Johnson}}, \bibinfo {author} {\bibfnamefont {T.~R.}\ \bibnamefont {Morris}}, \ and\ \bibinfo {author} {\bibfnamefont {A.}~\bibnamefont {Watterstam}},\ }\href@noop {} {\bibfield  {journal} {\bibinfo  {journal} {Mod. Phys. Lett.}\ }\textbf {\bibinfo {volume} {A7}},\ \bibinfo {pages} {2753} (\bibinfo {year} {1992}{\natexlab{c}})},\ \Eprint {http://arxiv.org/abs/hep-th/9206060} {hep-th/9206060} \BibitemShut {NoStop}%
\bibitem [{\citenamefont {Norbury}(2024)}]{norbury2024superweilpeterssonmeasuresmoduli}%
  \BibitemOpen
  \bibfield  {author} {\bibinfo {author} {\bibfnamefont {P.}~\bibnamefont {Norbury}},\ }\href {https://arxiv.org/abs/2312.14558} {\enquote {\bibinfo {title} {Super weil-petersson measures on the moduli space of curves},}\ } (\bibinfo {year} {2024}),\ \Eprint {http://arxiv.org/abs/2312.14558} {arXiv:2312.14558 [math.AG]} \BibitemShut {NoStop}%
\bibitem [{\citenamefont {Alexandrov}\ and\ \citenamefont {Norbury}(2024)}]{Alexandrov:2024kuj}%
  \BibitemOpen
  \bibfield  {author} {\bibinfo {author} {\bibfnamefont {A.}~\bibnamefont {Alexandrov}}\ and\ \bibinfo {author} {\bibfnamefont {P.}~\bibnamefont {Norbury}},\ }\href@noop {} {\  (\bibinfo {year} {2024})},\ \Eprint {http://arxiv.org/abs/2412.17272} {arXiv:2412.17272 [math.AG]} \BibitemShut {NoStop}%
\bibitem [{\citenamefont {Norbury}(2023)}]{Norbury:2017eih}%
  \BibitemOpen
  \bibfield  {author} {\bibinfo {author} {\bibfnamefont {P.}~\bibnamefont {Norbury}},\ }\href {\doibase 10.2140/gt.2023.27.2695} {\bibfield  {journal} {\bibinfo  {journal} {Geom. Topol.}\ }\textbf {\bibinfo {volume} {27}},\ \bibinfo {pages} {2695} (\bibinfo {year} {2023})},\ \Eprint {http://arxiv.org/abs/1712.03662} {arXiv:1712.03662 [math.AG]} \BibitemShut {NoStop}%
\bibitem [{\citenamefont {Mertens}\ \emph {et~al.}(2017)\citenamefont {Mertens}, \citenamefont {Turiaci},\ and\ \citenamefont {Verlinde}}]{Mertens:2017mtv}%
  \BibitemOpen
  \bibfield  {author} {\bibinfo {author} {\bibfnamefont {T.~G.}\ \bibnamefont {Mertens}}, \bibinfo {author} {\bibfnamefont {G.~J.}\ \bibnamefont {Turiaci}}, \ and\ \bibinfo {author} {\bibfnamefont {H.~L.}\ \bibnamefont {Verlinde}},\ }\href {\doibase 10.1007/JHEP08(2017)136} {\bibfield  {journal} {\bibinfo  {journal} {JHEP}\ }\textbf {\bibinfo {volume} {08}},\ \bibinfo {pages} {136} (\bibinfo {year} {2017})},\ \Eprint {http://arxiv.org/abs/1705.08408} {arXiv:1705.08408 [hep-th]} \BibitemShut {NoStop}%
\bibitem [{\citenamefont {Witten}(1990)}]{Witten:1989ig}%
  \BibitemOpen
  \bibfield  {author} {\bibinfo {author} {\bibfnamefont {E.}~\bibnamefont {Witten}},\ }\href {\doibase 10.1016/0550-3213(90)90449-N} {\bibfield  {journal} {\bibinfo  {journal} {Nucl. Phys. B}\ }\textbf {\bibinfo {volume} {340}},\ \bibinfo {pages} {281} (\bibinfo {year} {1990})}\BibitemShut {NoStop}%
\bibitem [{\citenamefont {Witten}(1991)}]{Witten:1990hr}%
  \BibitemOpen
  \bibfield  {author} {\bibinfo {author} {\bibfnamefont {E.}~\bibnamefont {Witten}},\ }\href {\doibase 10.4310/SDG.1990.v1.n1.a5} {\bibfield  {journal} {\bibinfo  {journal} {Surveys Diff. Geom.}\ }\textbf {\bibinfo {volume} {1}},\ \bibinfo {pages} {243} (\bibinfo {year} {1991})}\BibitemShut {NoStop}%
\bibitem [{\citenamefont {Kontsevich}(1992)}]{Kontsevich:1992ti}%
  \BibitemOpen
  \bibfield  {author} {\bibinfo {author} {\bibfnamefont {M.}~\bibnamefont {Kontsevich}},\ }\href {\doibase 10.1007/BF02099526} {\bibfield  {journal} {\bibinfo  {journal} {Commun. Math. Phys.}\ }\textbf {\bibinfo {volume} {147}},\ \bibinfo {pages} {1} (\bibinfo {year} {1992})}\BibitemShut {NoStop}%
\bibitem [{\citenamefont {Do}\ and\ \citenamefont {Norbury}(2018)}]{Do:2016odu}%
  \BibitemOpen
  \bibfield  {author} {\bibinfo {author} {\bibfnamefont {N.}~\bibnamefont {Do}}\ and\ \bibinfo {author} {\bibfnamefont {P.}~\bibnamefont {Norbury}},\ }\href {\doibase 10.4310/CNTP.2018.v12.n1.a2} {\bibfield  {journal} {\bibinfo  {journal} {Commun. Num. Theor. Phys.}\ }\textbf {\bibinfo {volume} {12}},\ \bibinfo {pages} {53} (\bibinfo {year} {2018})},\ \Eprint {http://arxiv.org/abs/1608.02781} {arXiv:1608.02781 [math-ph]} \BibitemShut {NoStop}%
\bibitem [{\citenamefont {Chidambaram}\ \emph {et~al.}(2025)\citenamefont {Chidambaram}, \citenamefont {Garcia-Failde},\ and\ \citenamefont {Giacchetto}}]{Chidambaram:2022cqc}%
  \BibitemOpen
  \bibfield  {author} {\bibinfo {author} {\bibfnamefont {N.~K.}\ \bibnamefont {Chidambaram}}, \bibinfo {author} {\bibfnamefont {E.}~\bibnamefont {Garcia-Failde}}, \ and\ \bibinfo {author} {\bibfnamefont {A.}~\bibnamefont {Giacchetto}},\ }\href {\doibase 10.1007/s00222-025-01351-y} {\bibfield  {journal} {\bibinfo  {journal} {Invent. Math.}\ }\textbf {\bibinfo {volume} {241}},\ \bibinfo {pages} {929} (\bibinfo {year} {2025})},\ \Eprint {http://arxiv.org/abs/2205.15621} {arXiv:2205.15621 [math.AG]} \BibitemShut {NoStop}%
\bibitem [{\citenamefont {Kazarian}\ and\ \citenamefont {Norbury}(2024)}]{Kazarian:2021zqh}%
  \BibitemOpen
  \bibfield  {author} {\bibinfo {author} {\bibfnamefont {M.}~\bibnamefont {Kazarian}}\ and\ \bibinfo {author} {\bibfnamefont {P.}~\bibnamefont {Norbury}},\ }\href {\doibase 10.1093/imrn/rnad061} {\bibfield  {journal} {\bibinfo  {journal} {Int. Math. Res. Not.}\ }\textbf {\bibinfo {volume} {2024}},\ \bibinfo {pages} {1825} (\bibinfo {year} {2024})},\ \Eprint {http://arxiv.org/abs/2112.11672} {arXiv:2112.11672 [math.AG]} \BibitemShut {NoStop}%
\bibitem [{\citenamefont {Johnson}\ and\ \citenamefont {Usatyuk}(2025)}]{Johnson:2024tgg}%
  \BibitemOpen
  \bibfield  {author} {\bibinfo {author} {\bibfnamefont {C.~V.}\ \bibnamefont {Johnson}}\ and\ \bibinfo {author} {\bibfnamefont {M.}~\bibnamefont {Usatyuk}},\ }\href {\doibase 10.1007/JHEP09(2025)164} {\bibfield  {journal} {\bibinfo  {journal} {JHEP}\ }\textbf {\bibinfo {volume} {09}},\ \bibinfo {pages} {164} (\bibinfo {year} {2025})},\ \Eprint {http://arxiv.org/abs/2407.17583} {arXiv:2407.17583 [hep-th]} \BibitemShut {NoStop}%
\bibitem [{\citenamefont {Delecroix}\ \emph {et~al.}(2021)\citenamefont {Delecroix}, \citenamefont {Schmitt},\ and\ \citenamefont {van Zelm}}]{Delecroix_2021}%
  \BibitemOpen
  \bibfield  {author} {\bibinfo {author} {\bibfnamefont {V.}~\bibnamefont {Delecroix}}, \bibinfo {author} {\bibfnamefont {J.}~\bibnamefont {Schmitt}}, \ and\ \bibinfo {author} {\bibfnamefont {J.}~\bibnamefont {van Zelm}},\ }\href {\doibase 10.2140/jsag.2021.11.89} {\bibfield  {journal} {\bibinfo  {journal} {Journal of Software for Algebra and Geometry}\ }\textbf {\bibinfo {volume} {11}},\ \bibinfo {pages} {89–112} (\bibinfo {year} {2021})}\BibitemShut {NoStop}%
\end{thebibliography}%

\end{document}